 \documentclass[preprint2]{aastex}

\slugcomment{Not to appear in Nonlearned J., 45.}

\shorttitle{electron-capture rates of nuclei}
\shortauthors{Jing-Jing et al.}

\begin{document}


\title{A new insight into neutrino energy loss by electron capture of iron group nuclei in magnetars surface}


\author{Jing-Jing Liu \altaffilmark{1,2} and Wei-Min Gu\altaffilmark{1}}
\email{syjjliu68@qzu.edu.cn; guwm@xmu.edu.cn}


\altaffiltext{1}{Department of Astronomy and Institute of
Theoretical Physics and Astrophysics, Xiamen University, Xiamen,
Fujian 361005, China} \altaffiltext{2}{College of Electronic and
Communication Engineering, Hainan Tropical Ocean University, Sanya,
572022, China}


\begin{abstract}

Based on the relativistic mean-field effective interactions theory,
and Lai dong model \citep{b37, b38, b39}, we discuss the influences
of superstrong magnetic fields (SMFs) on electron Fermi energy,
nuclear blinding energy, and single-particle level structure in
magnetars surface. By using the method of Shell-Model Monte Carlo
(SMMC), and the Random Phase Approximation (RPA) theory, we detailed
analyze the neutrino energy loss rates(NELRs) by electron capture
(EC) for iron group nuclei in SMFs. Firstly, when $ B_{12} <100$, we
find that the SMFs has a slight influence on the NELR for most
nuclides at relativistic low temperature(e.g., $T_9=0.233$),
nevertheless, the NELRs increases by more than four orders of
magnitude at relativistic high temperature(e.g., $T_9=15.53$). When
$B_{12}>100$, the NELRs decreases by more than three orders of
magnitude  (e.g., at $T_9=15.53$ for $^{52-61}$Fe, $^{55-60}$Co and
$^{56-63}$Ni). Secondly, for a certain value of magnetic field and
temperature, the NELRs increases by more than four orders of
magnitude when $\rho_7\leq 10^3$, but as the density increases(i.e.
when $\rho_7> 10^3$), there are almost not influence of density on
NELRs. For the density around $\rho_7= 10^2$, there is an abrupt
increase in NELRs when $ B_{12}\geq10^{3.5}$. Such jumps are an
indication that the underlying shell structure has changed due to
single-particle behavior by SMFs. Finally, we compare our NELRs with
those of \citet{b20, b21} (FFN), and \citet{b48} (NKK). For the case
without SMFs, one finds that our rates for certain nuclei are about
close to five orders magnitude lower than FFN, and NKK at
relativistic low temperature (e.g., $T_9=1$). However, at the
relativistic high temperature (e.g., $T_9=3$), our results are in
good agreement with NKK, but about one order magnitude lower than
FFN. For the case with SMFs, our NELRs for some iron group nuclei
can be about five orders of magnitude higher than those of FFN, and
NKK. (note $B_{12}, T_9$, and $\rho_7$ are in unit of
$10^{12}\rm{G}, 10^9\rm{K}$, and $10^7\rm{g/cm^3}$, respectively)

\end{abstract}


\keywords{stars: magnetic fields --- stars: neutron --- Physical
Date and Processes: nuclear reactions.}



\section{Introduction}

In the process of core-collapse supernova explosions, neutrino
processes and weak interaction (e.g. electron capture(hereafter EC)
and beta decay) play pivotal roles. At the late stages of evolution,
a large amount of energy from a stellar is lost mainly through
neutrinos and this process is fairly independent of the mass of
star. For instance, White dwarfs and supernovae, both have cooling
rates largely dominated by neutrino production. At high temperatures
and densities, the EC and the accompanying neutrino energy loss
rates (hereafter NELRs) are of prime importance in determining the
equation of state of supernova. An accurate determination of
neutrino emission rates is very necessary in order to perform a
careful analysis of the final branches of stellar evolutionary
tracks. The neutrino cooling rates can strongly influence on the
evolutionary time scale and the configuration of iron core at the
onset of the supernova explosion.

Some researches for cooling of white dwarfs, neutron stars, and
magnetars need us giving a accurate estimation of some weak
interaction rates and NELRs. Due to the importance of the EC and
NELRs in astrophysical environments, \citet{ b7, b8}; \citet{b3,
b16, b19, b20, b21, b27, b29, b32, b33, b42, b43, b44, b45, b46}
have done some pioneering works on the weak interaction reactions
and NELRs for some iron group nuclei. Recently, by using the
proton-neutron quasi-particle random phase approximation (pn-QRPA)
theory, \citet{b47, b48, b49} have detailed investigated the
neutrino and anti-neutrino energy loss rates. However, their works
seemed to pay no attention to the influence of superstrong magnetic
fields (hereafter SMFs)on NELRs.

The neutrino processes will play a crucial role in magnetars and
some neutron stars due to electron capture (EC) and beta decay. A
great deal of energies and messages set free with the escape of the
neutrino. Thus, the works on neutrino and the NELRs have been the
hotspot and former-border issue in magnetars and some neutron stars.
In order to understand the nature of the weak interaction process
and NELRs in magnetars, the study of matter in SMFs is obviously an
important component of magnetars in astrophysics research. Some
works show that the strengths of magnetic fields at the crust of
neutron stars are in the range of $10^{8}-10^{13}$G \citep{b6, b14}.
There has recently been growing evidence for the existence of
neutron stars possessing magnetic fields with strengths that exceed
the quantum critical field strength of $4.4 \times 10^{13}$ G. Such
evidence has been provided by new discoveries of radio pulsars
having very high spin-down rates and by observations of bursting
gamma-ray sources termed magnetars. In addition, some researches
show that magnetars possess magnetic fields as strong as from
$10^{13}$G to $10^{15}$G \citep{b22, b23, b26, b37, b38, b39, b53}.
Although it is not clear how strong magnetic fields in magnetars
could be, some calculations indicate that fields of
$10^{15}-10^{16}$ G are not impossible. Theoretical models display
that these magnetic fields might reach up to $10^{18}$ G, and even
larger values when one considers the limit imposed by the virial
theorem \citep{b50}. For such SMFs, the classical description of the
trajectories of a free electron is no longer valid and quantum
effect must be considered.

Previous works \citep{b40, b41} show that SMFs influences on
electron capture rate and NELRs greatly and decreases with the
increasing of the strength of magnetic field. Recent studies
\citep{b37, b38, b39} have found the magnetic field will make the
Fermi surface would elongate from a spherical surface to a Landau
surface along the magnetic field direction and its level is
perpendicular to the magnetic field direction and strongly
quantized. The properties of matter (such as atoms, molecules, and
condensed matter/plasma) are dramatically changed by the superstrong
magnetic fields in magnetars. Recently, due to the influence of
SMFs, the electron cyclotron kinetic energy will be greater than
electron-static energy in magnetars. The microscopic state number of
electron momentum along the field in the interval of $P_z\sim
P_z+dP_z$, is obtained via using the non-relativistic motion
equation. Based on the above mentioned conditions, we detailed
discuss some properties of matter of magnetar surface due to these
are significantly modified by strong magnetic fields. We also
analyze the influences of SMFs on NELRs from the EC reaction for
some iron group nuclei.

Our work differs from previous works \citep{b19, b20, b21} about the
discussion of electron capture process and NELRs. Firstly, their
works are based on the theory of Brink Hypothesis by FFN and
overlooked the influence of SMFs on EC and NELRs. Secondly, our work
also differs from the works of \citet{b48, b49}. they detailed
discussed the NELRs by using the quasi-particle random phase
approximation theory only in the case without SMFs. We analyze the
EC process and NELRs for iron group nuclei and derive new results in
SMFs according to the Shell-Model Monte Carlo (SMMC) method
\citep{b29, b32}. Although \citet{b29, b32} detailed discussed the
EC in presupernova surrounding, but they have lost sight of the
influence of SMFs on the EC process and NELR. We make detailed
comparison of our results in SMFs with those of FFN, and \citep{b48}
(hereafter NKK), which are in the case without SMFs. Finally, our
discussions also differs from recent works of \citet{b40, b41},
which analyzed the EC and NELRs by using the method of Brink
Hypothesis. The Brink Hypothesis is a very crude approximation,
which assumes that the Gamow-Teller strength distribution on excited
states is the same as for the ground state, only shifted by the
excitation energy of the state. By using the method of SMMC, and the
RPA theory, we discuss NELRs by EC of some typical iron group nuclei
basing on Lai dong model \citep{b37, b38, b39}in SMF. Based on the
relativistic mean-field effective interactions theory, we discuss
the influences of SMFs on electron Fermi energy, blinding energy per
nuclei, and single-particle level structure in magnetars surface. We
also detailed compare our results in SMFs with those of FFN, and
NKK. These NELRs of iron group nuclei may be universal, very
important and helpful for the researches of thermal and magnetic
evolution, particularly for study of cooling mechanisms, and
numerical simulation of the neutron stars and magnetars.

The present paper is organized as follows. In the next Section, we
analyze the influence of SMFs on the electron property and nuclear
energy in magnetars. The studies about the EC process and NELRs in
SMFs will be given in Section 3. Some numerical results and
discussion are given in Section 4. And some conclusions are
summarized in the last section.

\section[]{The influence of SMFs in magnetars}

\subsection[]{How will the SMFs influence on the electron properties in magnetar surface}

The properties of matter are significantly modified by strong
magnetic fields, such as the states equation, electron energy, the
outer crust structure and composition in neutron stars. Many works
\citep{b36, b11} detailed presented the quantum mechanics of a
charged particle in SMFs. When we consider the nonrelativistic
motion of a particle (charge $e_i$ and mass $m_i$) in a uniform
magnetic field, which is assumed to be along the z-axis,  the
circular orbit radius and (angular) frequency in the process of
particle gyrates are given by $r=m_icv_{\bot}/|e_i|B$,
$\omega_c=|e_i|B/m_ic$, respectively, here $v_{\bot}$ is the
velocity perpendicular to the magnetic field. The kinetic energy for
electron ($m_i\rightarrow m_e$, $e_i\rightarrow -e$)of the
transverse motion is quantized in Landau levels in non-relativistic
quantum mechanics, and is written by
\begin{equation}
E_{kin}=\frac{1}{2}m_iv^2_{\bot}\rightarrow
(n_l+\frac{1}{2})\hbar\omega_c
 \label{1}
\end{equation}
where $n_l=0,1,2...$ is  Landau levels number. The cyclotron energy
for a electron, which is the basic energy quantum is given by

\begin{equation}
E_{cyc}=\hbar\omega_c=\hbar\frac{eB}{ec}=11.577B_{12} \rm{keV},
 \label{2}
\end{equation}
where $B_{12}=B/10^{12}\rm{G}$ is the magnetic field strength in
units of $10^{12}$ Gauss, The total electron energy, which is
included the kinetic energy associated with the z-momentum ($p_z$)
and the spin energy in non-relativistic quantum surrounding can be
written as \citep{b36, b11}
\begin{equation}
E_{n}=\nu \hbar\omega_{c}+\frac{p_z^2}{2m_e},
 \label{3}
\end{equation}
where $\nu=n_l+(1+\sigma_z)/2$. $\sigma_z=-1$, $\sigma_z=\pm 1$ are
the spin degeneracy for the ground Landau level($n_l=0$), and
excited levels, respectively.

We define a critical magnetic field strength $B_{cr}$ from the
relation of $\hbar\omega_{c}=m_ec^2$ (i.e.
$B_{cr}=m_e^2c^3/e\hbar=4.414\times 10^3$G). The transverse motion
of the electron becomes relativistic when $\hbar\omega_c\geq m_ec^2$
(i.e. $B\geq B_{cr}$) for extremely strong magnetic fields. The
energy eigenstates of electron must obey the relativistic Dirac
equation and is given by \citep{b11, b31}
\begin{equation}
E_n=[c^2p_z^2+m_e^2c^4(1+2\nu \frac{B}{B_{cr}})]^{1/2},
 \label{4}
\end{equation}
where the shape of the Landau wavefunction in the relativistic
theory is the same as in the nonrelativistic theory due to the fact
the cyclotron radius is independent of the particle mass.
\citep{b53}

In SMFs the number density $n_{\rm{e}}$ of electrons is related to
the chemical potential $U_{\rm{F}}$ by \citep{b37, b38, b39, b26}
\begin{equation}
n_{\rm{e}}^{\rm{B}}=\frac{1}{(2\pi\widehat{\rho})^2\hbar}\sum_0^\infty
g_{n0} \int_{-\infty} ^{+\infty} f dp_{\rm{z}},
 \label{5}
\end{equation}
where $\widehat{\rho}=(\hbar
c/eB)^{1/2}=2.5656\times10^{-10}B_{12}^{1/2}\rm{cm}$ is the
cyclotron radius (the characteristic size of the wave packet), and
$g_{n0}$ is the spin degeneracy of the Landau level, $g_{00}=1$ and
$g_{n0}=2$ for $n\geqslant 1$, and $f=[1+\exp((E_n-U_F)/kT)]^{-1}$
is the Fermi-Dirac distribution.

According to the relation of the usual relativistic energy and
momentum from Eq.(4), the interaction energy term, which is
proportional to the quantum number $\nu$, and cannot exceed the
electron chemical potential, will appears due to the electron
interaction with the magnetic field. Thus the maximum number of
Landau levels $\nu_{max}$, related to the highest value of the
allowed interaction energy should be satisfied with $E_n(\nu_{max},
p_z=0)=U_F$. So we have
\begin{equation}
\nu_{max}=\frac{1}{2}\frac{B_{cr}}{B}(\frac{U_F^2}{m_e^2c^4}-1),
\label{6}
\end{equation}

However, in the general case (i.e. $0\leq \nu \leq \nu_{max}$), when
the maximum electron momentum is equaled to the Fermi momentum $P_F$
for different Landau level value $\nu$, the electron chemical
potential from Eq.(4) can be computed as follows
\begin{equation}
E_n(\nu)=[c^2p_F^2+m_e^2c^4(1+2\nu \frac{B}{B_{cr}})]^{1/2}=U_F,
 \label{7}
\end{equation}

When we define a nondimensional Fermi momentum $x_e(\nu) =
p_F/m_ec$, and Fermi energy $\gamma_e = U_F/m_ec^2$, the electronic
density, electron energy, and pressure can be written as \citep{b37}
\begin{equation}
n_{e}^B=\frac{B}{2\pi^2B_{cr}\lambda_e^3}\sum_{\nu=0}^{\nu_{max}}g_{n0}x_e(\nu),
 \label{8}
\end{equation}
\begin{equation}
\varepsilon_{e}=\frac{Bm_ec^2}{2\pi^2B_{cr}\lambda_e^3}\sum_{\nu=0}^{\nu_{max}}g_{n0}(1+2\nu\frac{B}{B_{cr}})\vartheta
_{+}[\frac{x_e(\nu)}{(1+2\nu B/B_{cr})^{1/2}}],
 \label{9}
\end{equation}
\begin{equation}
P_{e}=\frac{Bm_ec^2}{2\pi^2B_{cr}\lambda_e^3}\sum_{\nu=0}^{\nu_{max}}g_{n0}(1+2\nu\frac{B}{B_{cr}})\vartheta_{-}
[\frac{x_e(\nu)}{(1+2\nu B/B_{cr})^{1/2}}],,
 \label{10}
\end{equation}
where
$\vartheta_{\pm}(x)=\frac{1}{2}x\sqrt{1+x^2}\pm\frac{1}{2}\ln(x+\sqrt{1+x^2})$,
$\lambda_{\rm{e}}=\hbar/m_{\rm{e}}c$ is the electron Compton
wavelength.

\subsection[]{How will the SMFs influence on the energy of nucleus in magnetar surface}
The matter in the outer crust of a cold ($T=0$K) magnetar consists
of a Coulomb lattice of completely ionized atoms and a uniform Fermi
gas of relativistic electrons. The Gibbs free energy per baryon
$g(A, Z, P)$ at a constant pressure and zero temperature can be
given by
\begin{equation}
 g(A, Z, P)=\frac{E(A,Z,P)+PV}{A}=\varepsilon+\frac{P}{n},
 \label{11}
\end{equation}
where $\varepsilon$ is the corresponding energy per nucleon, $n=A/V$
is the baryon density in a cell, and $V$ is the volume occupied by a
unit cell of the Coulomb lattice. The energy per nucleon
$\varepsilon$, which consists of three different contributions from
nuclear, electronic, and lattice is given by
\begin{equation}
 \varepsilon=\varepsilon_n(A, Z)+\varepsilon_e(A, Z, P)+\varepsilon_l(A, Z, n),
 \label{12}
\end{equation}
where the nuclear contribution to the total energy per nucleon is
simple and independent of the density and is written by
\begin{equation}
 \varepsilon_n(A, Z)=\frac{M(A, Z)}{A}=\frac{1}{A}[zm_p+(A-Z)m_n-\varepsilon_{bind}(A, Z)],
 \label{13}
\end{equation}
where $M(A, z)$ is the nuclear mass, $\varepsilon_{bind}(A, Z)$ is
the corresponding binding energy, and $m_n$ and $m_p$ are neutron
and proton masses, respectively.

Based on the relativistic mean-field effective interactions theory
of NL3 \citep{b34} and DD-ME2 \citep{b35}, \citet{b51} detailed
discussed the influence of SMFs on the nuclear binding energies.
Based on a covariant density functional, an effective Lagrangian
with nucleons and mesons is given by simple and independent of the
density and is written by \citep{b24, b59}
\begin{equation}
 L=L_N+L_m+L_{int}+L_{BO}+L_{BM},
 \label{14}
\end{equation}
where $L_N$, $L_m$, and $L_{int}$ are the Lagrangian of the free
nucleon, the free meson fields and the electromagnetic field
generated by the proton, and the Lagrangian describing the
interactions, respectively. These Lagrangian are represented by
\begin{equation}
 L_N=\bar{\psi}(i\gamma ^\mu \partial_{\mu}-m_{nu})\psi,
 \label{15}
\end{equation}
\begin{eqnarray}
 L_m=\frac{1}{2}\partial_{\mu} \sigma \partial^{\mu}\sigma-\frac{1}{2}m_{\sigma}\sigma^2-\frac{1}{4}\Omega_{\mu
 \nu}\Omega^{\mu\nu}+\frac{1}{2}m^2_{\omega}\omega_{\mu}\omega^{\mu} \nonumber\\
 -\frac{1}{4}\vec{R}_{\mu \nu}\vec{R}^{\mu\nu}+\frac{1}{2}m^2_{\rho}\vec{\rho}_{\mu}\vec{\rho}^{\mu}-\frac{1}{4}F_{\mu
 \nu}F^{\mu\nu}-U(\sigma),
 \label{16}
\end{eqnarray}
\begin{equation}
 L_{int}=-g_{\sigma}\bar{\psi} \sigma\psi-g_{\omega}\bar{\psi}\gamma^{\mu}\omega_{\mu}
 \sigma\psi-g_{\rho}\bar{\psi} \gamma^{\mu} \vec{\tau} \vec{\rho}^{\mu}\psi-e \bar{\psi}\gamma^{\mu}A_{\mu} \psi,
 \label{17}
\end{equation}
where $\psi$ is the Dirac spinor. $m_{nu}$, and $m_{\sigma},
m_{\omega}, m_{\rho},$ are the nucleon and meson masses,
respectively. $U(\sigma)=(g_2/3)\sigma^3+g_3/4)\sigma^4$ is the
standard form for the nonlinear coupling of the $\sigma$ meson
field. $g_{\sigma}, g_{\omega}, g_{\rho}, e$ are the  coupling
constants for the $\sigma, \omega, \rho,$ and the electric charge
for photon, respectively.

The coupling of the proton orbital motion with the external magnetic
field, and the coupling of protons and neutrons intrinsic dipole
magnetic moments with the external magnetic field can be respective
expressed as \citep{b10}
\begin{equation}
 L_{BO}=e \bar{\psi}\gamma^{\mu}A_{\mu}^{(ext)} \psi,
 \label{18}
\end{equation}
\begin{equation}
 L_{BM}=-\bar{\psi}\chi_{\tau_3}^{(ext)} \psi,
 \label{19}
\end{equation}
where $\chi_{\tau_3}^{(ext)}=\kappa_{\tau_3}\mu
N\frac{1}{2}\sigma_{\mu\nu}F^{(ext)\mu\nu}$, $F^{(ext)\mu\nu}$ is
the external field strength tensor.
$\sigma_{\mu\nu}=\frac{i}{2}[\gamma^{\mu}, \gamma^{\nu}]$,
$\mu_N=e\hbar/2m$ is the nuclear magneton, $\kappa_n=g_n/2$,
$\kappa_p=g_p/2-1$ (here $g_n=-3.8263, g_p=5.5856$)are the intrinsic
magnetic moments of protons and neutrons, respectively.

The electronic contribution is modeled as a degenerate free Fermi
gas and is given by
\begin{eqnarray}
 \varepsilon_{e}(A, z, P)=\frac{1}{n\pi^2}\int_0^{p_F}p^2\sqrt{p^2+m_e^2c^4}dp\nonumber\\
=\frac{m_e^4c^8}{8n\pi^2}[x_e(\nu)\gamma_e(x_e^2(\nu)+\gamma^2_e)-\ln(x_e(\nu)+\gamma_e)]
 \label{20}
\end{eqnarray}

The lattice energy per baryon $\varepsilon_l(A, Z, n)$ may be
written as \citep{b56}
\begin{equation}
 \varepsilon_l(A, Z, n)=-1.81962\frac{(ze)^2}{a}=-C_{bcc}\frac{z^2}{A^{4/3}}p_F,
 \label{21}
\end{equation}
where $C_{bcc}=3.40665\times10^{-3}$, and $a$ is the lattice
constant. Similar calculations such as faced-centered cubic or
simple cubic ones can be carried out evaluating different lattice
configurations.

\section[]{The NELRs due to EC in magnetars}

\subsection{The SMMC method and Gamow-Teller response functions}

The Gamow-Teller (GT) properties of nuclei in the medium mass region
of the periodic table are crucial determinants in the  process of
electron capture. Some works \citep{b17, b11} demonstrated that the
 GT transition matrix elements for electron capture and beta
decay don't depend on the magnetic fields. Thus we will neglect the
effect of SMFs on the GT properties of nuclei in this paper. The
SMMC method is based on a statistical formulation of the nuclear
many-body problem. We use the shell model Monte Carlo (SMMC)
approach to find the GT strength distributions. SMMC has the added
advantage that it treats nuclear temperature exactly. Based on a
statistical formulation of the nuclear many-body problem, in the
finite-temperature version of this approach, an observable is
calculated as the canonical expectation value of a corresponding
operator $\hat{A}$ by the SMMC method at a given temperature $T$,
and is written by \citep{b28, b50, b1, b30}
\begin{equation}
\hat{A}=\frac{\rm{Tr_A[\hat{A}e^{-\beta\hat{\emph{H}}}]}}{\rm{Tr_A[e^{-\beta\hat{\emph{H}}}]}},
 \label{eq:22}
\end{equation}

The problem of shell model Hamiltonian $\hat{\emph{H}}$ have been
detailed investigated by \citet{b1}. When some many-body Hamiltonian
$\hat{\emph{H}}$ is given, a tractable expression for the imaginary
time evolution operator is written by
\begin{equation}
 \hat{U}=\exp^{-\beta \hat{\emph{H}}},
 \label{eq:23}
\end{equation}
where $\beta=1/T_{\rm{N}}$, $T_{\rm{N}}$ is the nuclear temperature
in units of Mev. $\rm{Tr_A}\hat{\emph{U}}$ is the canonical
partition function for $\emph{A}$ nucleons. In terms of a spectral
expansion, the total strength of a transition operator $\hat{A}$ is
then given by the following expectation value:
\begin{equation}
 B(\emph{A})\equiv\langle \hat{A}^{\dagger}\hat{A} \rangle=\frac{\sum_{i,f}e^{-\beta E_i}|\langle f|\hat{A}|i\rangle|^2}{\sum_{i}e^{-\beta E_i}},
 \label{eq:24}
\end{equation}
here $|i\rangle$, $|f\rangle$ are the many-body states of the
initial (final) nucleus with energy $E_i, E_f$, respectively.

The SMMC method is used to calculate the response function
$R_{\rm{A}}(\tau)$ of an operator $\hat{{\rm{A}}}$ at an
imaginary-time $\tau$. By using a spectral distribution of initial
and final states $|i\rangle$ and $|f\rangle$ with energies $E_i$ and
$E_f$. $R_{\rm{A}}(\tau)$ is given by \citep{b13, b32, b33}
\begin{eqnarray}
R_{\rm{A}}(\tau)&\equiv \langle \hat{A}^{\dagger}(\tau)\hat{A}(0)
\rangle= \frac{\rm{Tr_A[e^{-(\beta-\tau)\hat{\emph{H}}}}\hat{A}^\dag e^{-\tau \hat{\emph{H}}}\hat{A}]}{\rm{Tr_A[e^{-\beta\hat{\emph{H}}}]}} \nonumber\\
 &=\frac{\sum_{if}(2J_i+1)e^{-\beta E_i}e^{-\tau
(E_f-E_i)}|\langle f|\hat{A}|i\rangle|^2}{\sum_i (2J_i+1)e^{-\beta
E_i}},
 \label{eq:25}
\end{eqnarray}

Note that the total strength for the operator is given by
$R(\tau=0)$. $S_{{\rm{GT}}^+}$ is the total amount of GT strength
available for an initial state is given by summing over a complete
set o final states in GT transition matrix elements
$|M_{{\rm{GT}}}|^{2}_{if}$. The strength distribution is given by
\citep{b13}
\begin{small}
\begin{eqnarray}
 S_{{\rm{GT}}^+}(E) &=& \frac{\sum_{if}\delta
(E-E_f+E_i)(2J_i+1)e^{-\beta
E_i}|\langle f|\hat{A}|i\rangle|^2}{\sum_i (2J_i+1)e^{-\beta E_i}} \nonumber\\
&=& S_{{\rm{A}}}(E),
 \label{eq:26}
\end{eqnarray}
\end{small}
which is related to $R_{\rm{A}}(\tau)$ by a Laplace Transform,
$R_{\rm{A}}(\tau)=\int_{-\infty}^{\infty}S_{\rm{A}}(E)e^{-\tau
E}dE$. Note that here $E$ is the energy transfer within the parent
nucleus, and that the strength distribution $S_{{\rm{GT}}^+}(E)$ has
units of $\rm {Mev^{-1}}$.

\subsection{The NELRs and EC process in the case without SMFs}
Based on the RPA theory with a global parameterization of the single
particle numbers, the stellar electron capture rates which is
related to the electron capture cross-section for the $k$ th nucleus
(Z, A) in thermal equilibrium at temperature $T$ is given by a sum
over the initial parent states $i$ and the final daughter states $f$
in the case without SMFs \citep{b13, b32, b33}
\begin{equation}
\lambda_{ec}^0=\frac{1}{\pi^2\hbar^3}\sum_{if}\int^{\infty}_{\varepsilon_0}
p^2_e\sigma_{ec}(\varepsilon_e,\varepsilon_i,\varepsilon_f)f(\varepsilon_e,U_F,T)d\varepsilon_e
\label{eq.27}
\end{equation}
where $\varepsilon_0=\max(Q_{if}, m_ec^2)$.
$p_e=\sqrt{\varepsilon_e^2-m_e^2c^4}$ is the momenta of the incoming
electron, and
 $\varepsilon_e$ is the total rest mass and kinetic energies of the
incoming electron, $U_{F}$ is the electron chemical potential, $T$
is the electron temperature. The electron Fermi-Dirac distribution
is defined as
\begin{equation}
  f=f(\varepsilon_{e},U_F,T)=[1+\exp(\frac{\varepsilon_{e}-U_F}{kT})]^{-1}
\label{eq.28}
\end{equation}

Due to the energy conservation, the electron, proton and neutron
energies are related to the neutrino energy, and $\rm{Q}$-value for
the capture reaction \citep{b12, b29}
\begin{equation}
  Q_{i,f}=\varepsilon_{e}-\varepsilon_{\nu}=\varepsilon_{n}-\varepsilon_{\nu}=\varepsilon^{n}_{f}-\varepsilon^{p}_{i}
\label{eq.29}
\end{equation}
and we have
\begin{equation}
  \varepsilon^{n}_{f}-\varepsilon^{p}_{i}=\varepsilon^{\ast}_{if}+\hat{\mu}+\Delta_{np}
\label{eq.30}
\end{equation}
where $\hat{\mu}=\mu_{n}-\mu_p$, the difference between neutron and
proton chemical potentials in the nucleus and
$\Delta_{np}=M_{n}c^2-M_{p}c^2=1.293Mev$, the neutron and the proton
mass difference. $Q_{00}=M_{f}c^2-M_{i}c^2=\hat{\mu}+\Delta_{np}$,
with $M_{i}$ and $M_{f}$ being the masses of the parent nucleus and
the daughter nucleus respectively; $\varepsilon^{\ast}_{if}$
corresponds to the excitation energies in the daughter nucleus at
the states of the zero temperature.

The electron chemical potential is found by inverting the expression
for the lepton number density \citep{b4, b19, b20, b21}
\begin{equation}
  n_e=\frac{8\pi}{(2\pi)^3}\int^\infty_0 p^2_e(f_{-e}-f_{+e})dp_e
\label{eq.31}
\end{equation}
where $f_{-e}=[1+\exp((\varepsilon_{e}-U_{F})/kT)]^{-1}$ and
$f_{+e}=[1+\exp((\varepsilon_{e}+U_{F})/kT)]^{-1}$ are the electron
and positron  distribution functions respectively, $k$ is the
Boltzmann constant.

According to the Shell-Model Monte Carlo method, which discussed the
GT strength distributions, the total cross section by EC is given by
\citep{b13, b29}
\begin{eqnarray}
\sigma_{ec}&=& \sigma_{ec}(E_e)=\sum_{if}\frac{(2J_{i}+1)\exp(-\beta E_i)}{Z_A}\sigma_{fi}(E_e)\nonumber\\
&=& 6g^{2}_{wk}\int d\xi(E_{e}-\xi)^2 \frac{G^2_A}{12\pi}
S_{GT^+}(\xi) F(Z,\varepsilon_e)\nonumber\\
\label{eq.32}
\end{eqnarray}
where $\beta=1/T_N$ is the inverse temperature, $T_N$ is the nuclear
temperature and in unit of Mev, and $E_e=\varepsilon_e$ is the
electron energy. $S_{GT^+}$ is the GT strength distribution, which
is as a function of the transition energy $\xi$. The
$g_{wk}=1.1661\times 10^{-5}\rm{Gev^{-2}}$ is the weak coupling
constant and $G_A$ is the axial vector form-factor which at zero
momentum is $G_A=1.25$. $F(Z, \varepsilon_e)$ is the Coulomb wave
correction which is the ratio of the square of the electron wave
function distorted by the coulomb scattering potential to the square
of wave function of the free electron.

By folding the total cross section with the flux of a degenerate
relativistic electron gas, the NELRs due to EC in the case without
SMFs is given by
\begin{eqnarray}
\lambda_{\rm{NEL}}^0 =\frac{\ln2}{6163}\int^{\infty}_{0}d\xi S_{GT}\frac{c^3}{(m_{e}c^2)^5}\nonumber\\
 \int^{\infty}_{p_0}dp_{e}p^2_e(-\xi+\varepsilon_e)^3
F(Z,\varepsilon_e)f(\varepsilon_e, U_F, T)~~~(\rm{s}^{-1})
 \label{eq.33}
\end{eqnarray}
where the $\xi$ is the transition energy of the nucleus, and
$f(\varepsilon_n, U_F, T)$ is the electron distribution function.
The $p_0$ is defined as
\begin{equation}
p_0=\left\lbrace \begin{array}{ll}~\sqrt{Q^2_{if}-m_e^2c^4}~~~~~~( Q_{if}<-m_ec^2)\\
                                  ~0 ~~~~~~(\rm{otherwise}).
                             \end{array} \right.
\label{eq.34}
\end{equation}

\subsection{The NELRs due to EC process in the case with SMFs}

The NELRs due to EC in an SMFs from one of the initial states to all
possible final states is given by
\begin{equation}
\lambda_{\rm{NEL}}^{\rm{B}}=\frac{\ln2}{6163}\int^{\infty}_{0}d\xi
S_{GT}\frac{c^3}{(m_{e}c^2)^5} f_{if}^{\rm{B}}.
 \label{eq.35}
\end{equation}

According to the method of SMMC and RPA theory, we can find the
phase space factor $f_{if}^B$ in SMFs, and it is defined as
\begin{eqnarray}
&f_{if}^{\rm{B}}=\frac{c^3}{(m_ec^2})^5\frac{b}{2}\sum_{0}^{\infty}\theta_n\nonumber\\
&=\frac{c^3}{(m_ec^2})^5\frac{b}{2}\sum_{0}^{\infty}
g_{no}\int^{\infty}_{p_0}dp_{e}p^2_e(-\xi+\varepsilon_n)^3
F(Z,\varepsilon_n)f,\nonumber\\
\label{eq.36}
\end{eqnarray}
where $b=B/B_{cr}$, and the  $\varepsilon_n$ is the total rest mass
and kinetic energies; $F(Z, \varepsilon_n)$ is the Coulomb wave
correction which is the ratio of the square of the electron wave
function distorted by the coulomb scattering potential to the square
of wave function of the free electron. We assume that a SMFs will
have no effect on $F(Z, \varepsilon_n)$, which is valid only under
the condition that the electron wave-functions are locally
approximated by the plane-wave functions. \citep{b14} The condition
requires that the Fermi wavelength $\lambda_{\rm{F}}\sim
\hbar/P_{\rm{F}}$ ($P_{\rm{F}}$ is the Fermi momentum without a
magnetic field) be smaller than the radius $\sqrt{2} \zeta$ (where
$\zeta=\lambda_e/b$ ) of the cylinder which corresponds to the
lowest Landau level\citep{b6}.

The $p_0$ is defined as
\begin{equation}
p_0=\left\lbrace \begin{array}{ll}~\sqrt{Q^2_{if}-\Theta},~~~~~~( Q_{if}<\Theta^{1/2})\\
                                  ~0 ~~~~~~~~~~~~~~~~~~~({\rm{otherwise}}),
                             \end{array} \right.
\end{equation}
\label{eq.37}
where $\Theta=m_e^2c^4(1+2\nu
B/B_{cr})=m_e^2c^4(1+2\nu b)$.

\section{Some numerical results and discussion}

An SMFs can significantly affect the cooling properties and thermal
structure of a neutron star crust. In general, the thermal
insulation can be decreased by the magnetic field due to Landau
quantization of electron motion. However, the thermal insulation of
the envelope may be increased by the tangential magnetic field,
which parallel to the stellar surface due to the fact that the
Larmor rotation of the electron significantly reduces the transverse
thermal conductivity. An SMFs also strongly affects the cooling
curve of a neutron star and magnetar. This is because for a given
core temperature, the time evolution largely depends on neutrino
emission from the surface and core of a neutron star and magnetar.

Figures 1-4 display the NELRs of some iron group nuclei as a
function of the magnetic field $B_{12}$ at relatively low, and
medium density (i.e. $\rho_7=5.86, 14.5$) and some typical
temperature surrounding (i.e. $ T_9=0.233, 15.53$). One finds that
when $ B_{12} <100$ and at relatively low temperature (e.g.
$T_9=0.233$) the magnetic field has a slight effect on the NELRs for
most nuclides from Figure 1 and 3. But the NELR of most nuclides are
influenced greatly at relatively high temperature (e.g.
$T_9=15.53$). For example, for most iron group nuclei
(e.g.$^{52-61}$Fe, $^{55-60}$Co and $^{56-63}$Ni), the NELRs
increases by by more than four orders of magnitude at $T_9=15.53$
when $ B_{12} <100$. However, the NEL rates decrease by more than
three orders of magnitude when $B_{12}>100$ in Figure 1, but
$B_{12}>200$ in Figure 2.

From Figure 1 to 4, we detailed discuss the NELRs due to EC process
according to SMMC method, especially for the contribution for EC due
to the GT transition base on RPA theory. One finds that the
influences of SMFs on NELR are significant. It is due to the fact
that we have more available phase space for electron in the higher
the magnetic field at a given density. On the other hand, the
electron is very relativistic in the crust of magnetar, where the
matter density is higher and magnetic field strength may greatly
exceed the surface value. The mean Fermi energy of an electron will
exceed its rest-mass energy at sufficiently high density. The
electron also is very relativistic when the cyclotron energy of an
electron also higher than its rest-mass energy at sufficiently high
magnetic field. The electron capture will rapid occur when the
electron energy becomes larger than the difference between the
neutron and proton rest-mass energy(about 1.3MeV). This EC process
will destroy electrons and emit massive neutrino, thereby changing
the composition of matter and softening the state equation of
magnetar surface.

The cross sections(hereafter ECCS) are very important parameter in
electron capture process. We find the influence of SMFs on ECCS at
different temperature is very significant for some nuclides due to
the difference of Q-value. Given the significant energy dependence
of cross sections for a process like electron capture on nuclei, it
is clear that in some cases the rates will be increased at the same
temperature and density as the magnetic field increases. With
increasing of electron energy, the ECCS increases according to our
investigations. The higher the temperature, the faster the changes
of ECCS becomes. It is because that the higher the temperature, the
larger the electron energy becomes. Thus even more electrons will
join in the EC process due to their energy is greater than the
Q-values. Furthermore, the GT transition may be dominated at high
temperature surroundings. On the other hand, the trigger mechanism
of electron capture process requires a minimum electron energy given
by the mass splitting between parent and daughter (i.e.
$Q_{\rm{if}}$). The EC threshold energy is lowered by the internal
excitation energy at finite temperature. The GT strength for
even-even parent nuclei centered at daughter excitation energies of
order of 2Mev at low temperatures. Therefore, the ECCS for these
parent nuclei increase drastically within the first couple of MeV of
electron energies above threshold, which will reflecting the GT
distribution. But the GT distribution for odd-A nuclei will peak at
noticeably higher daughter excitation energies at low temperatures.
So the ECCS are shifted to higher electron energies for odd-A nuclei
in comparison to even-even parent nuclei by about 3 MeV.

\begin{figure}
\centering
    \includegraphics[width=4cm,height=4cm]{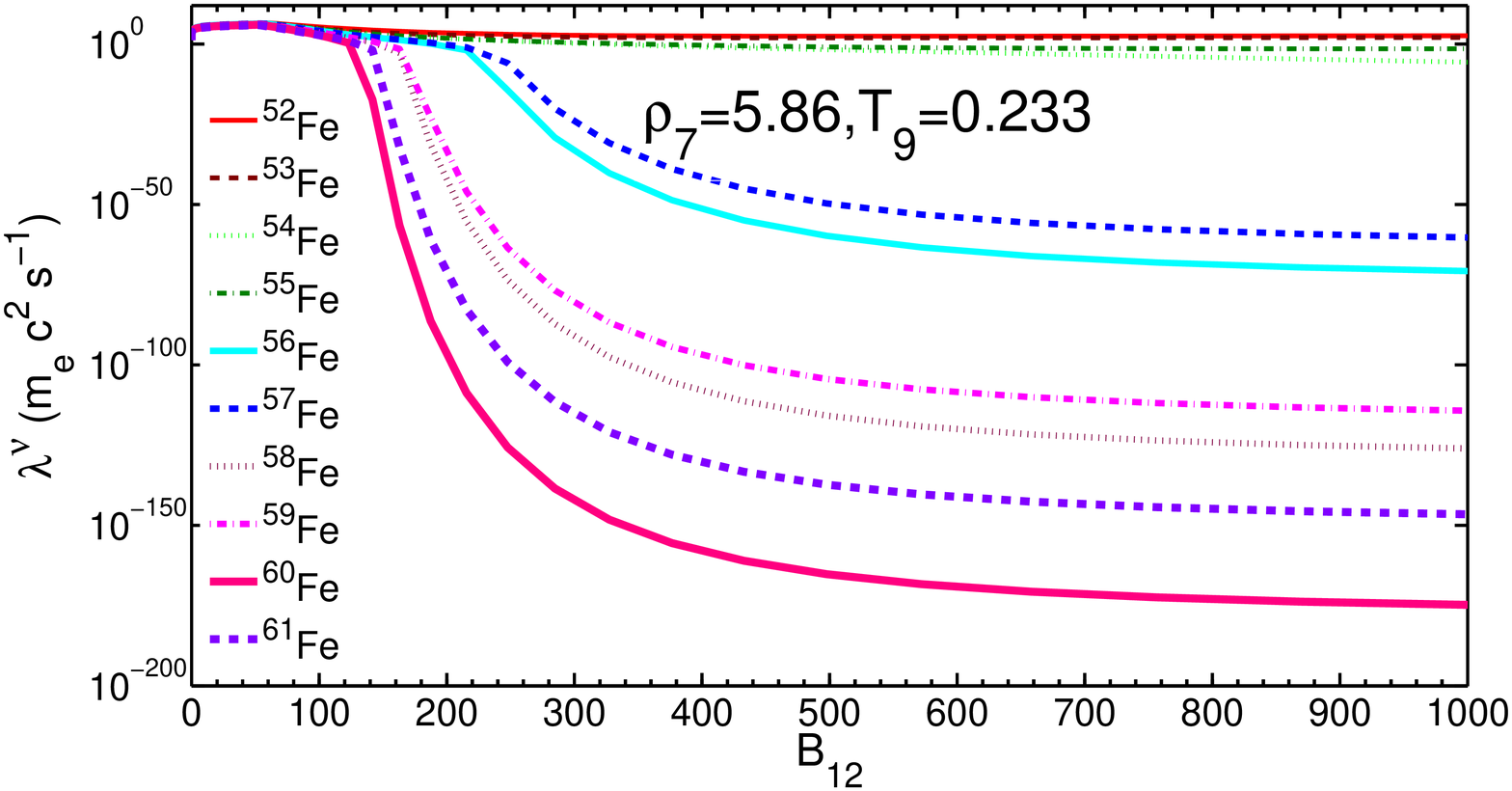}
    \includegraphics[width=4cm,height=4cm]{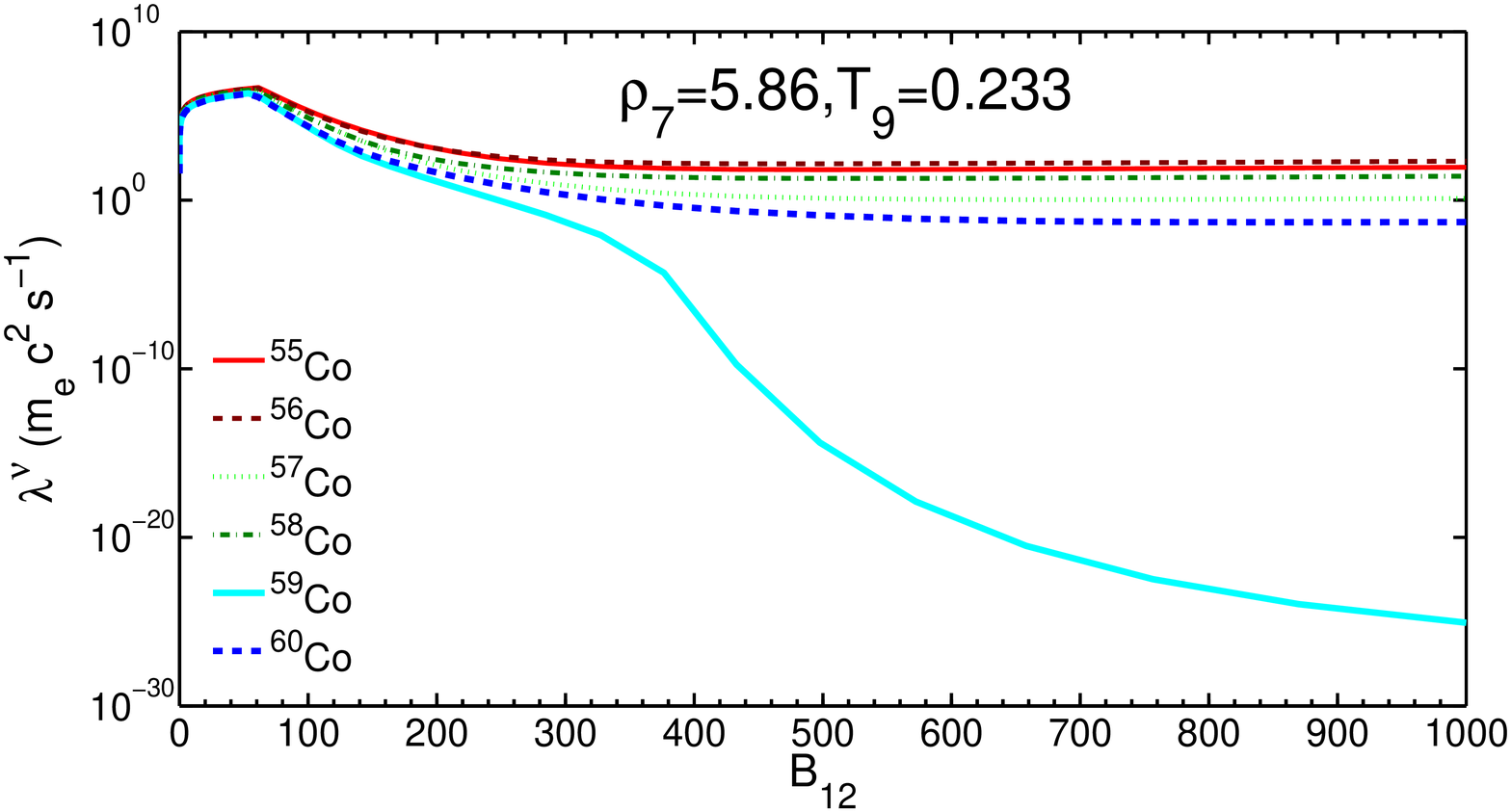}
    \includegraphics[width=4cm,height=4cm]{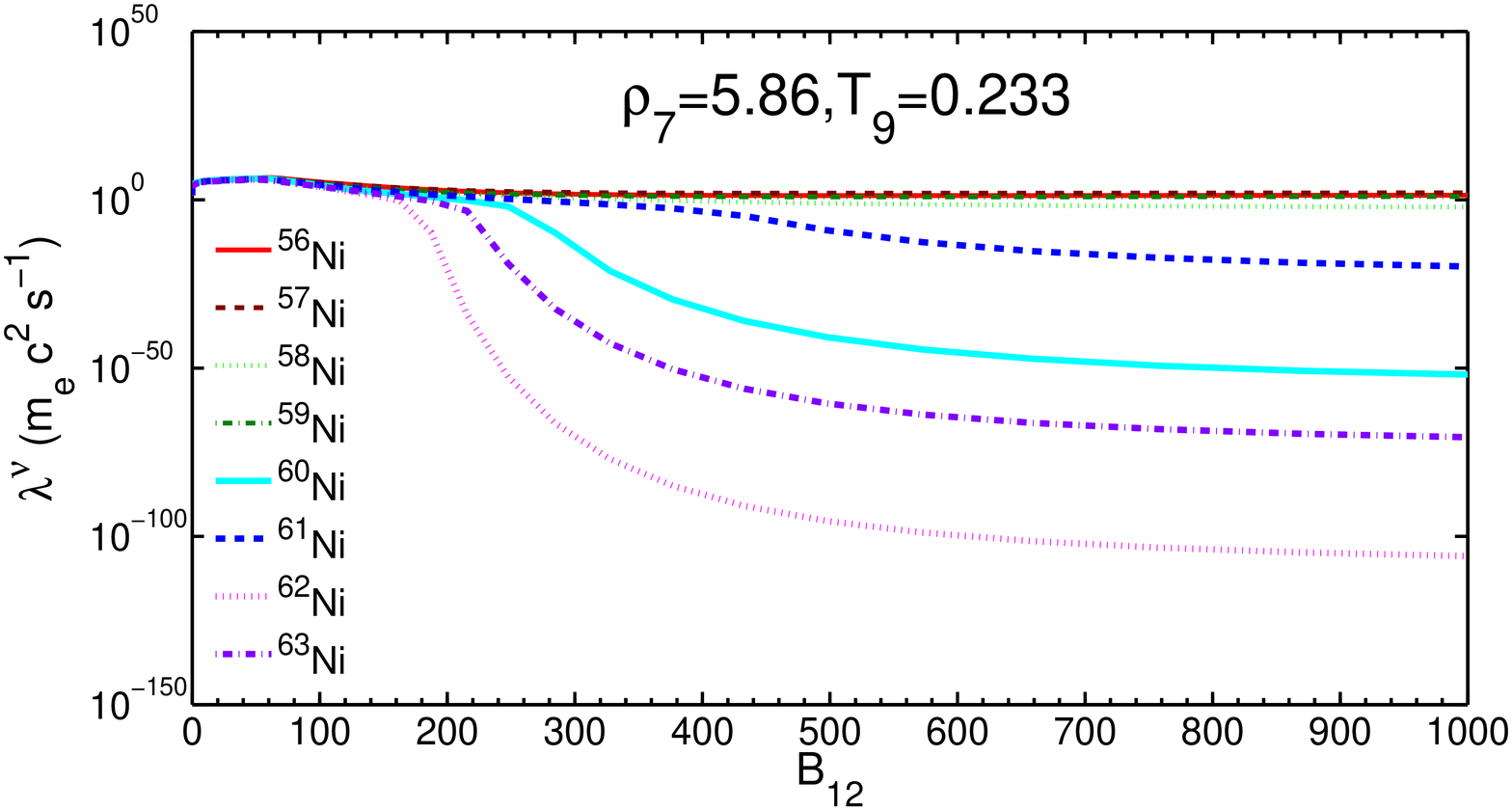}
    \includegraphics[width=4cm,height=4cm]{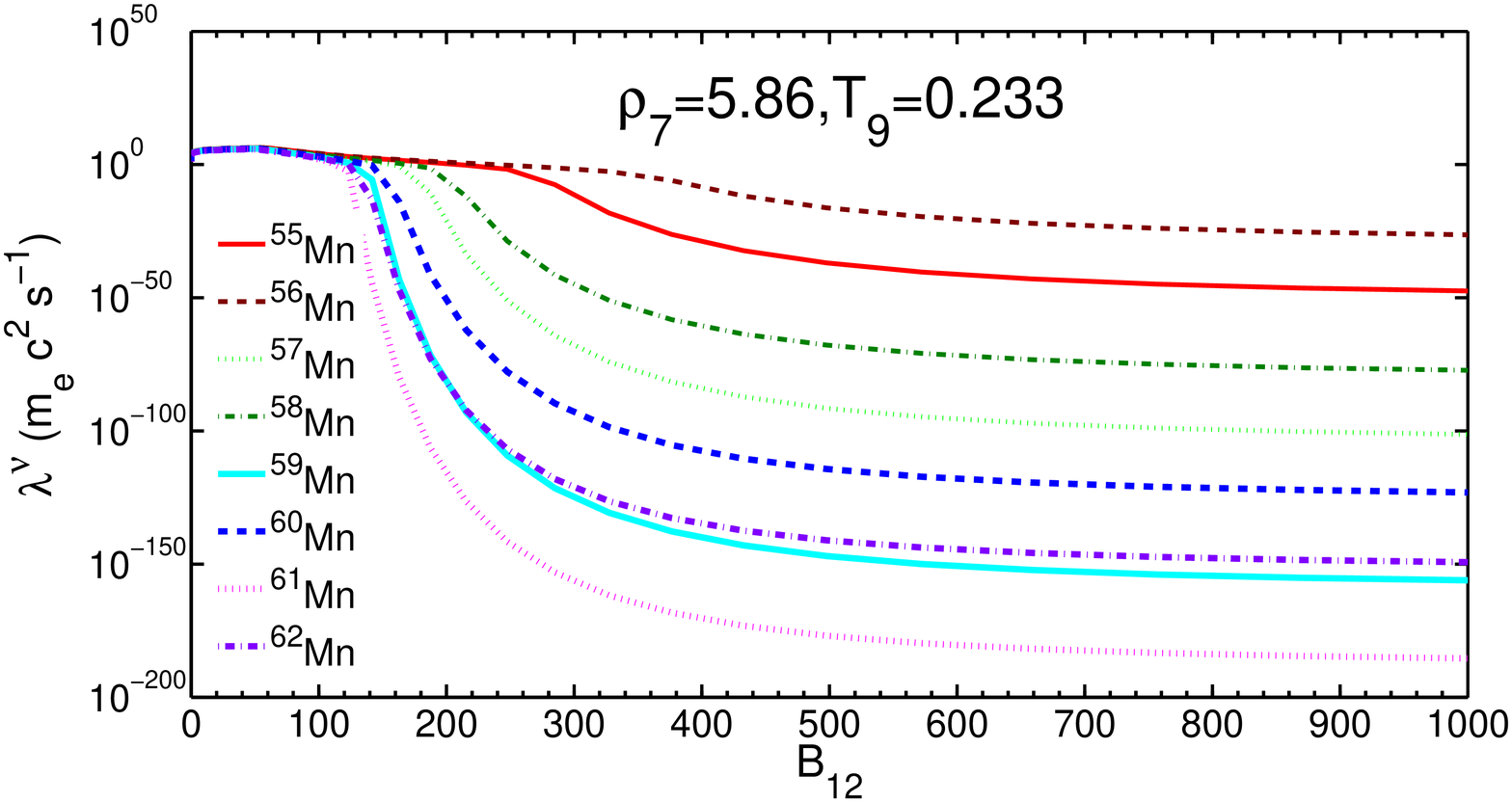}
    \includegraphics[width=4cm,height=4cm]{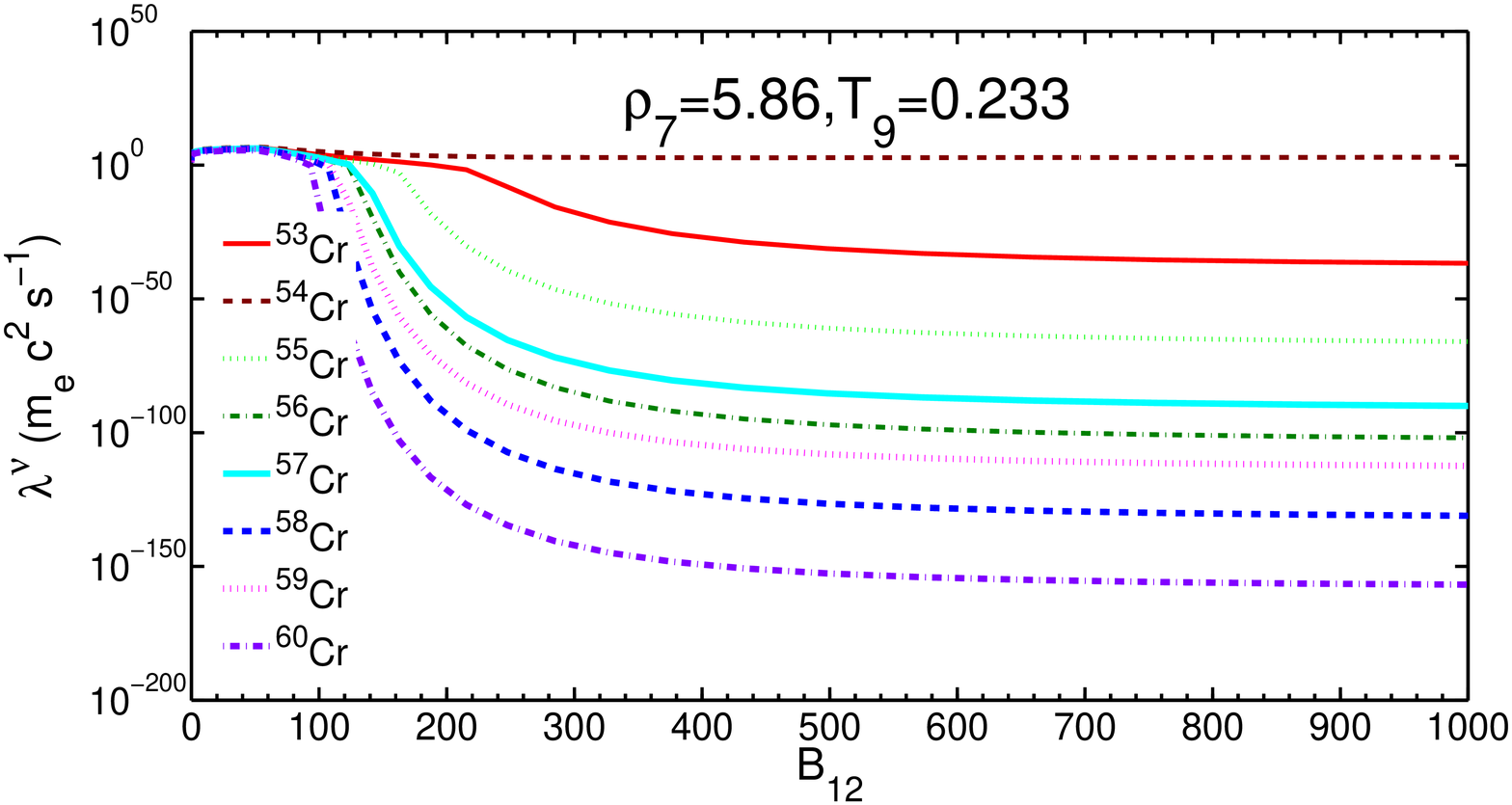}
    \includegraphics[width=4cm,height=4cm]{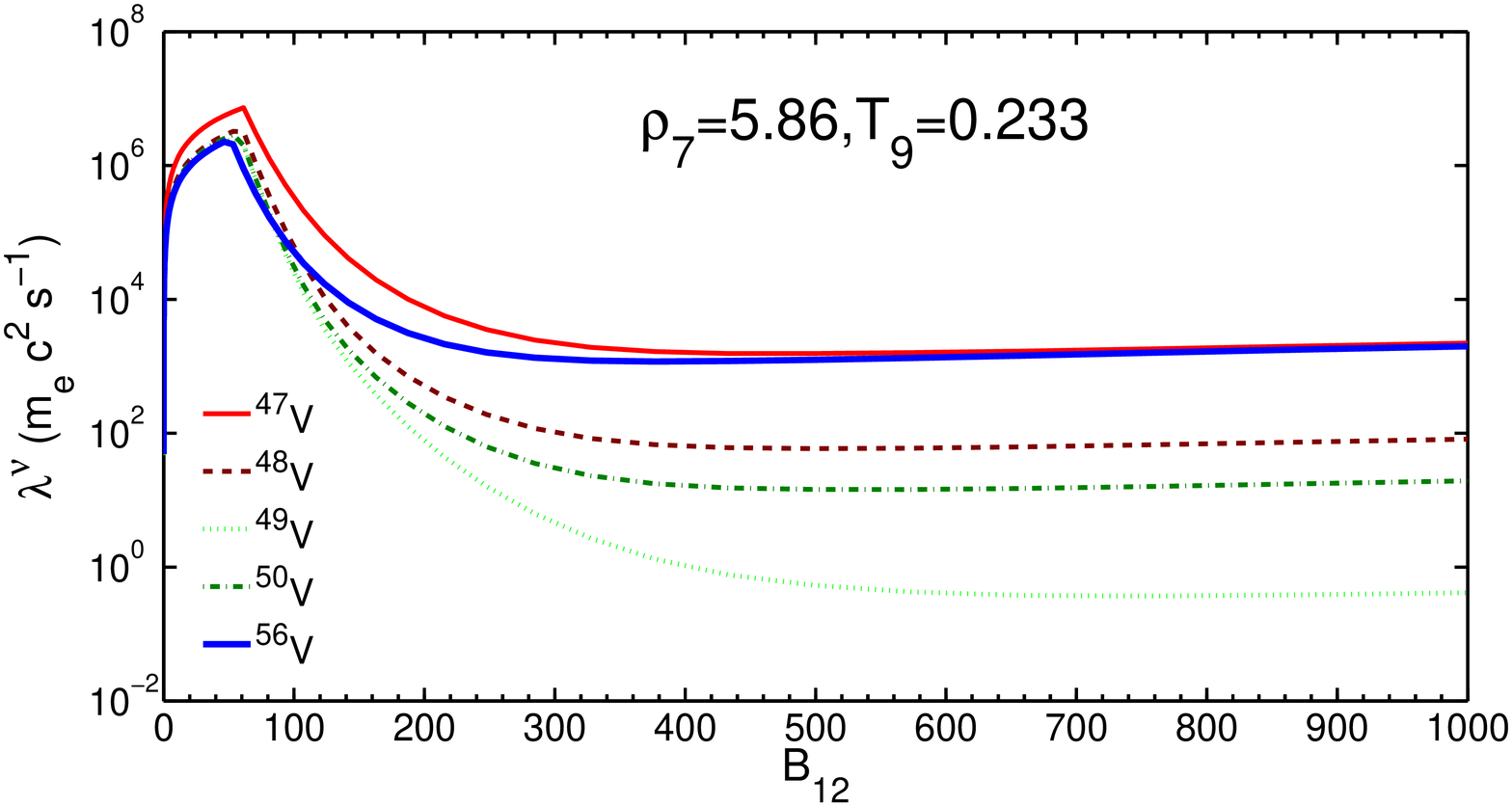}
   \caption{The NELRs for some typical iron group nuclei as a function of $B_{12}$ at $\rho_7=5.86, T_9=0.233$}
   \label{Fig:1}
\end{figure}

%

\begin{figure}
\centering
    \includegraphics[width=4cm,height=4cm]{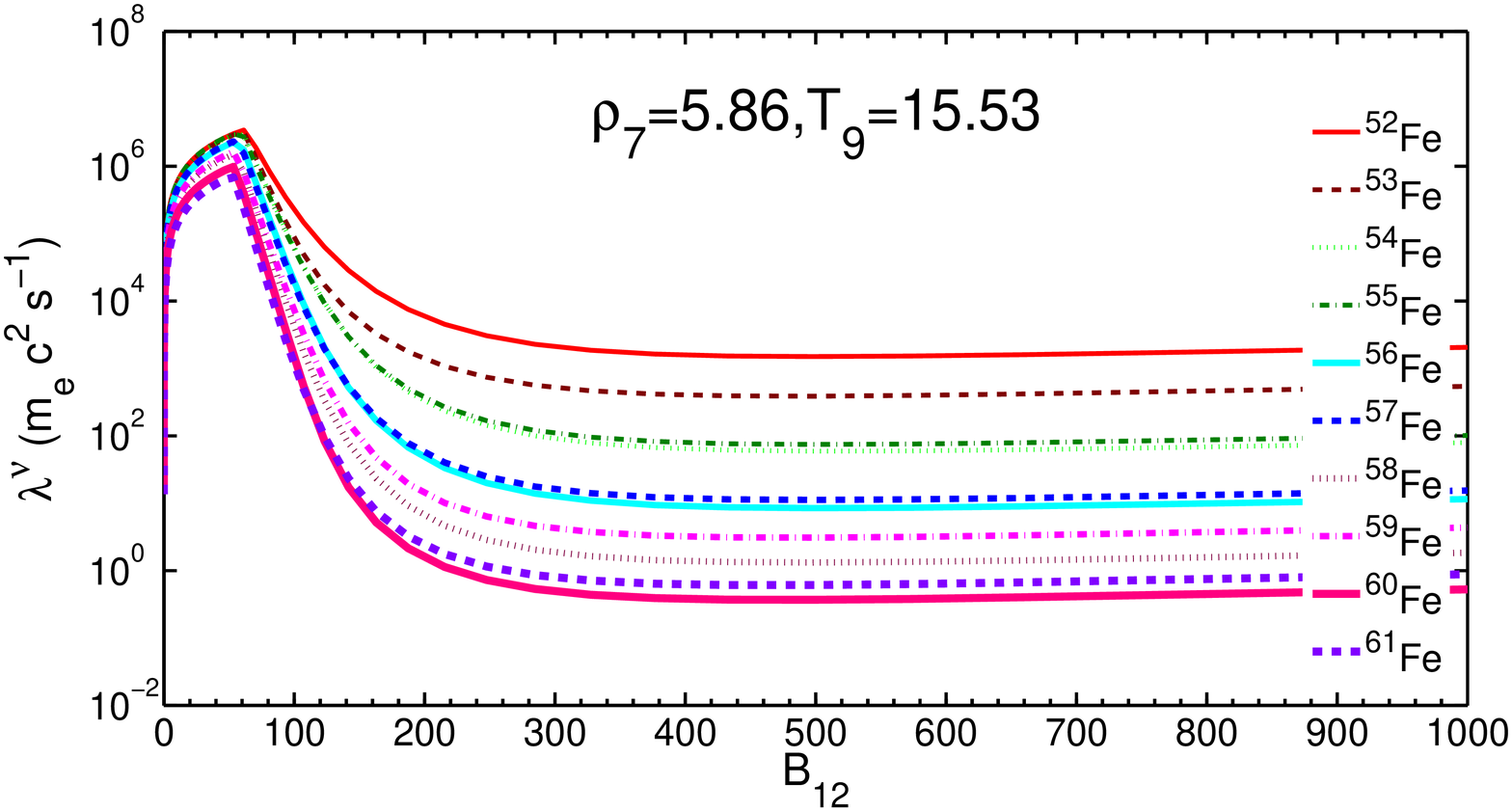}
    \includegraphics[width=4cm,height=4cm]{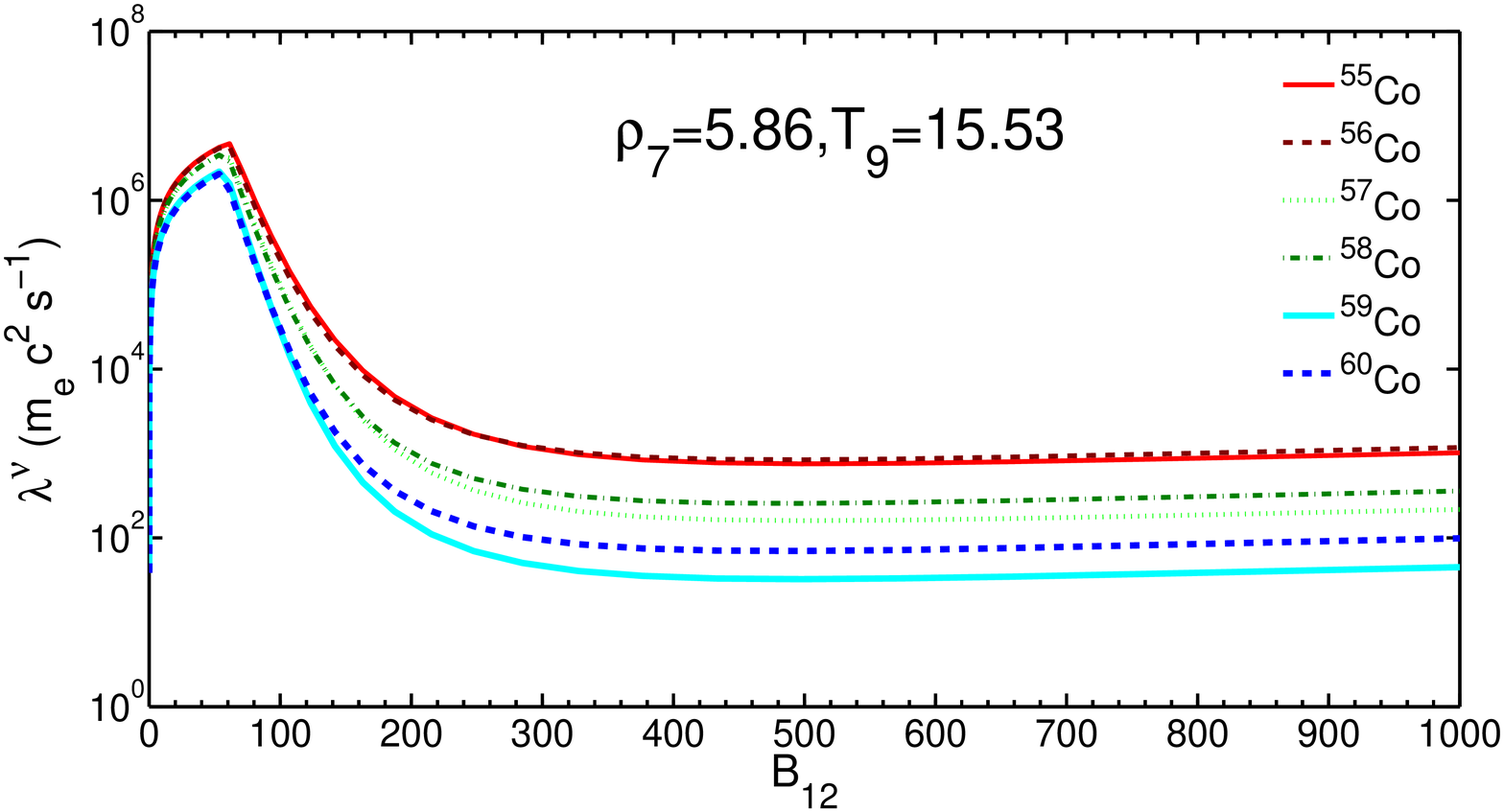}
    \includegraphics[width=4cm,height=4cm]{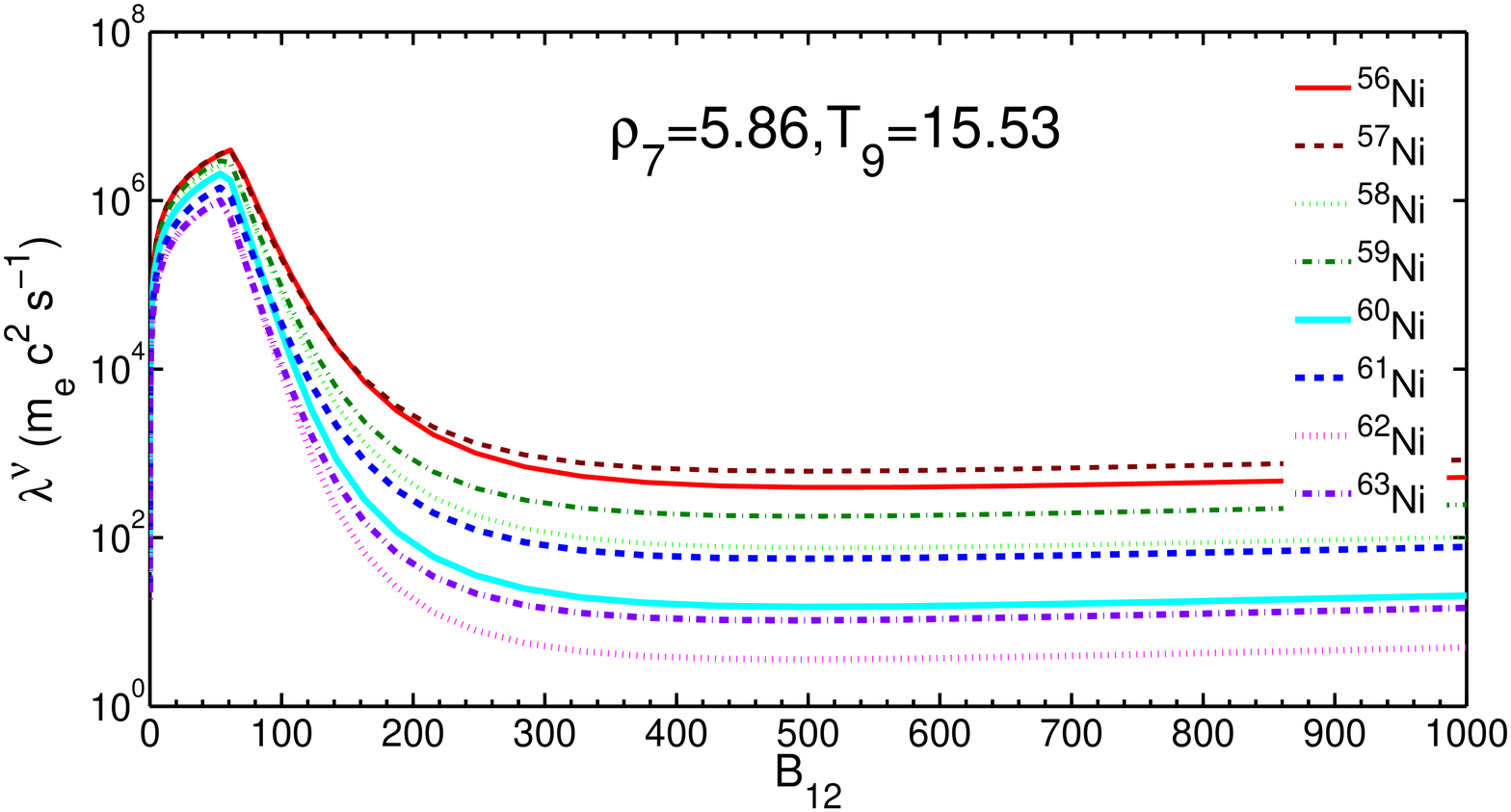}
    \includegraphics[width=4cm,height=4cm]{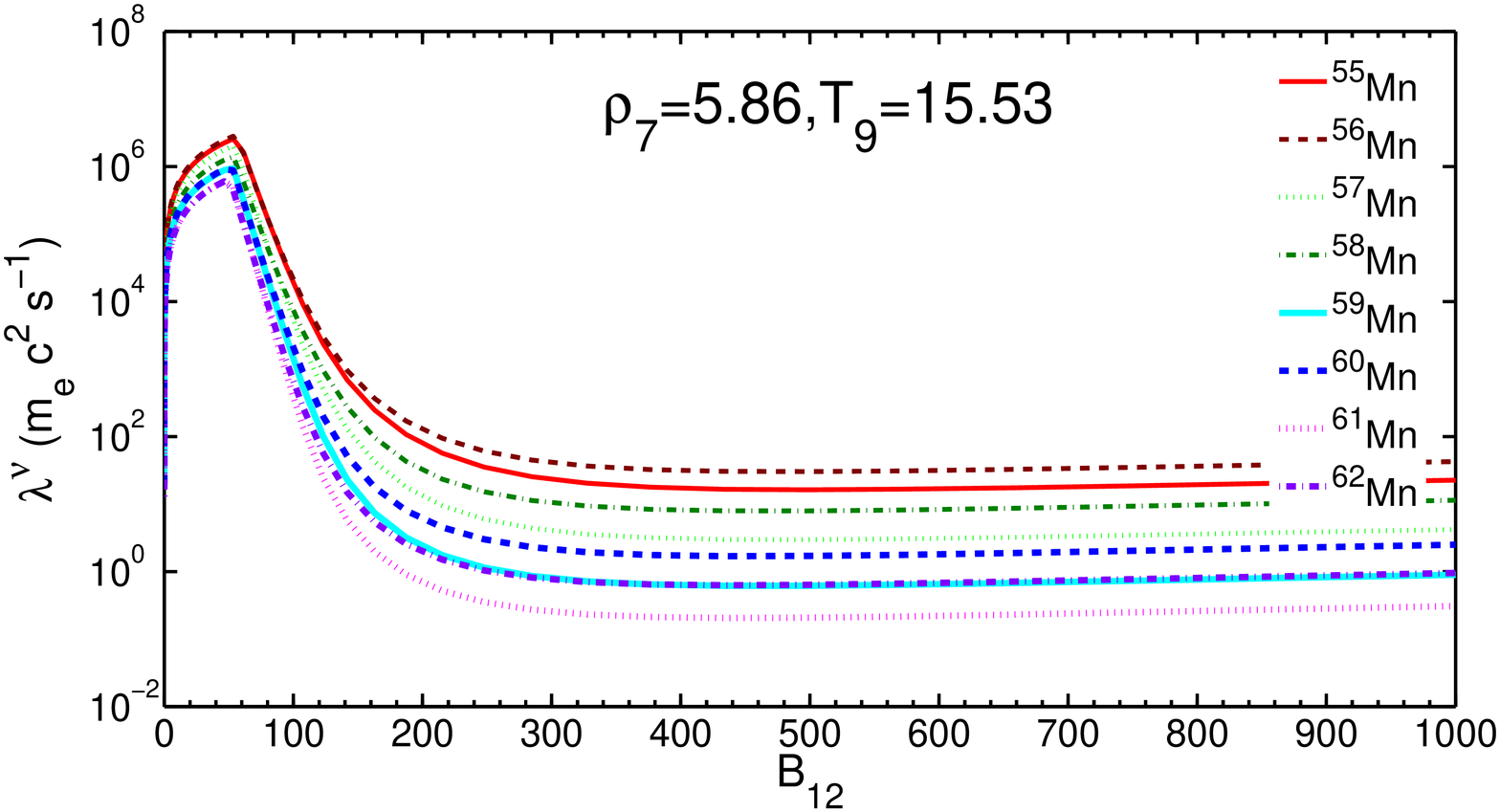}
    \includegraphics[width=4cm,height=4cm]{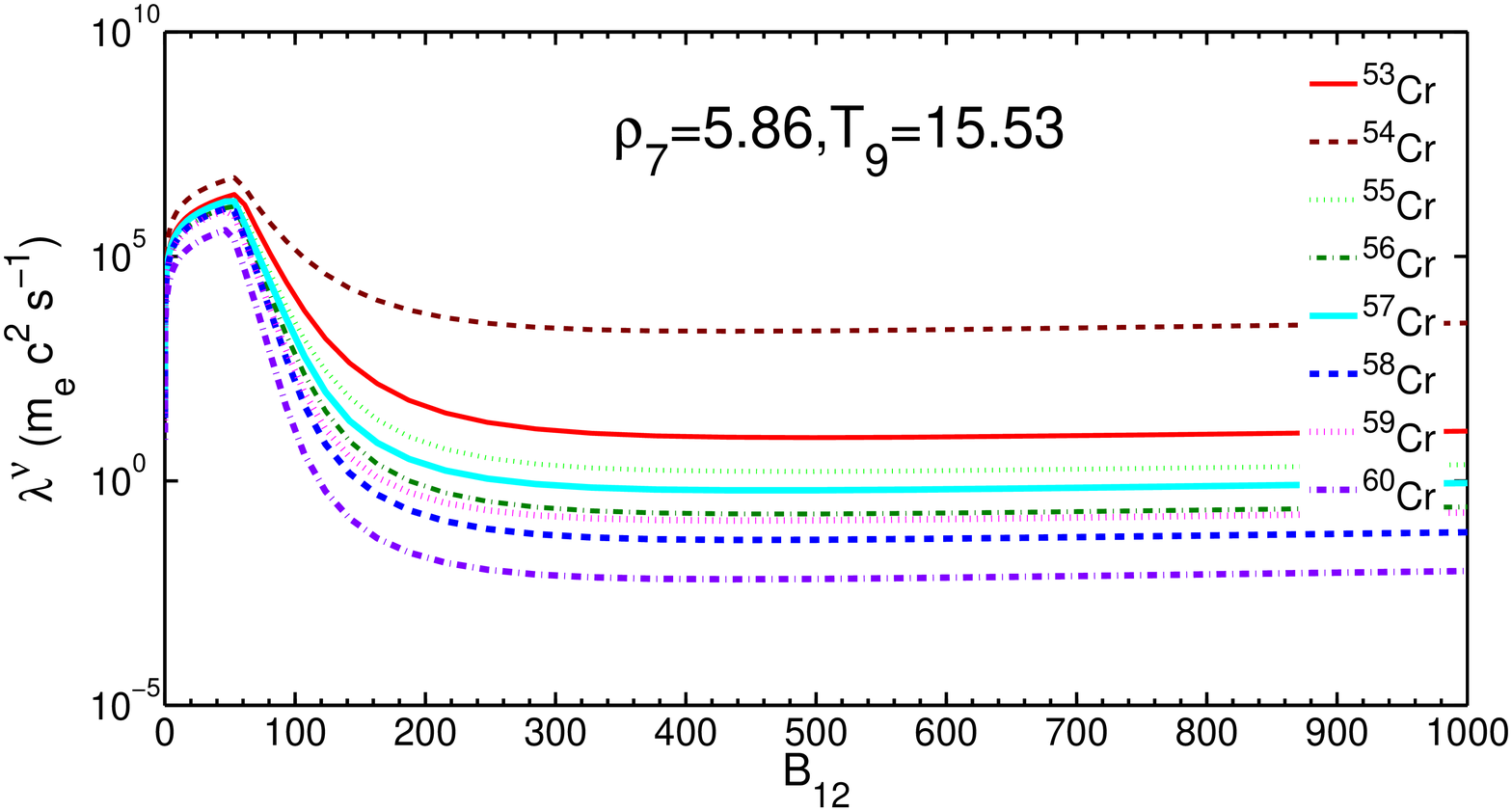}
    \includegraphics[width=4cm,height=4cm]{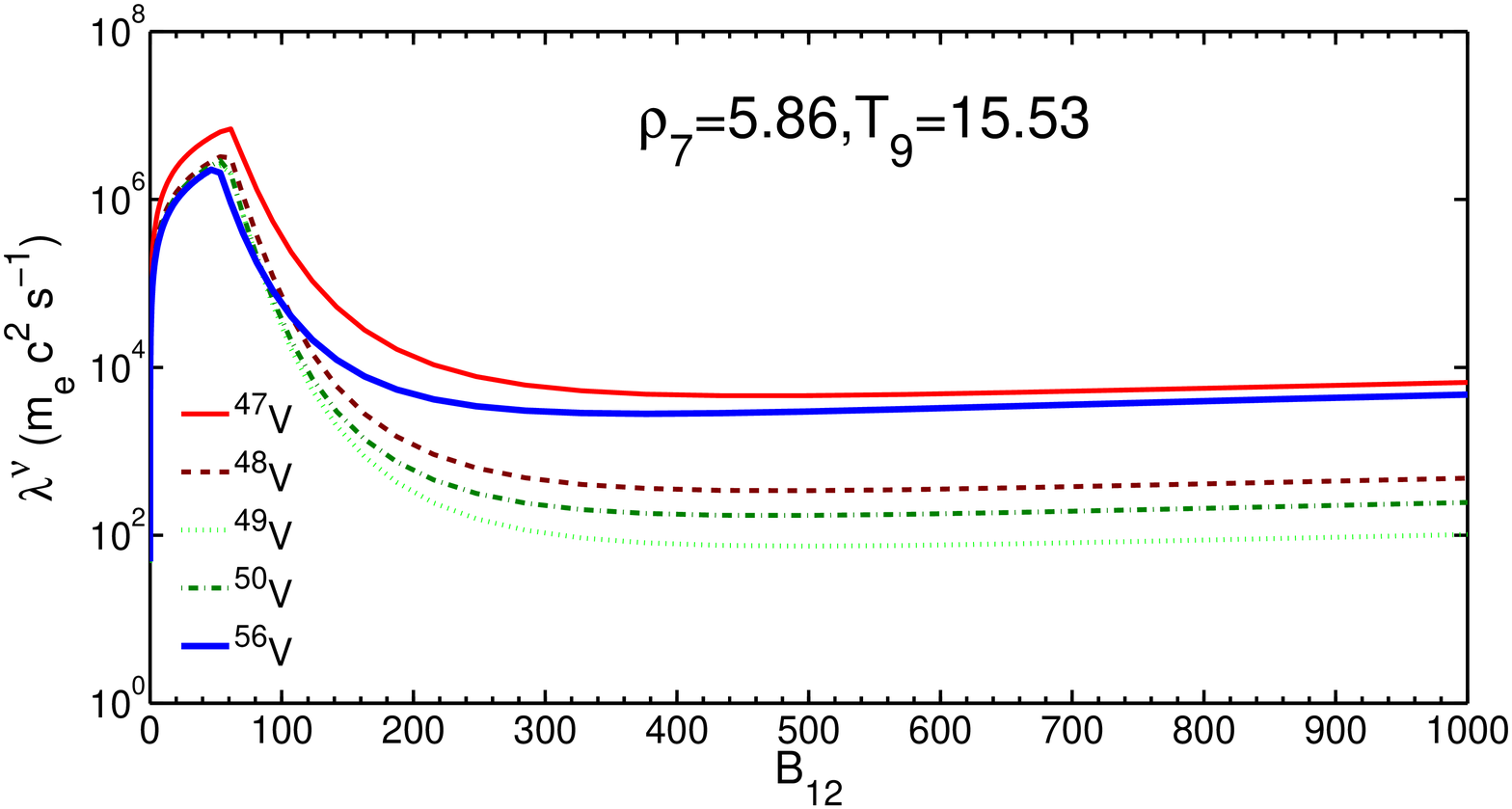}
   \caption{The NELRs for some typical iron group nuclei as a function of $B_{12}$ at $\rho_7=5.86, T_9=15.53$}
   \label{Fig:3}
\end{figure}

\begin{figure}
\centering
    \includegraphics[width=4cm,height=4cm]{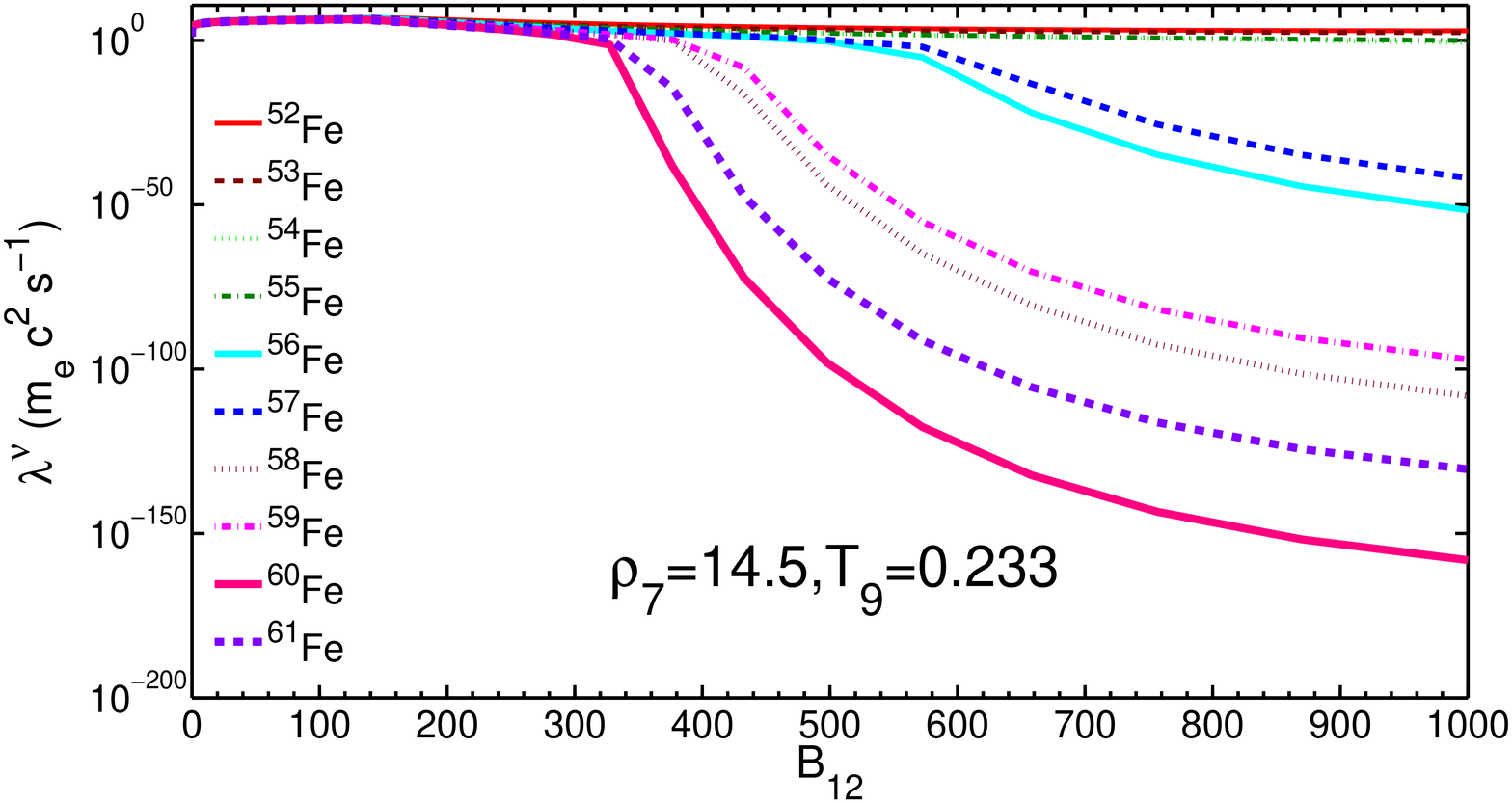}
    \includegraphics[width=4cm,height=4cm]{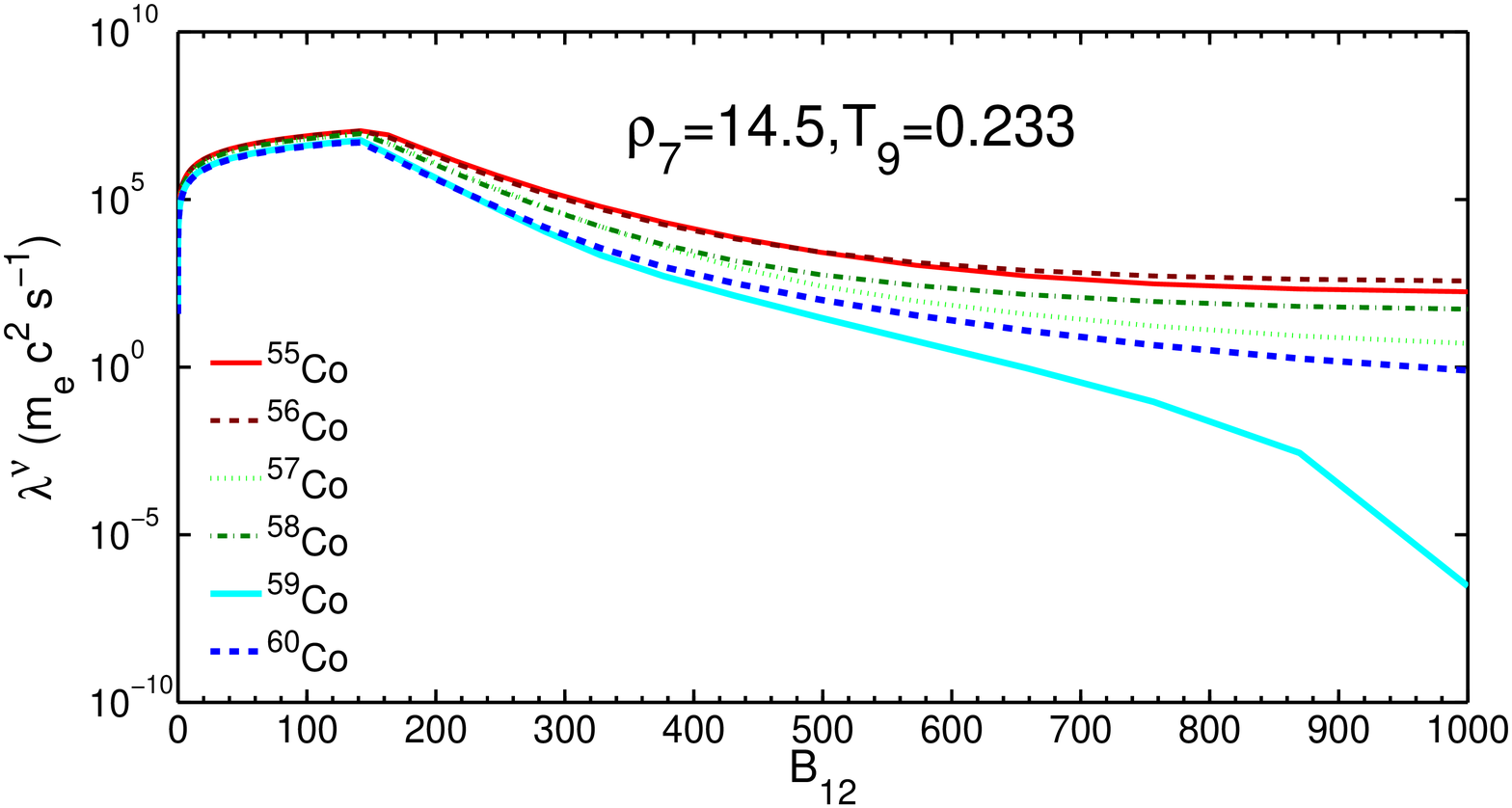}
    \includegraphics[width=4cm,height=4cm]{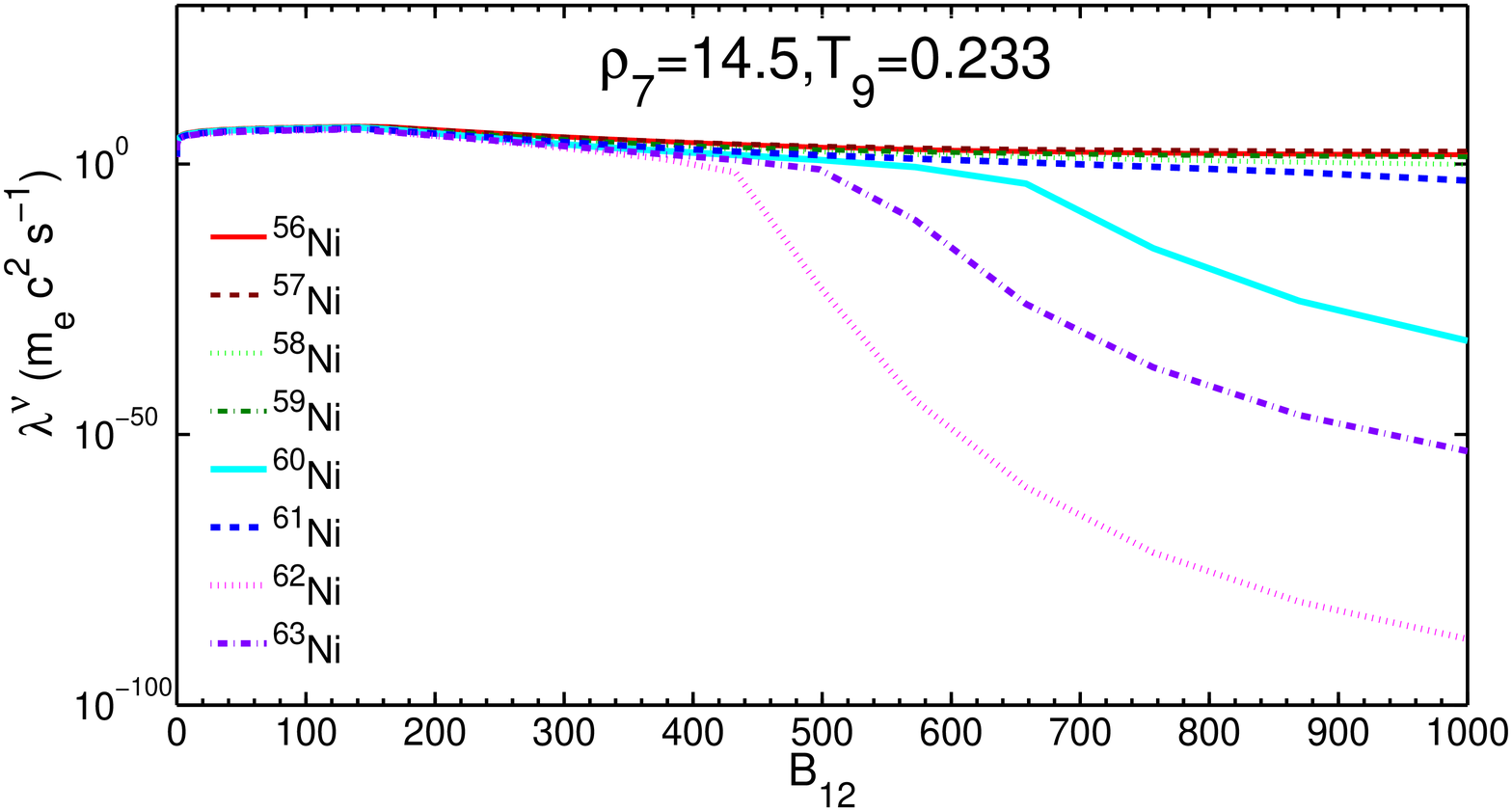}
    \includegraphics[width=4cm,height=4cm]{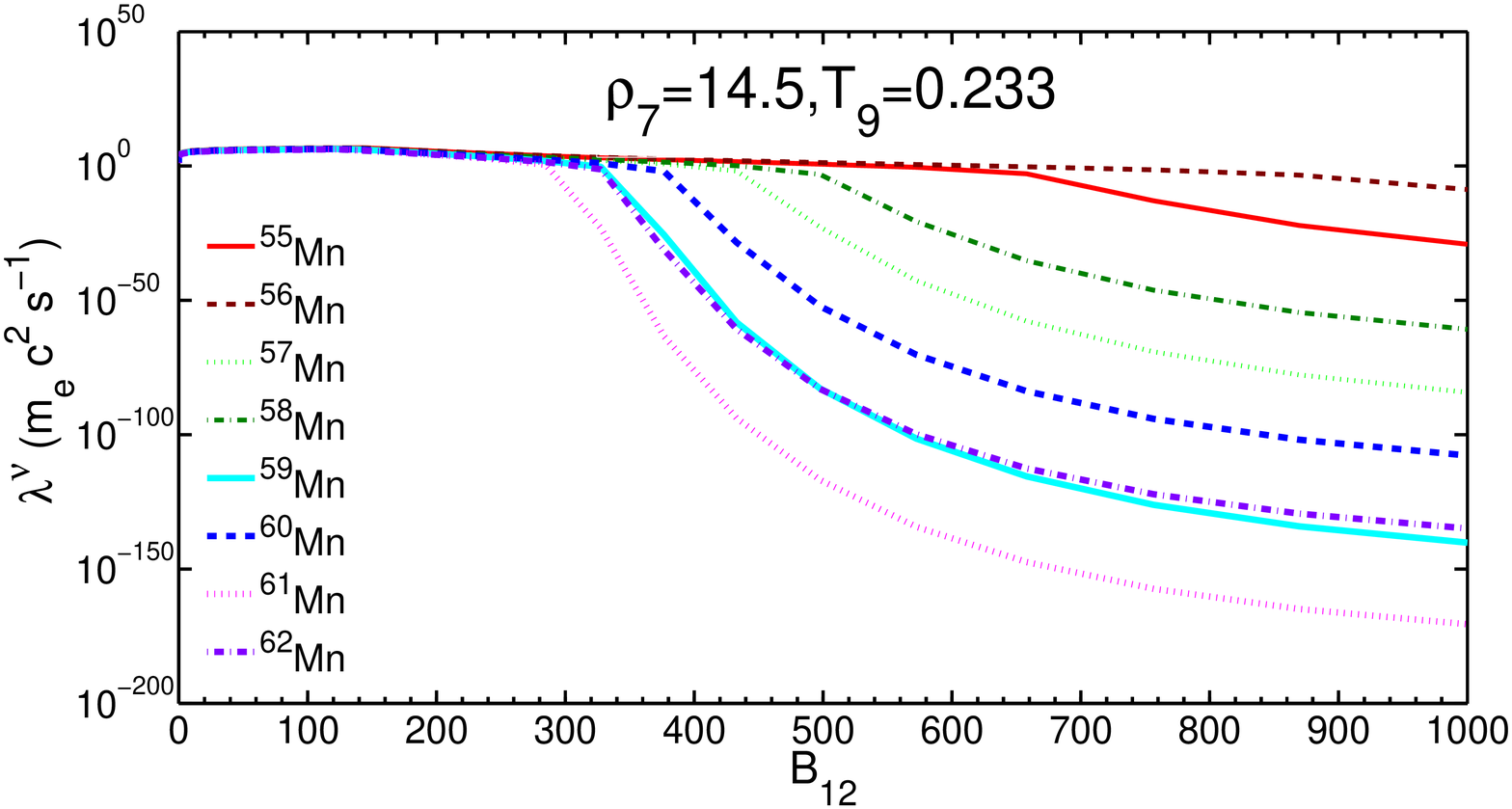}
    \includegraphics[width=4cm,height=4cm]{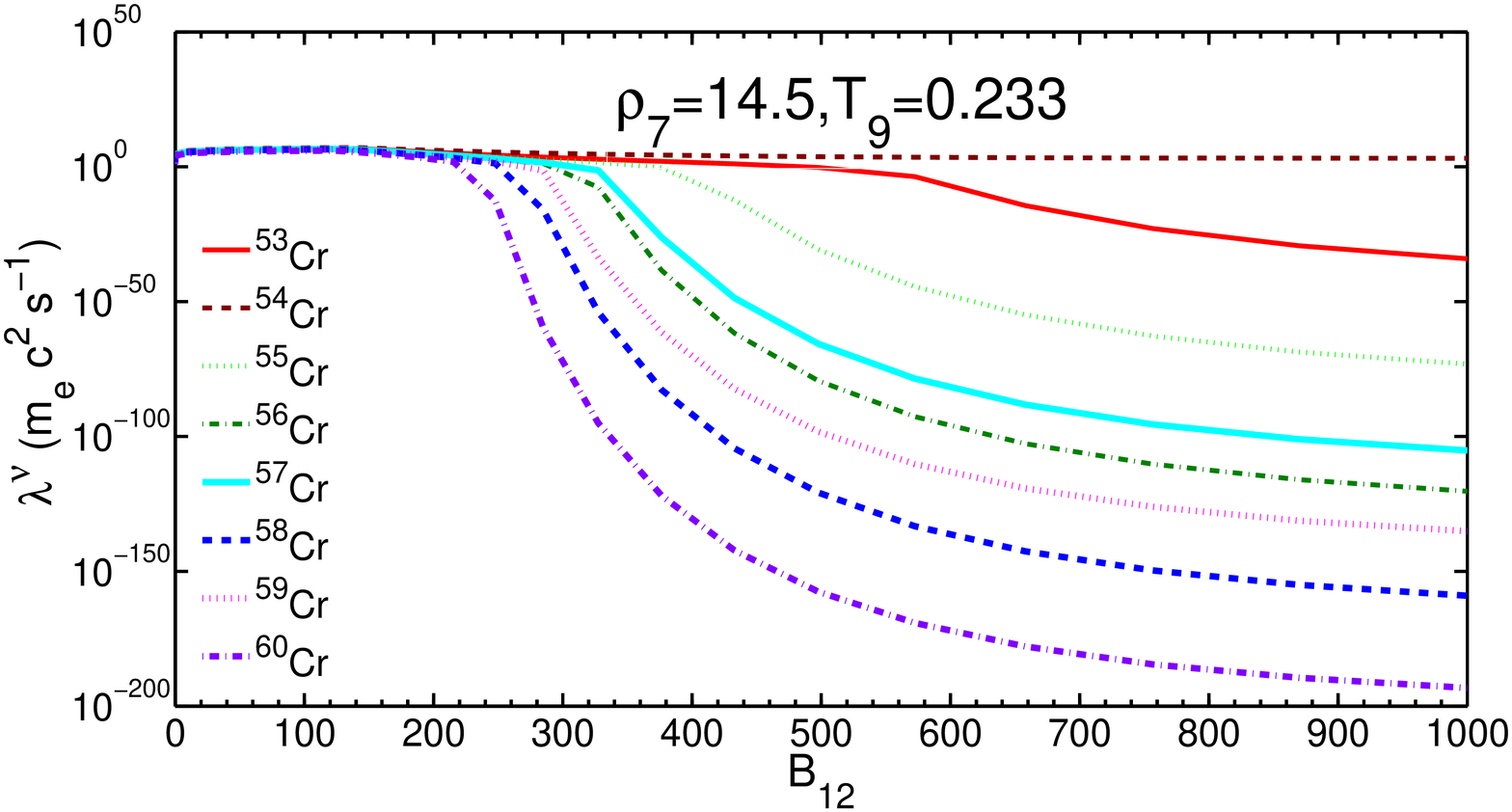}
    \includegraphics[width=4cm,height=4cm]{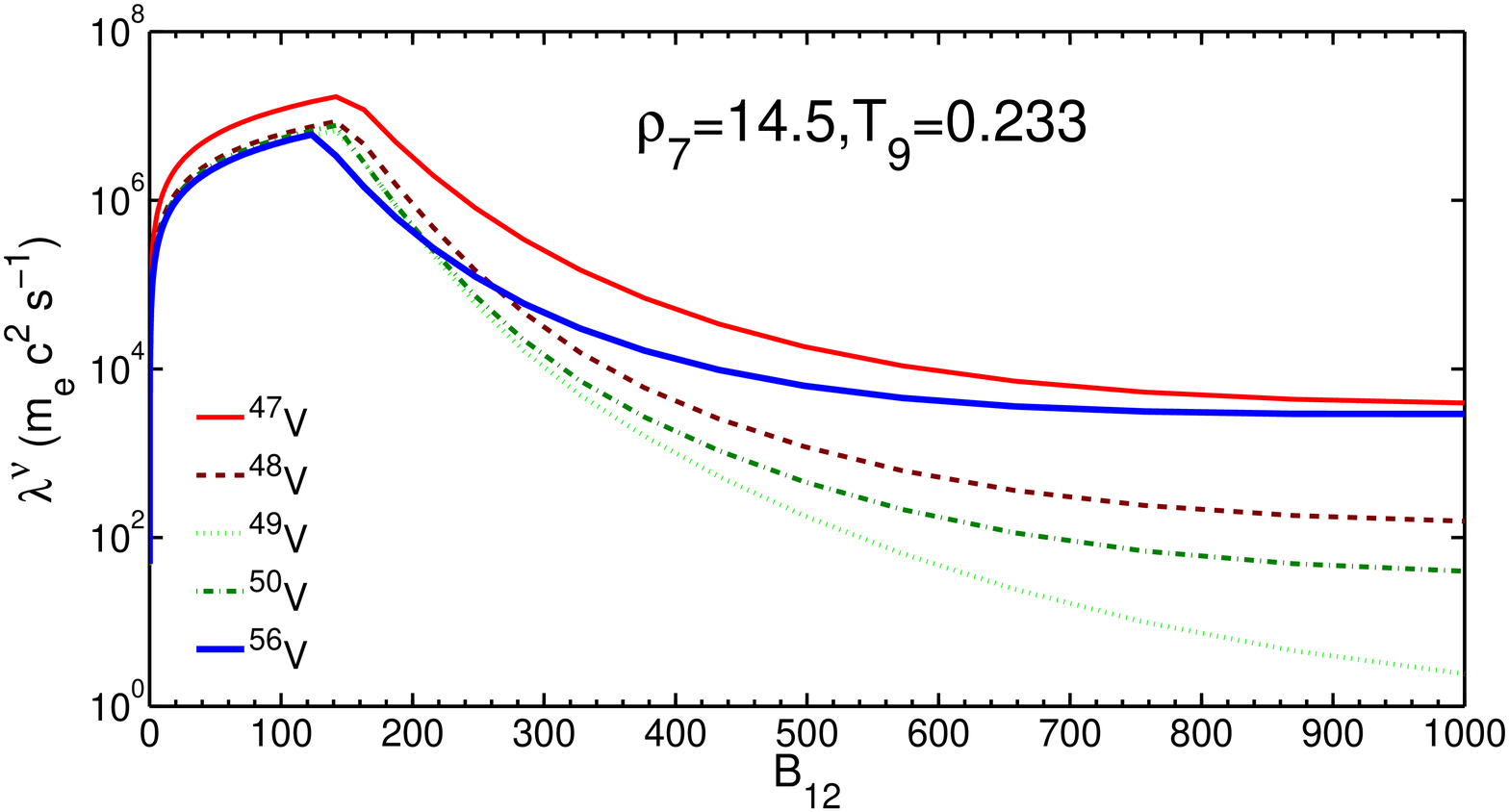}
   \caption{The NELRs for some typical iron group nuclei as a function of $B_{12}$ at $\rho_7=14.5, T_9=0.233$}
   \label{Fig:4}
\end{figure}

\begin{figure}
\centering
    \includegraphics[width=4cm,height=4cm]{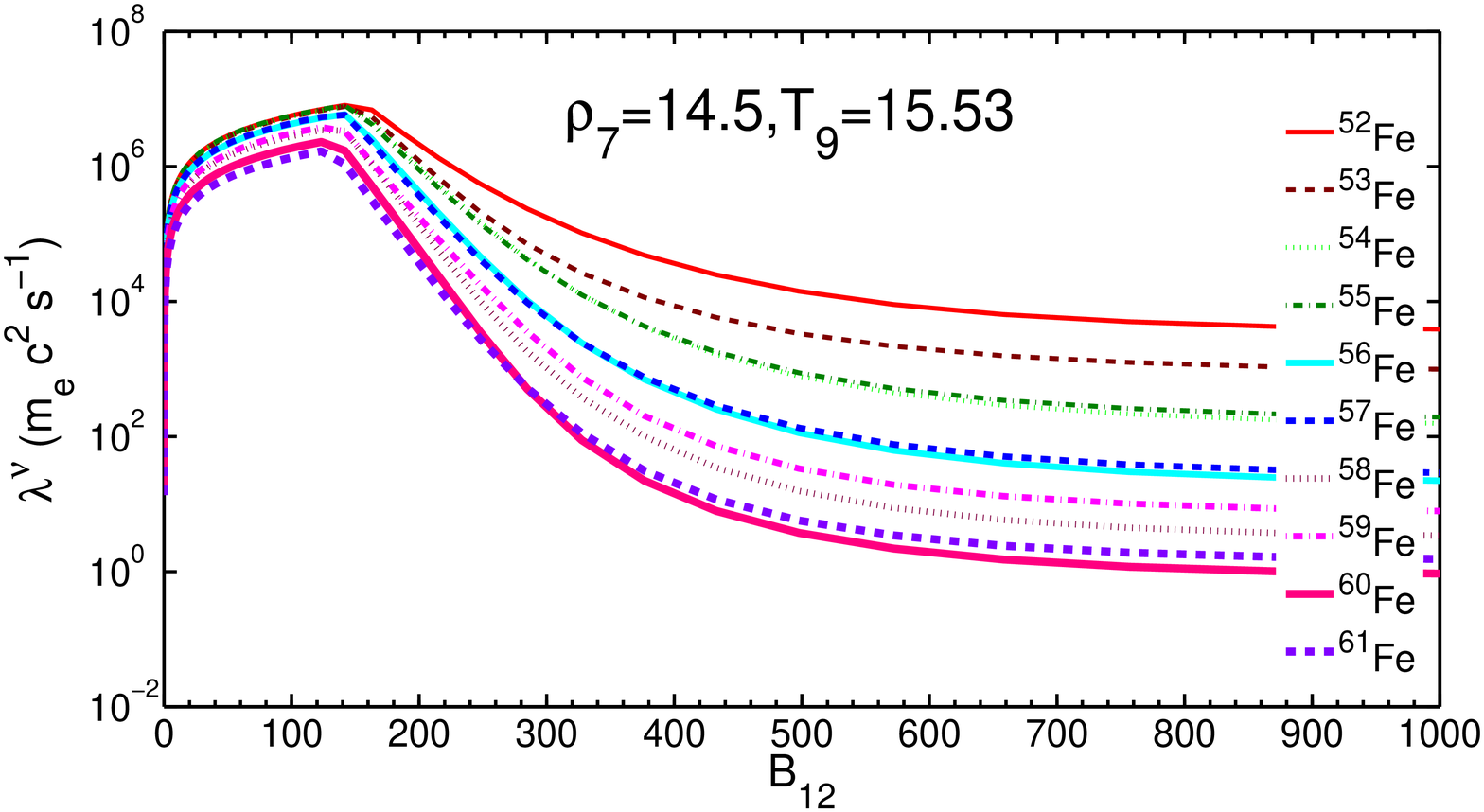}
    \includegraphics[width=4cm,height=4cm]{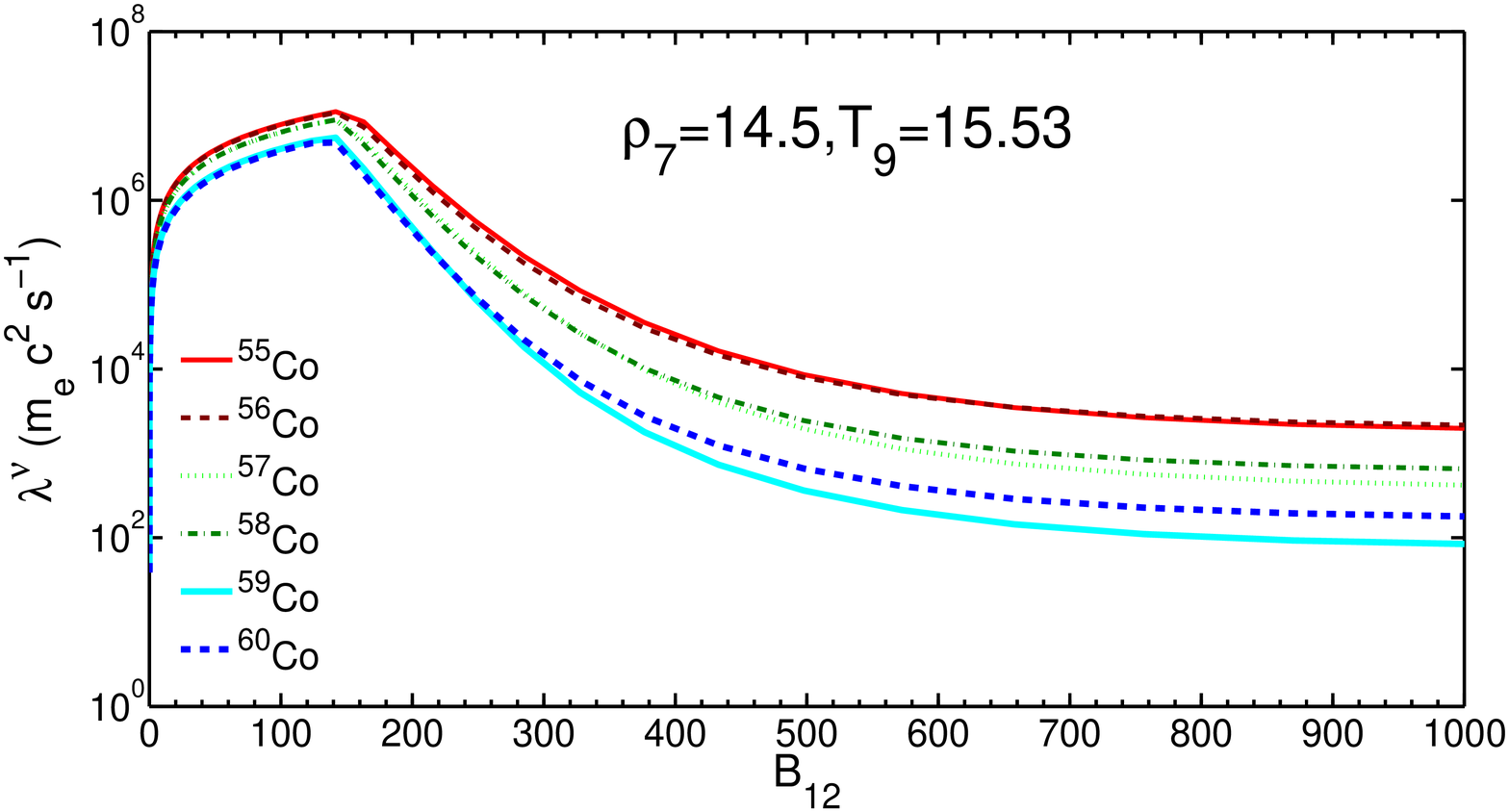}
    \includegraphics[width=4cm,height=4cm]{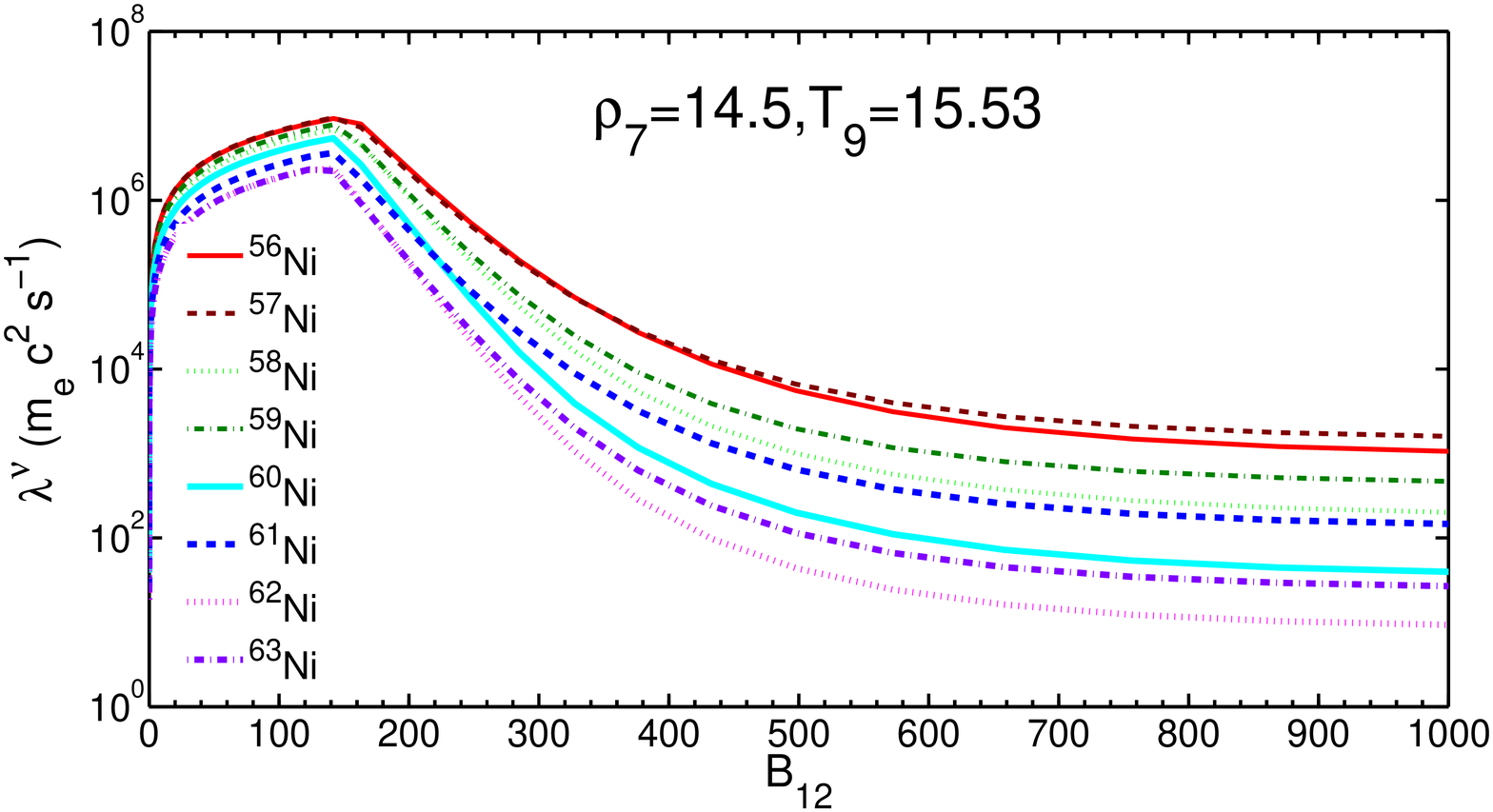}
    \includegraphics[width=4cm,height=4cm]{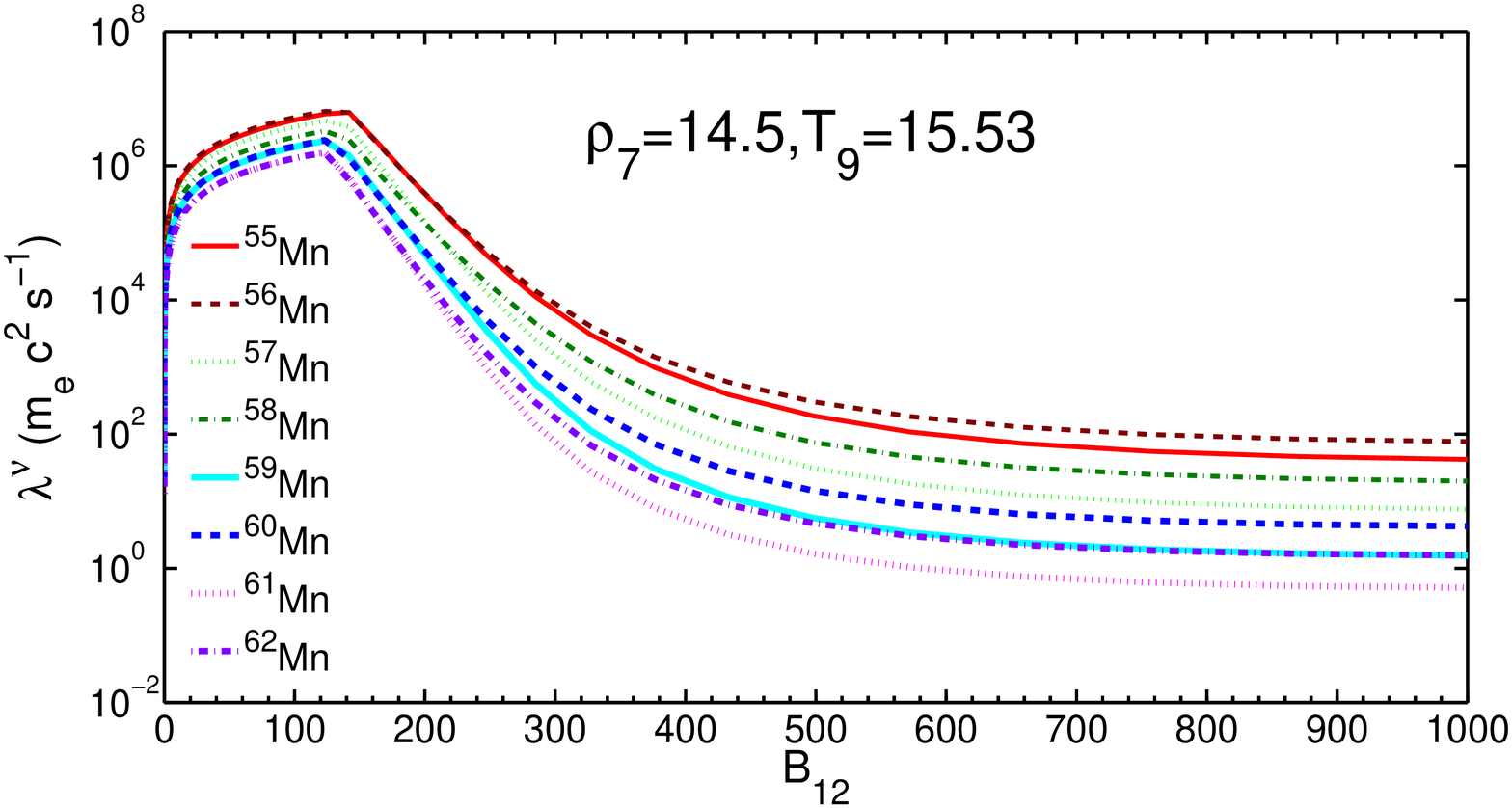}
    \includegraphics[width=4cm,height=4cm]{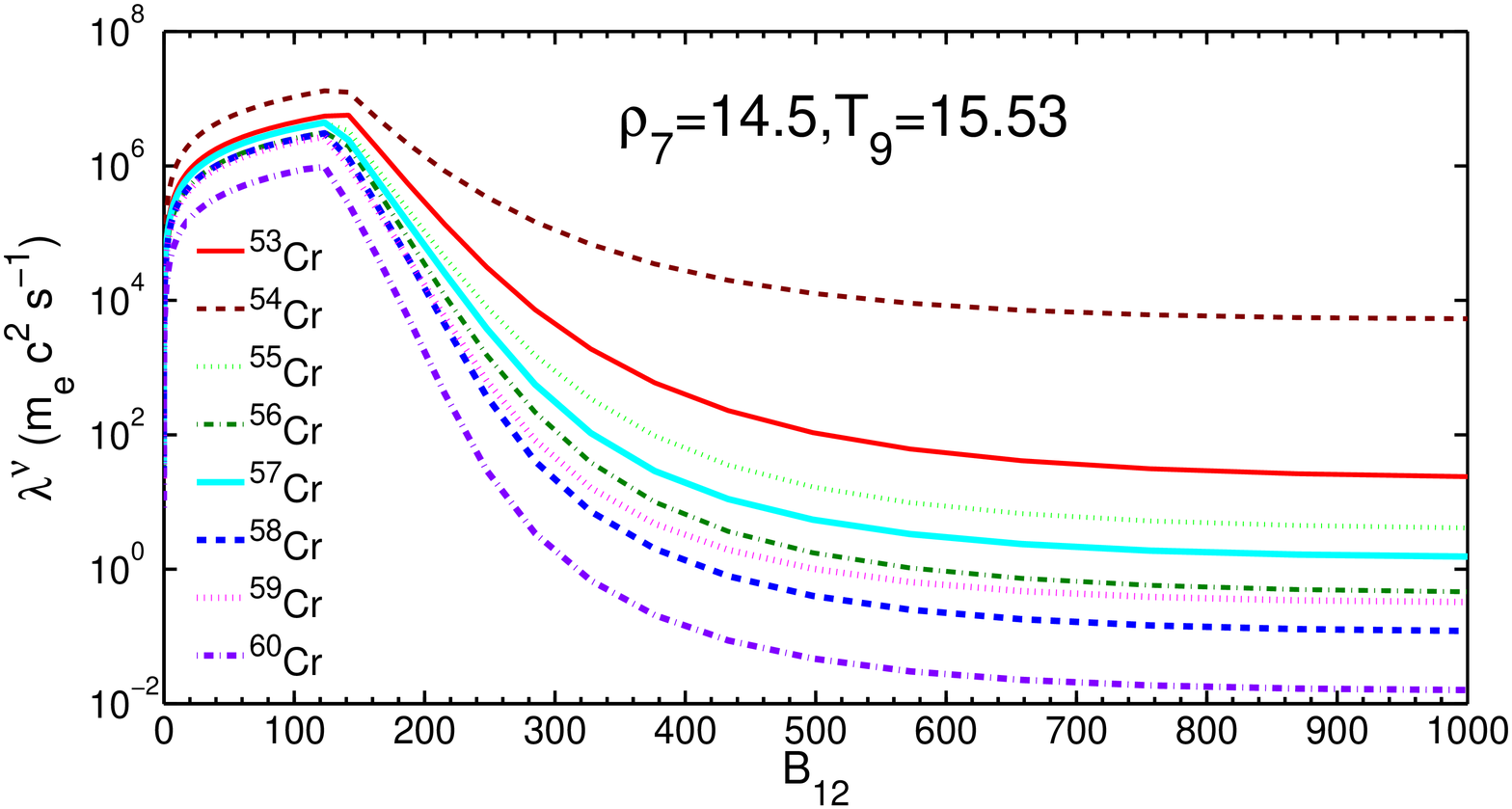}
    \includegraphics[width=4cm,height=4cm]{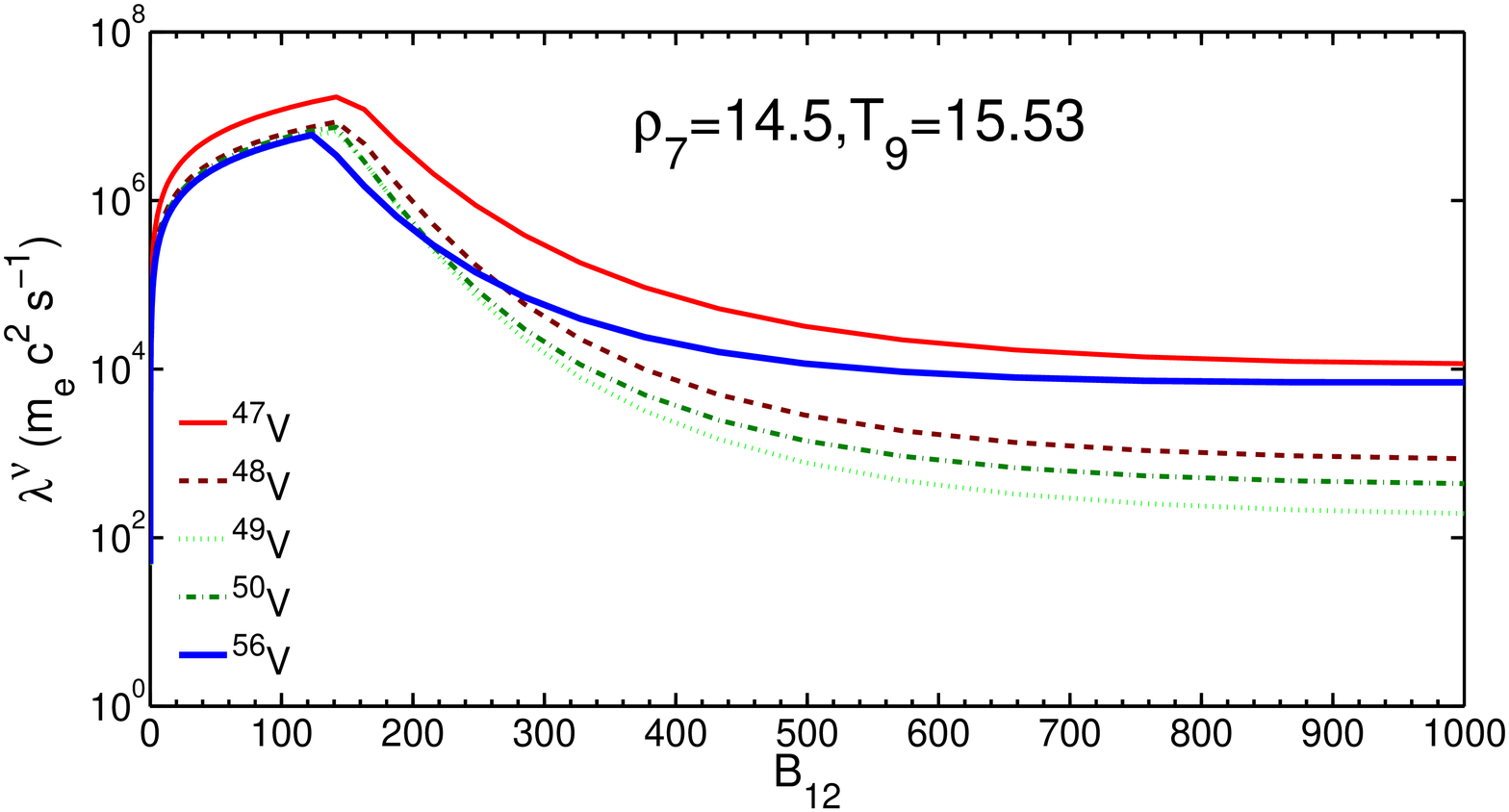}
   \caption{The NELRs for some typical iron group nuclei as a function of $B_{12}$ at $\rho_7=14.5, T_9=15.53$}
   \label{Fig:6}
\end{figure}

\begin{figure}
\centering
    \includegraphics[width=4cm,height=4cm]{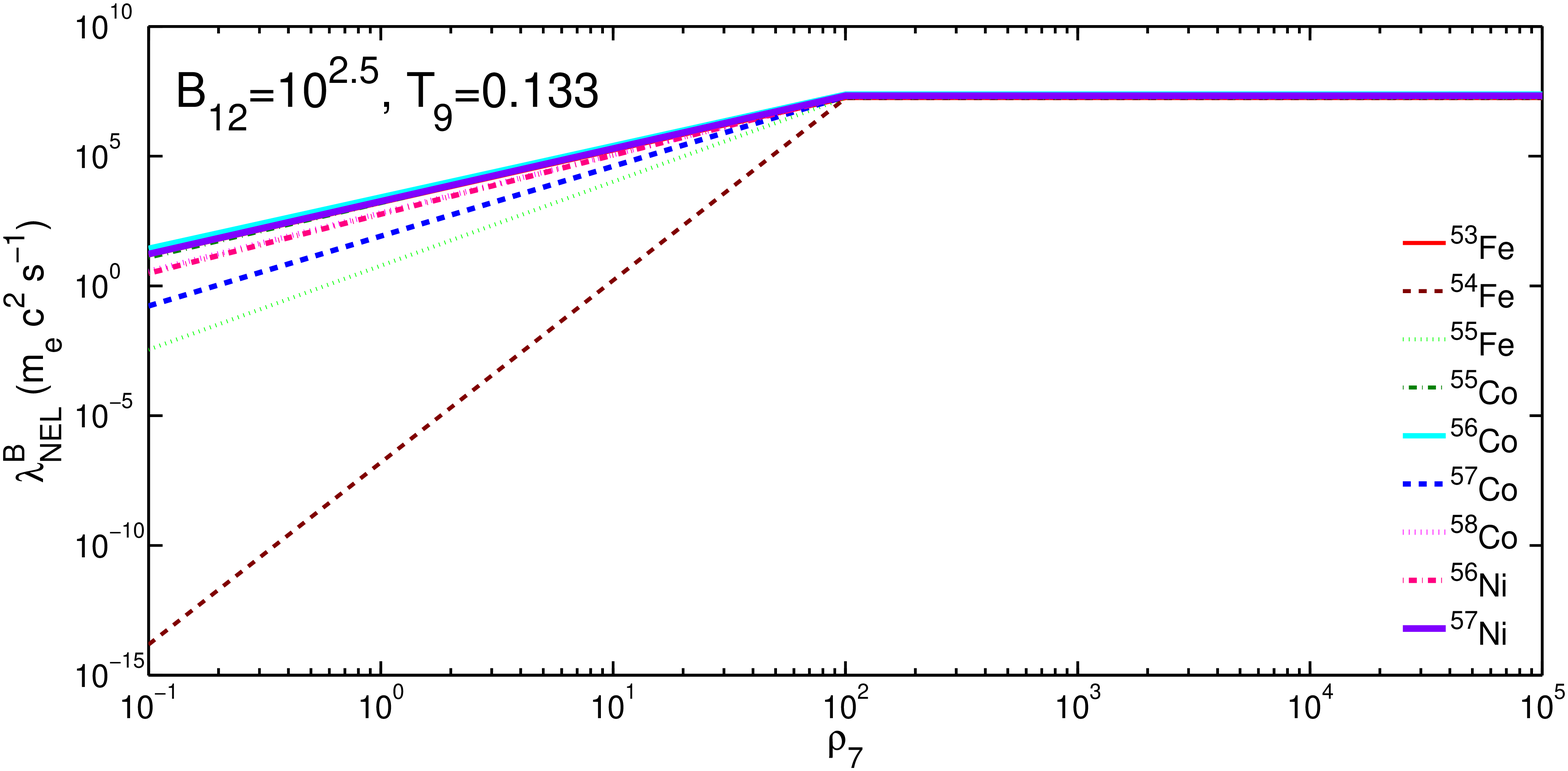}
    \includegraphics[width=4cm,height=4cm]{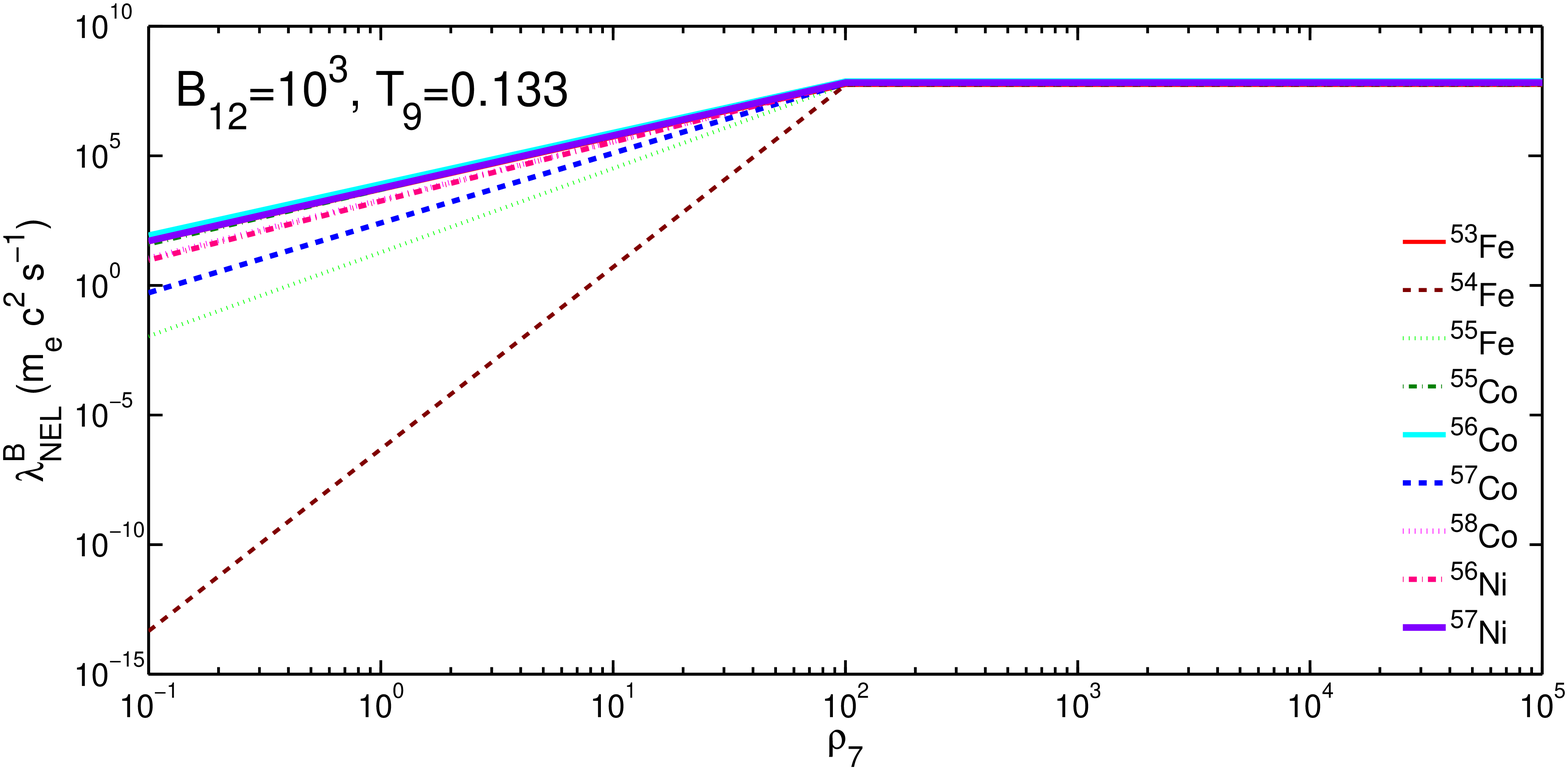}
    \includegraphics[width=4cm,height=4cm]{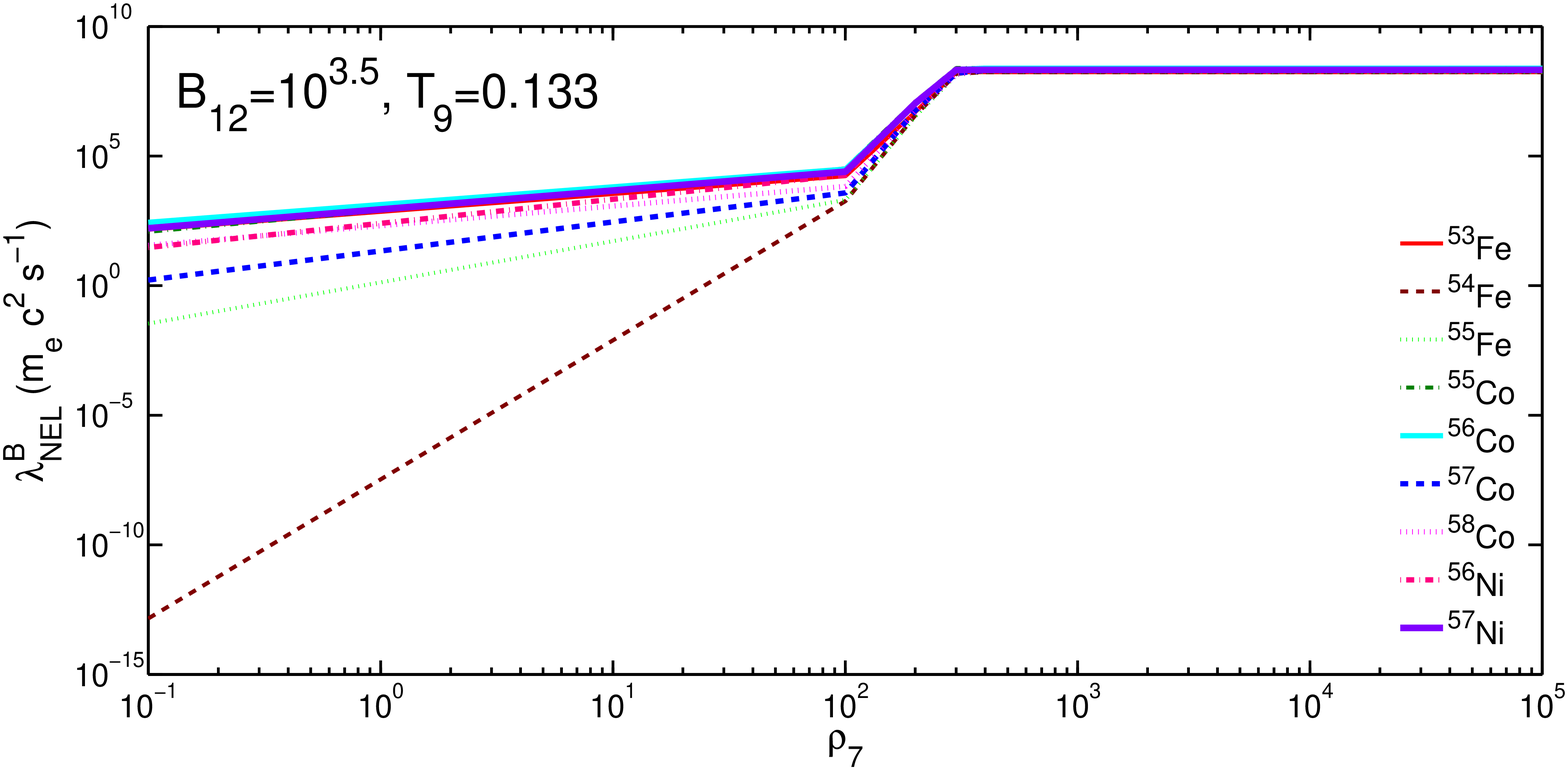}
    \includegraphics[width=4cm,height=4cm]{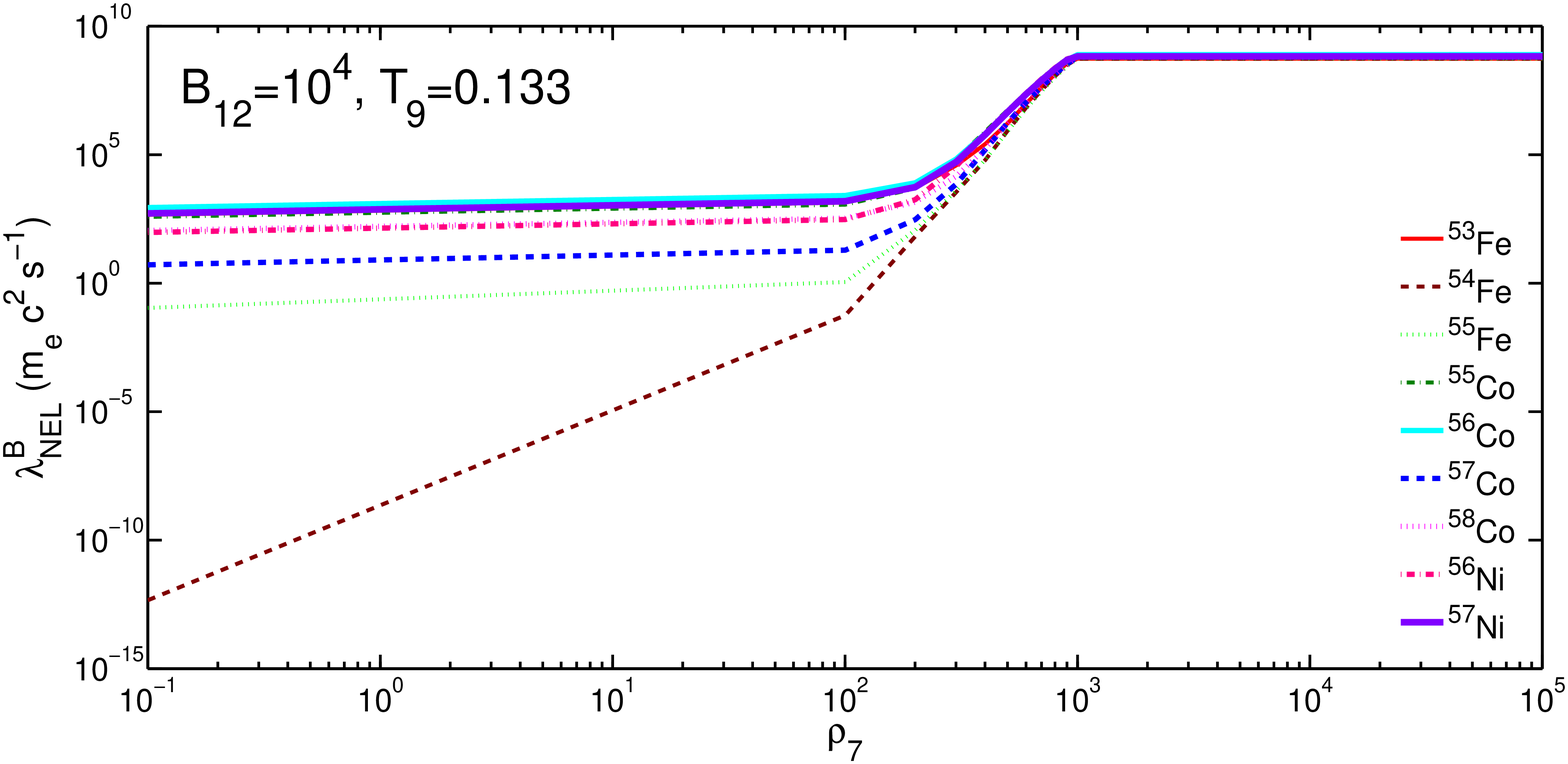}
    \includegraphics[width=4cm,height=4cm]{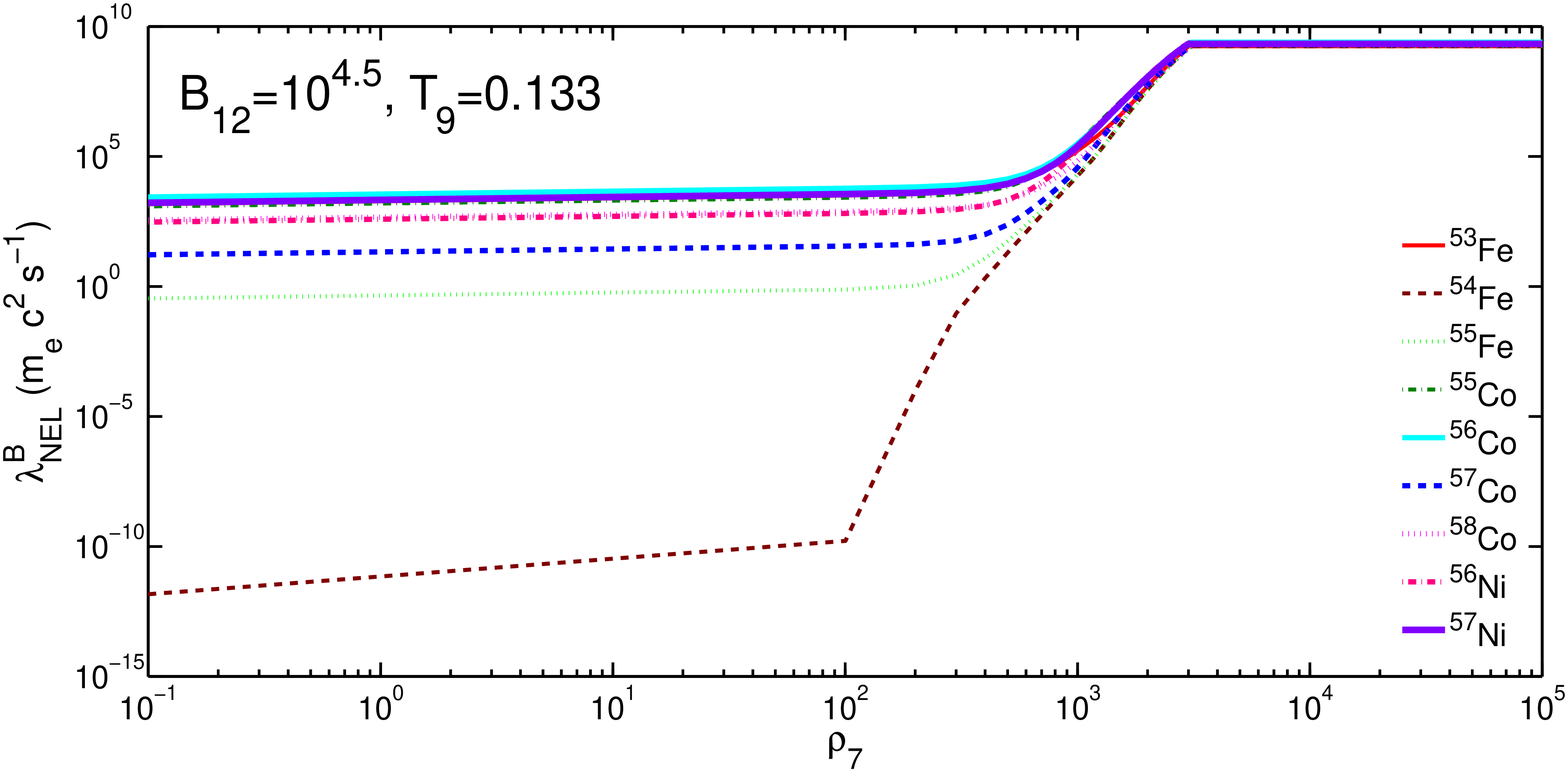}
    \includegraphics[width=4cm,height=4cm]{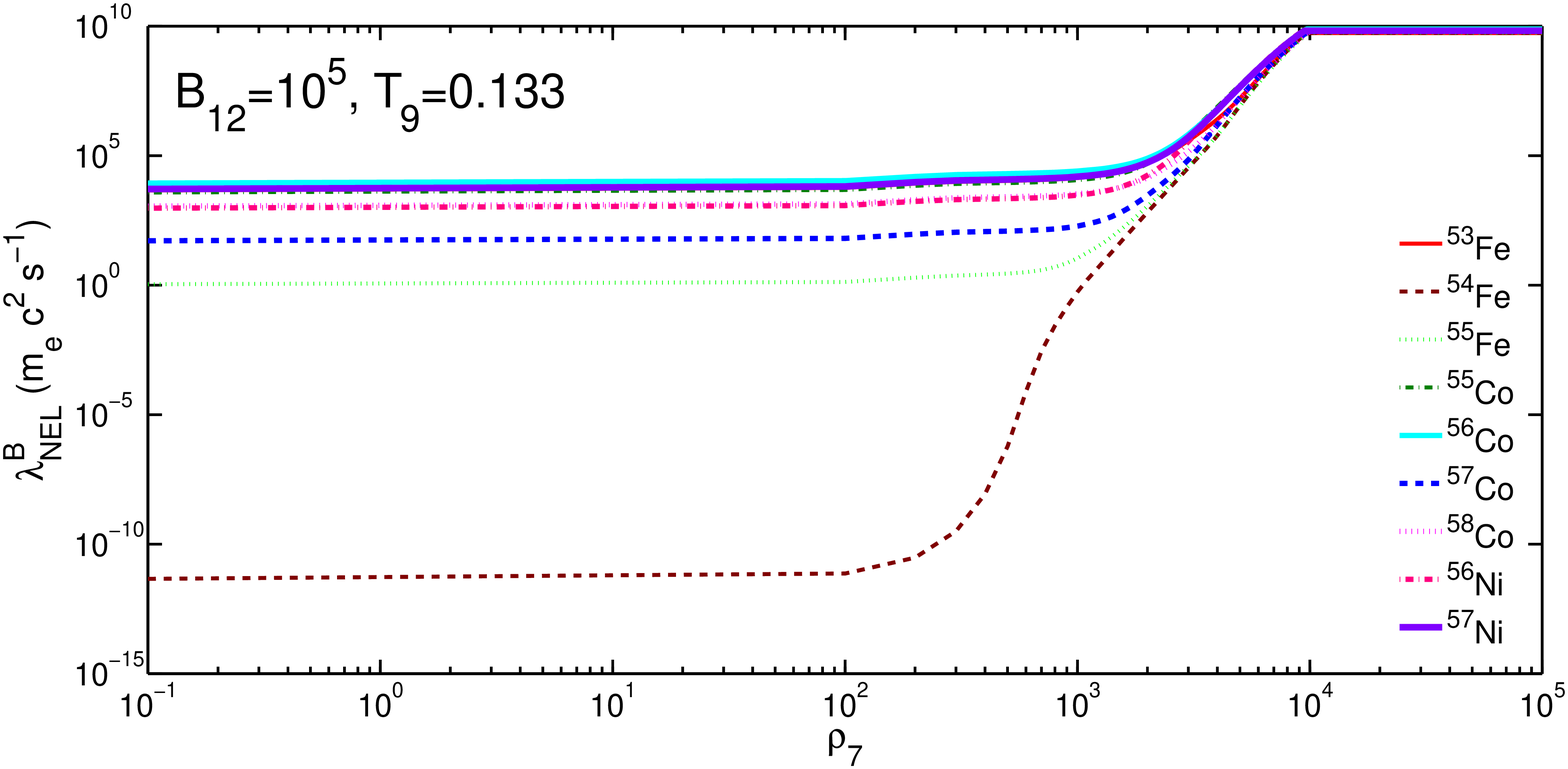}
   \caption{The NELRs for some typical iron group nuclei as a function of $\rho_{7}$ at $B_{12}=10^{2.5},10^{3},10^{3.5},10^{4},10^{4.5},10^{5}$, and $T_9=0.133$}
   \label{Fig:12}
\end{figure}

\begin{figure}
\centering
    \includegraphics[width=4cm,height=4cm]{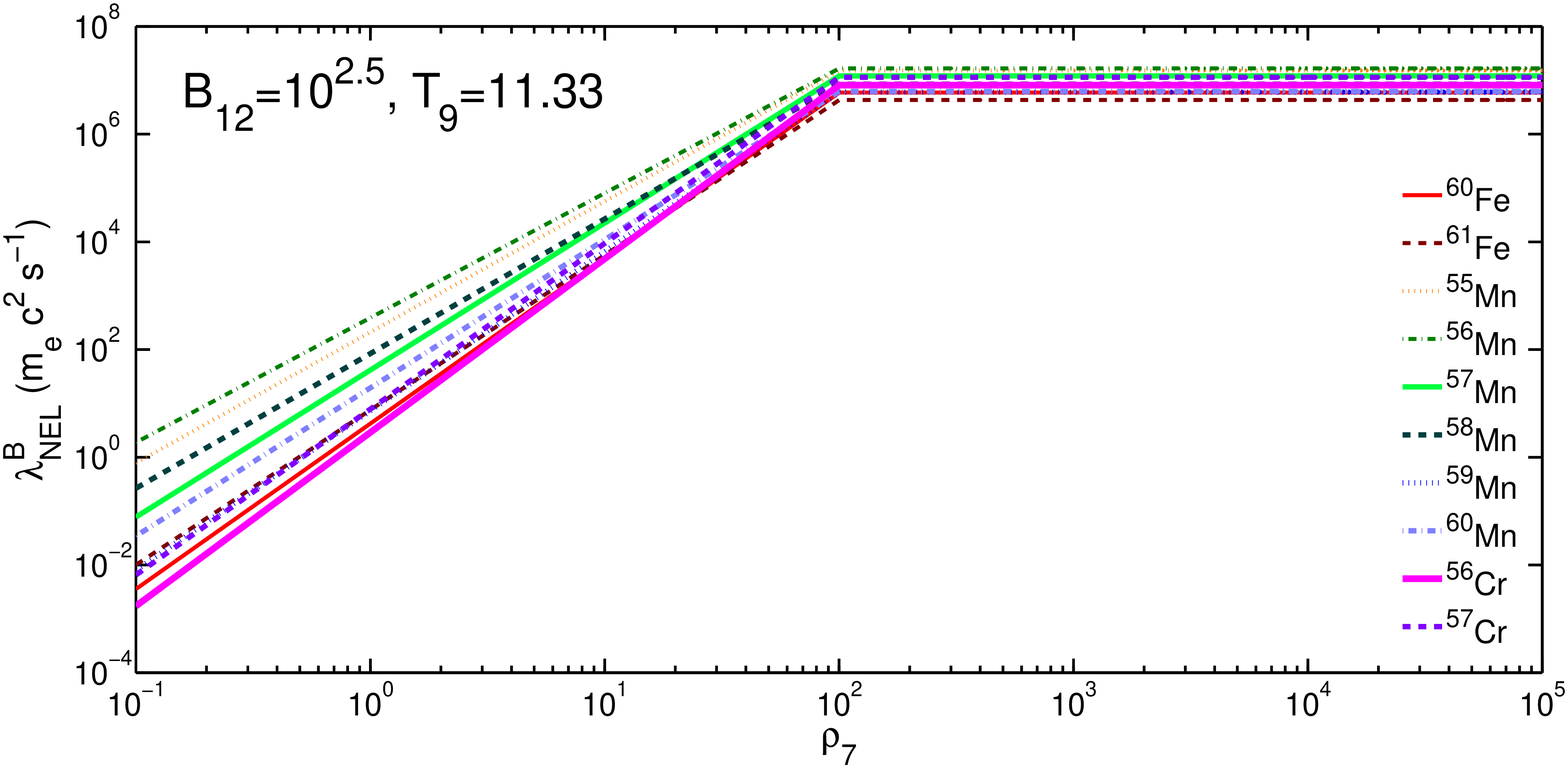}
    \includegraphics[width=4cm,height=4cm]{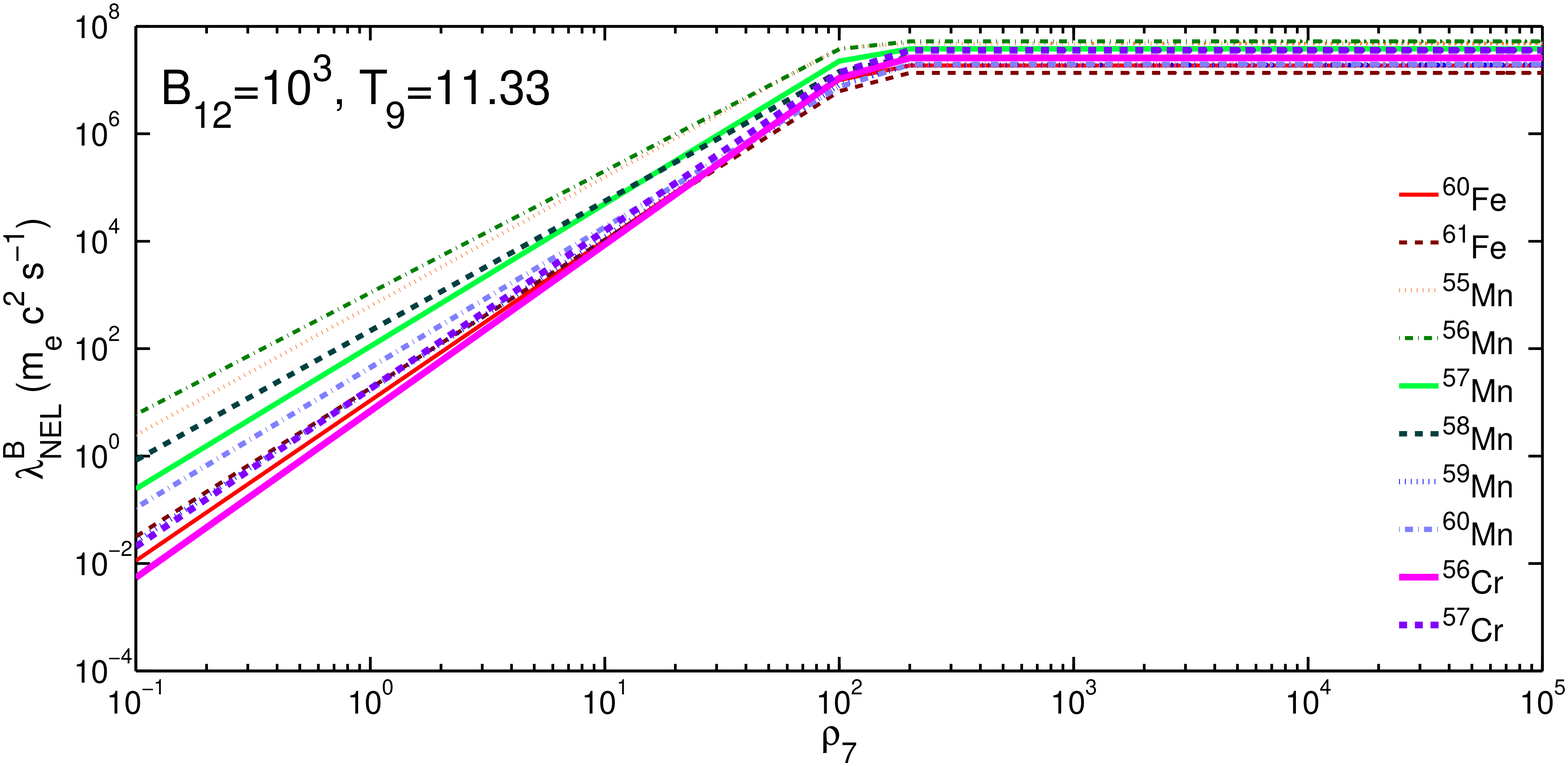}
    \includegraphics[width=4cm,height=4cm]{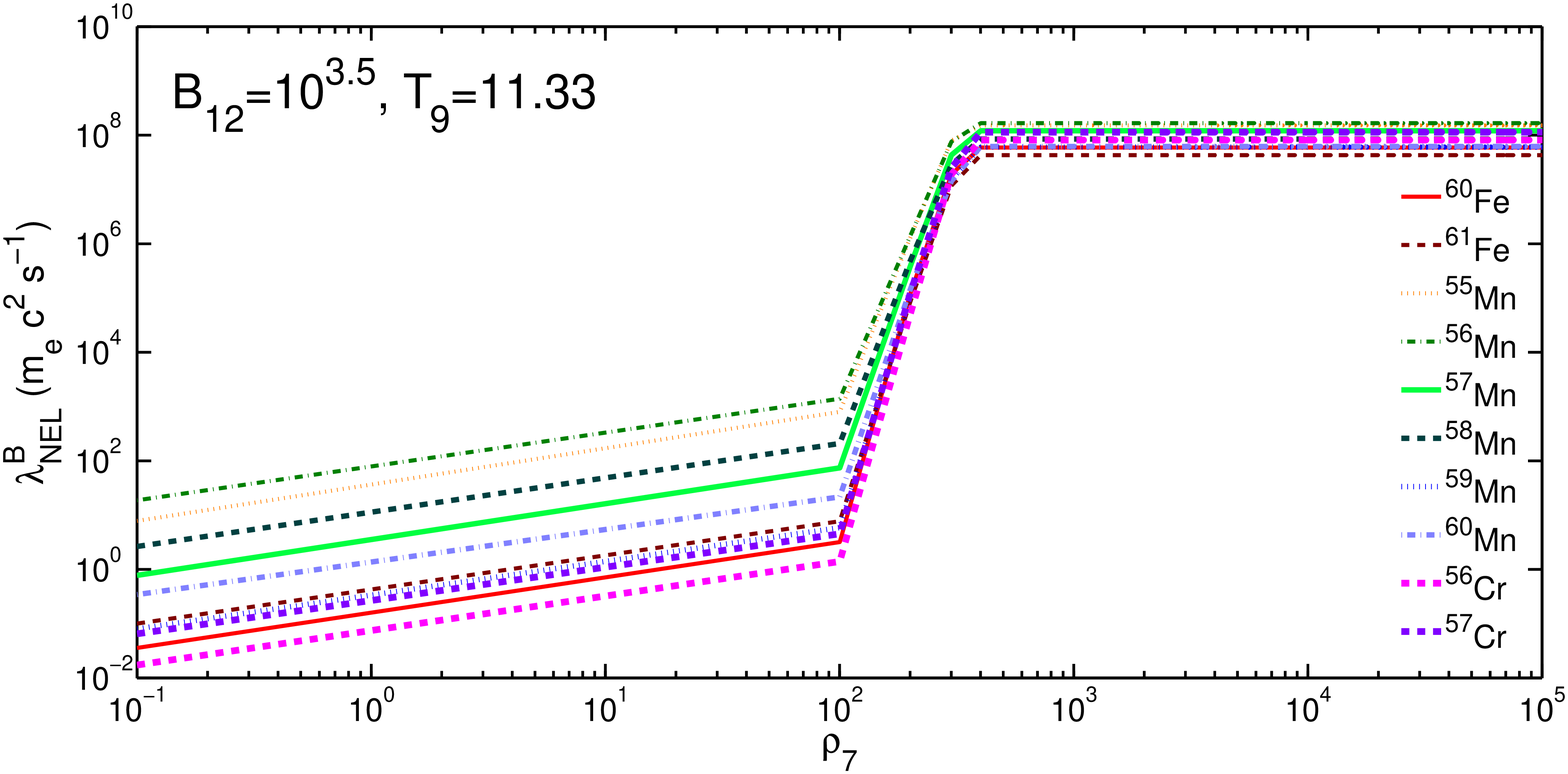}
    \includegraphics[width=4cm,height=4cm]{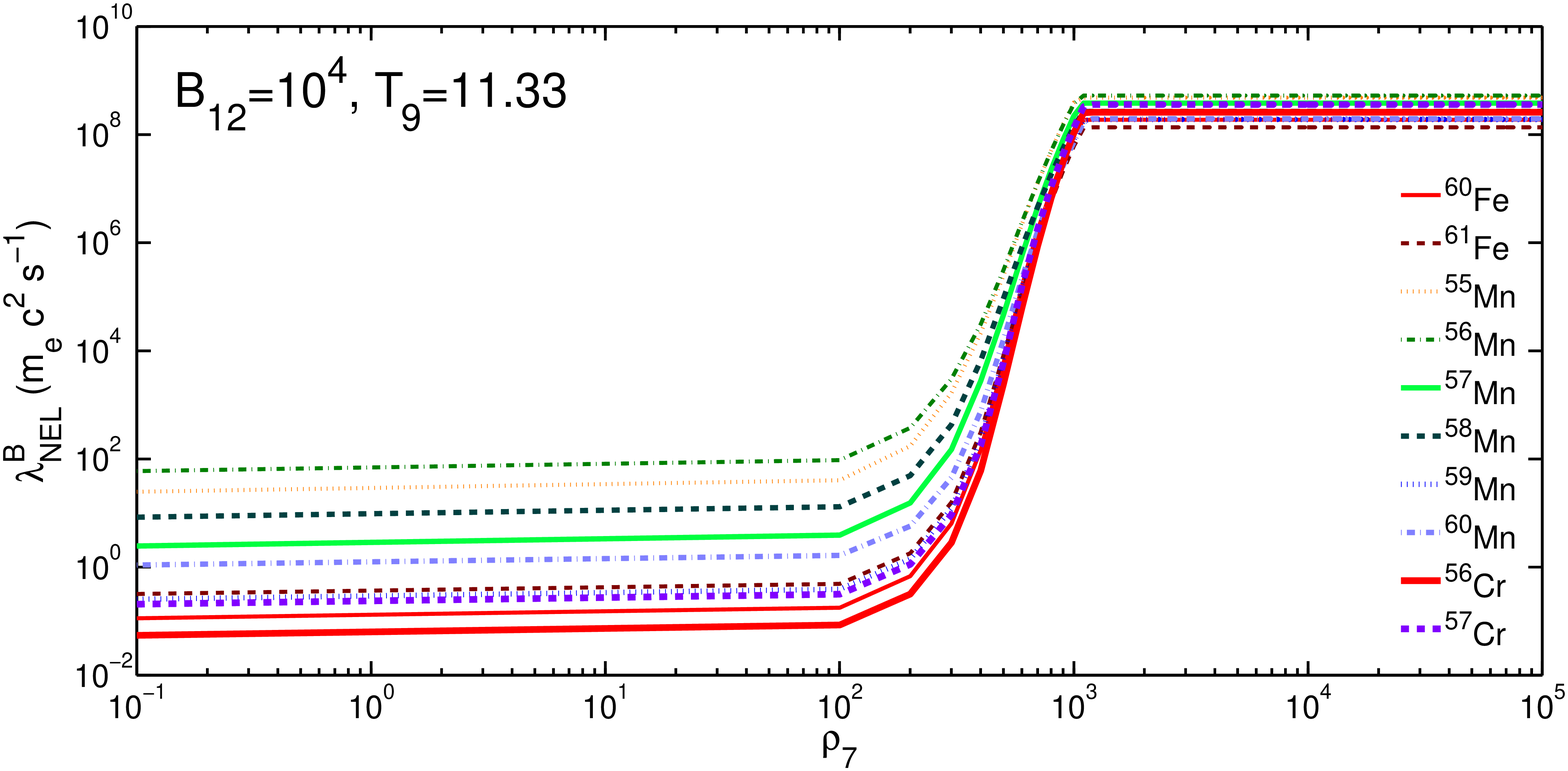}
    \includegraphics[width=4cm,height=4cm]{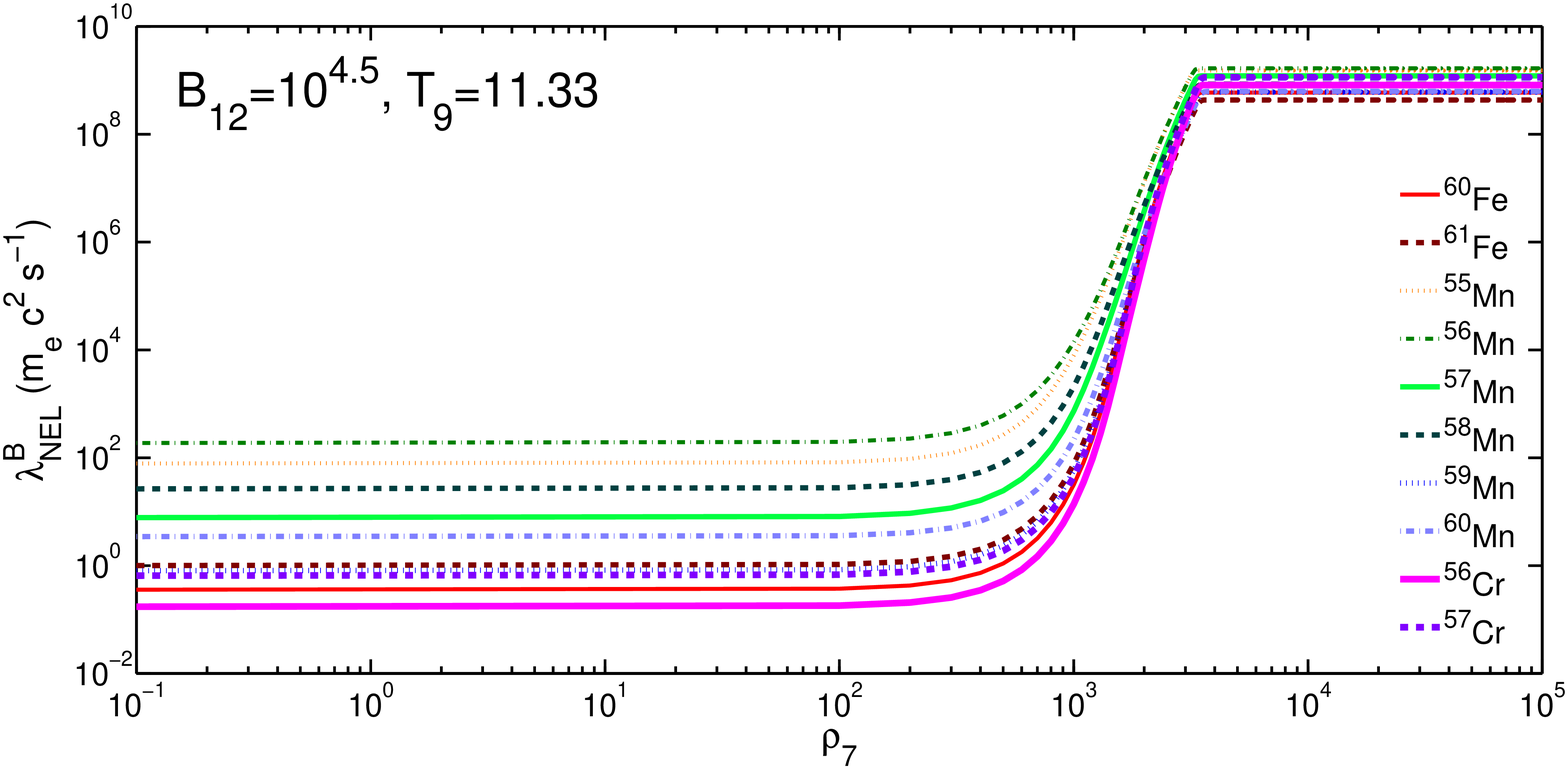}
    \includegraphics[width=4cm,height=4cm]{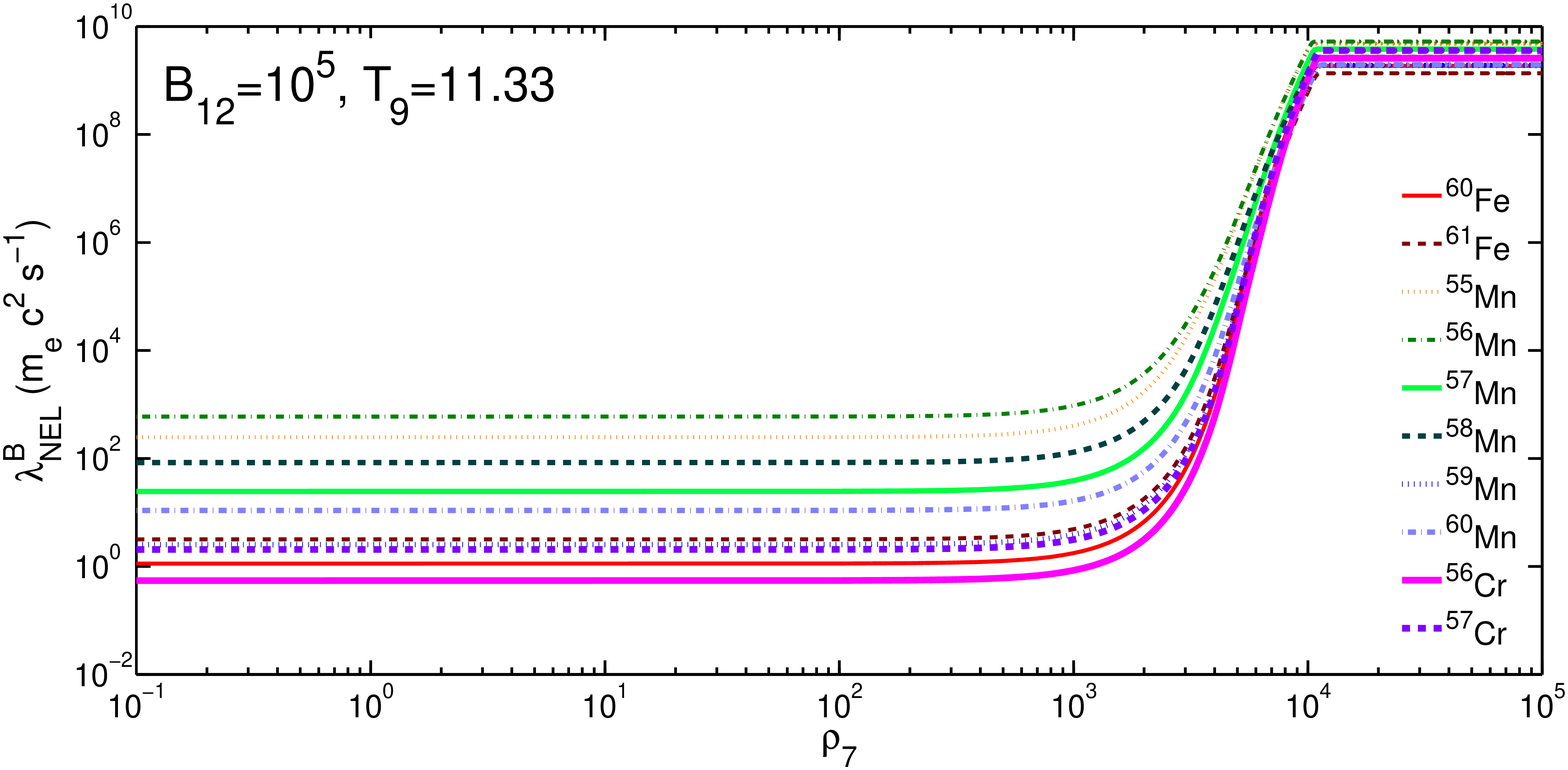}
   \caption{The NELRs for some typical iron group nuclei as a function of $\rho_{7}$ at $B_{12}=10^{2.5},10^{3},10^{3.5},10^{4},10^{4.5},10^{5}$, and $T_9=11.33$}
   \label{Fig:17}
\end{figure}

\begin{figure}
\centering
    \includegraphics[width=4cm,height=4cm]{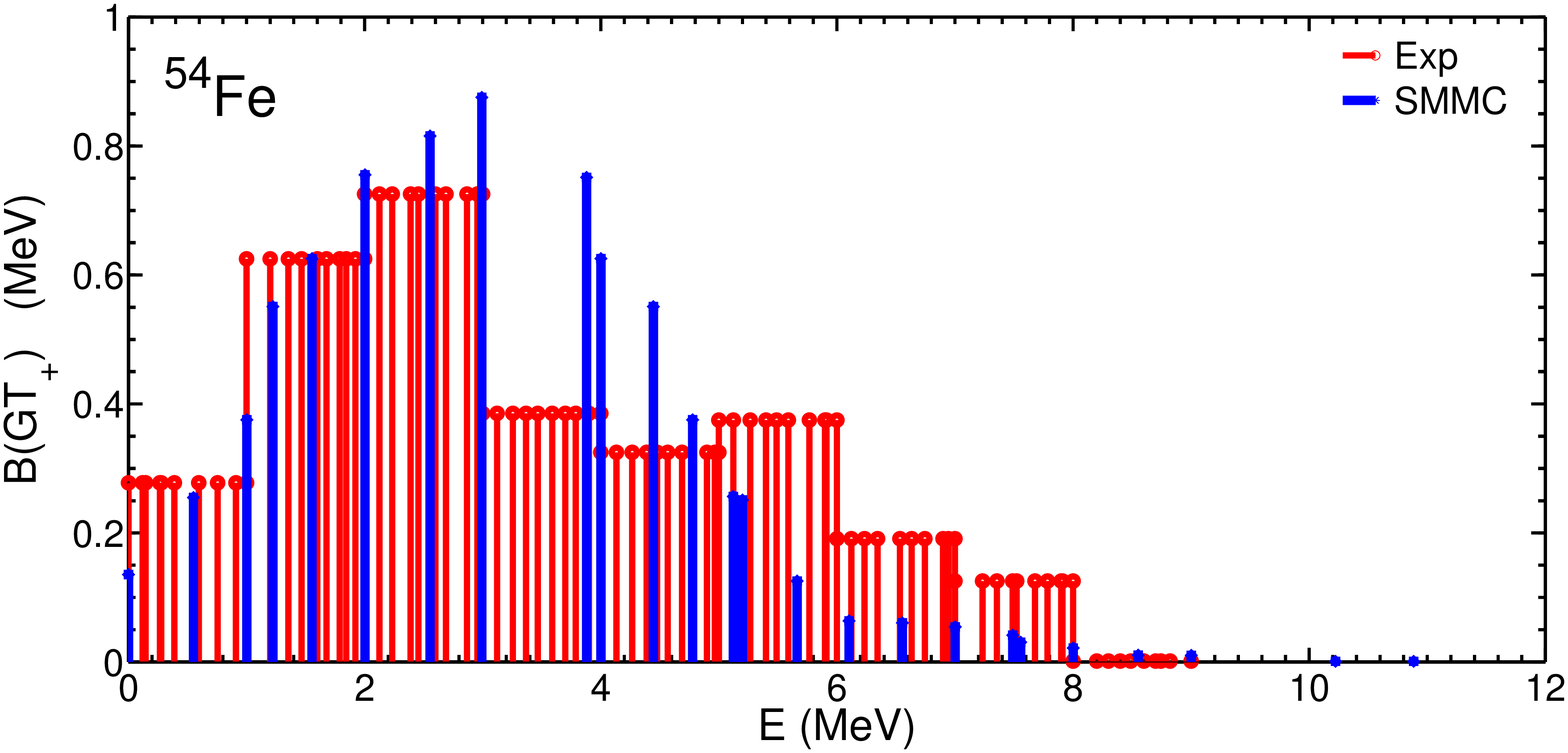}
    \includegraphics[width=4cm,height=4cm]{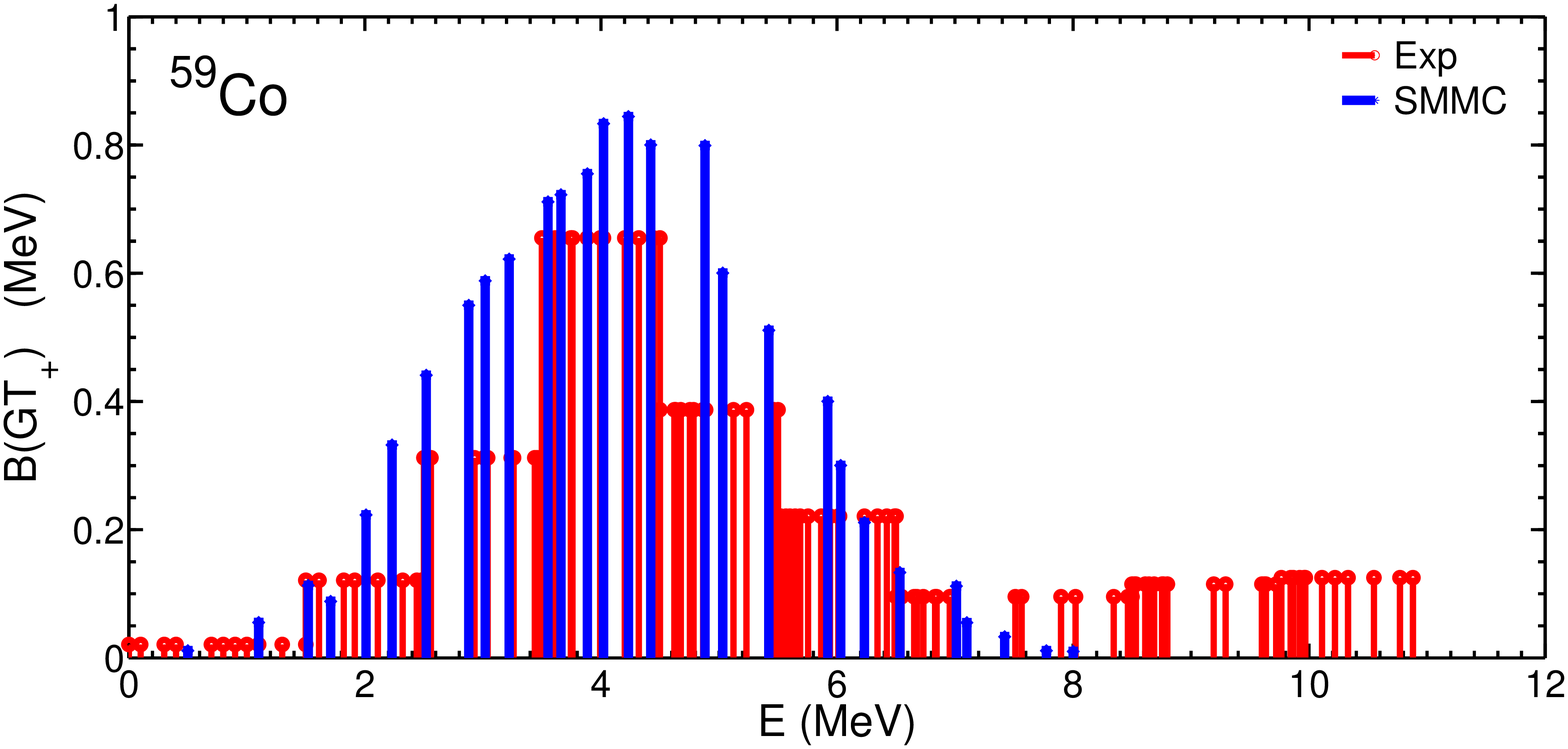}
    \includegraphics[width=4cm,height=4cm]{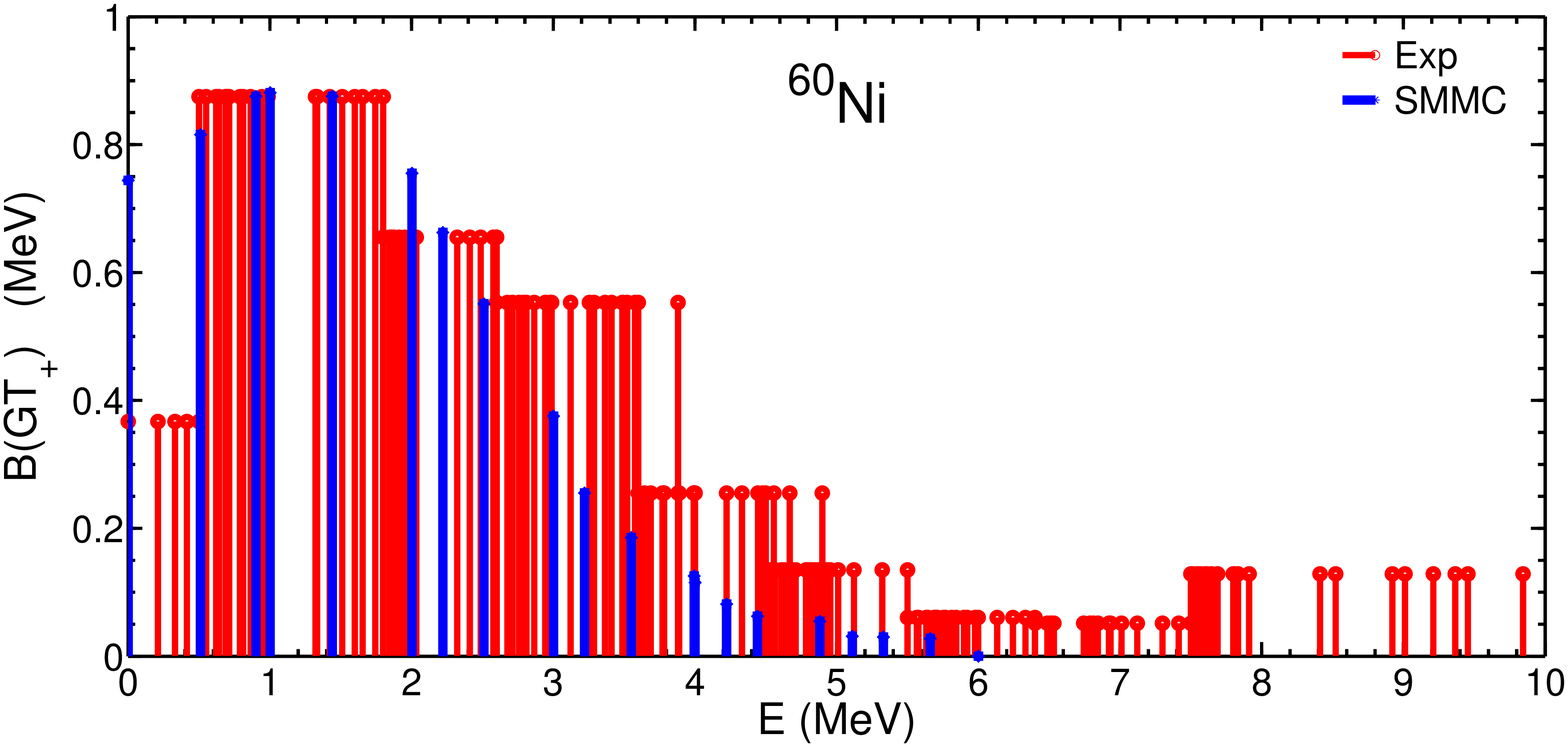}
    \includegraphics[width=4cm,height=4cm]{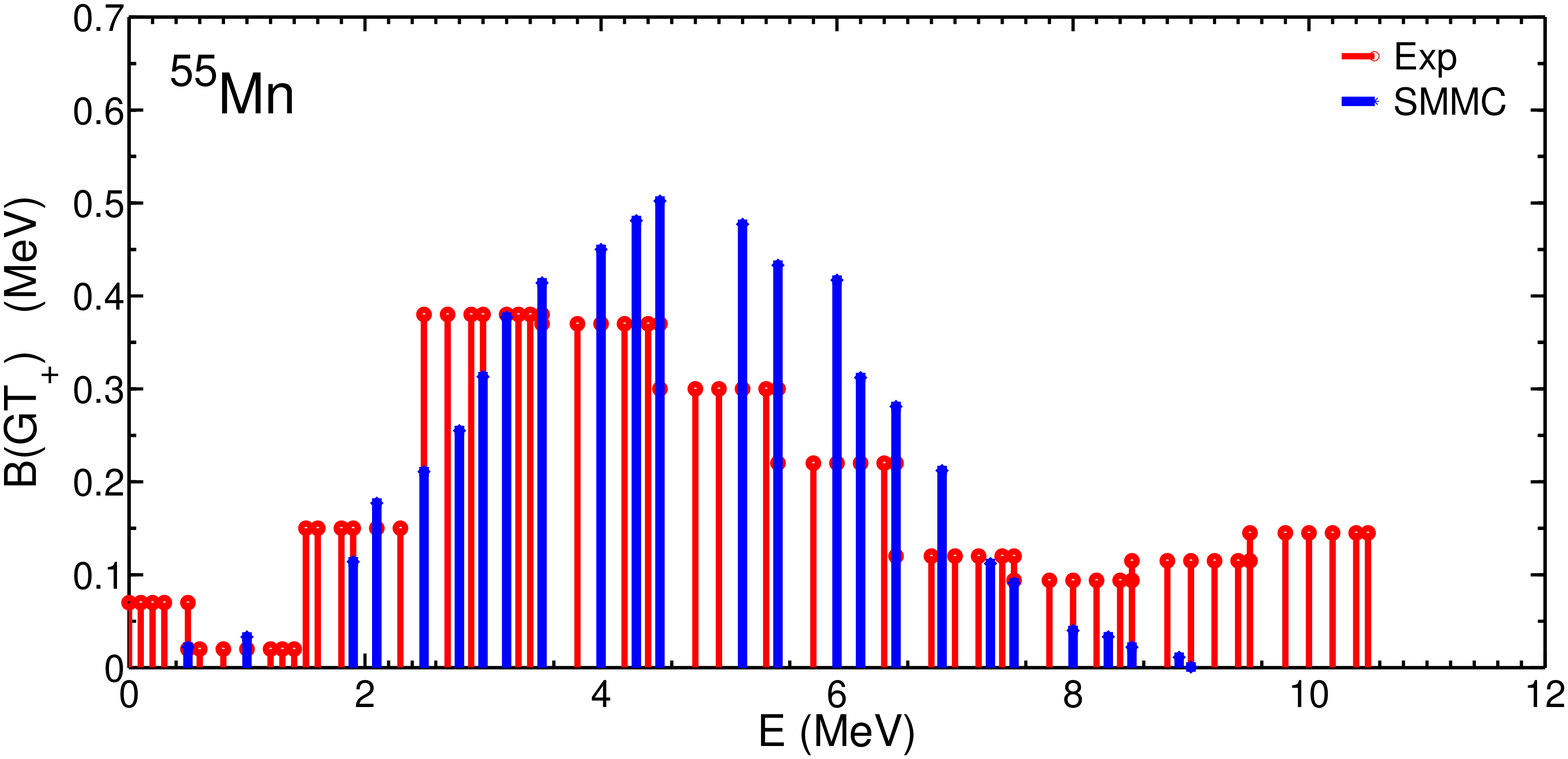}

\caption{The comparison of calculated B($\rm{GT}_+$) strength
distribution  against experiment\citep{b2, b60, b15, b55} for some
typical iron group nuclei as a function of excitation energy in the
corresponding daughter nuclei at temperature $T=0.8$MeV}
   \label{Fig:17}
\end{figure}

\begin{figure}
\centering
    \includegraphics[width=4cm,height=4cm]{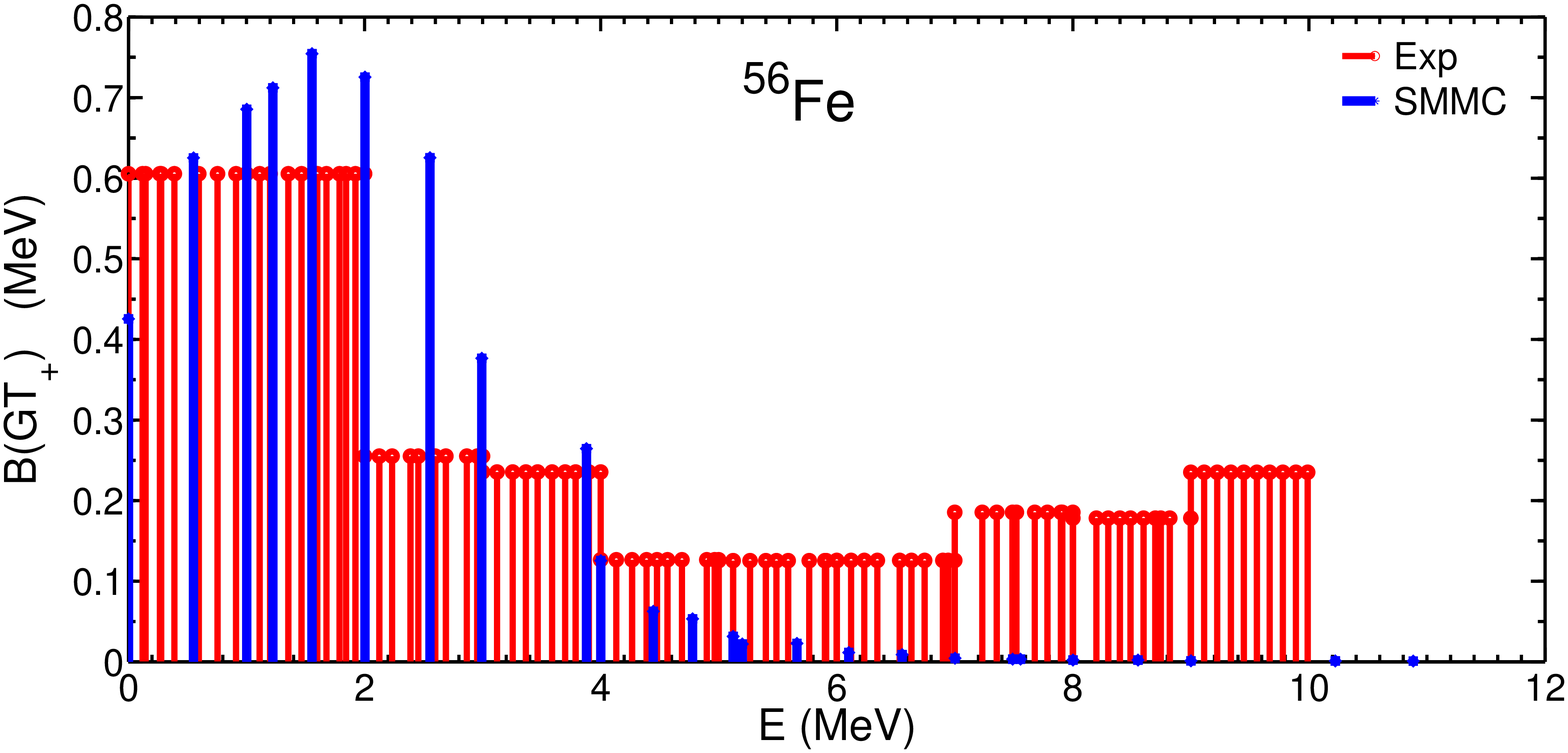}
    \includegraphics[width=4cm,height=4cm]{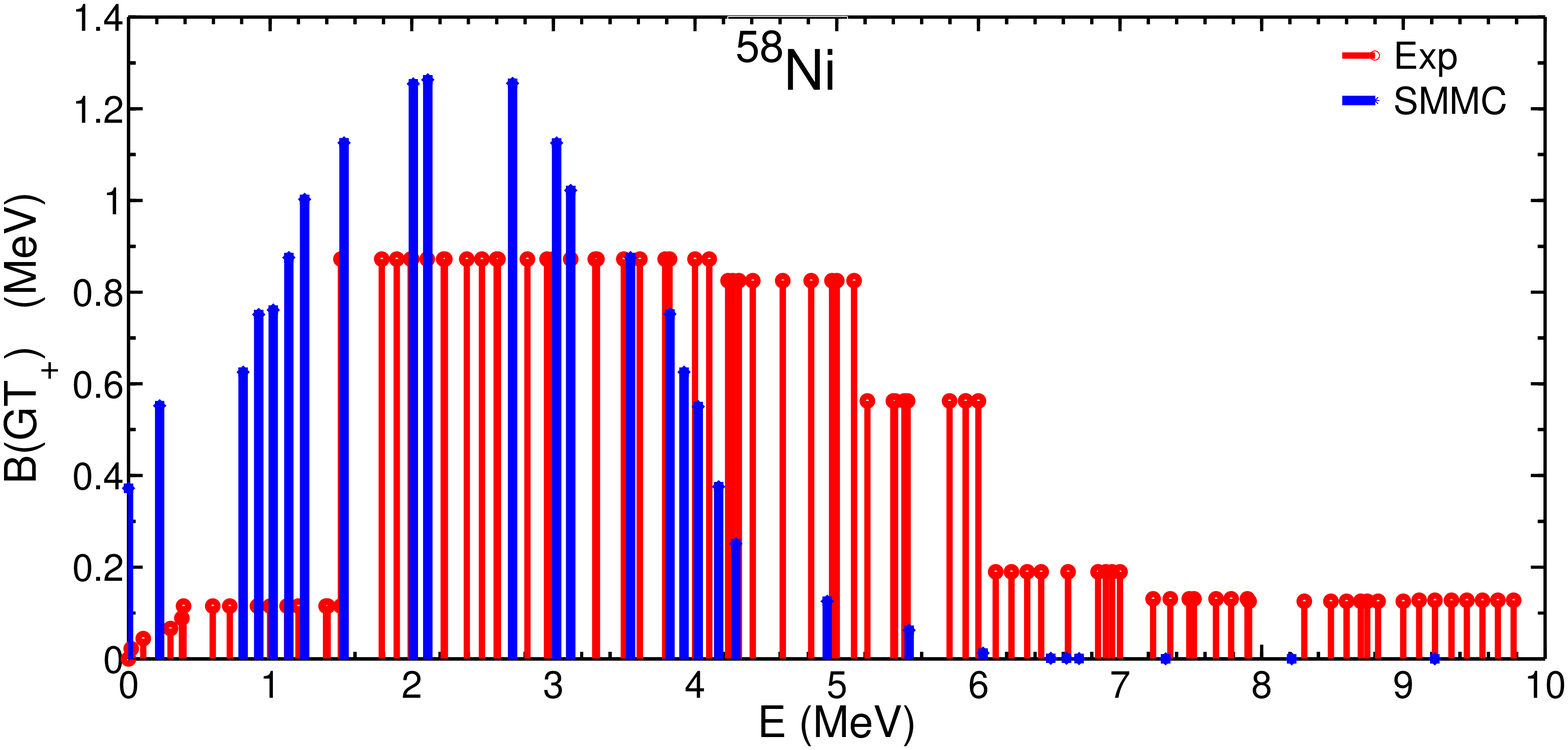}
    \includegraphics[width=4cm,height=4cm]{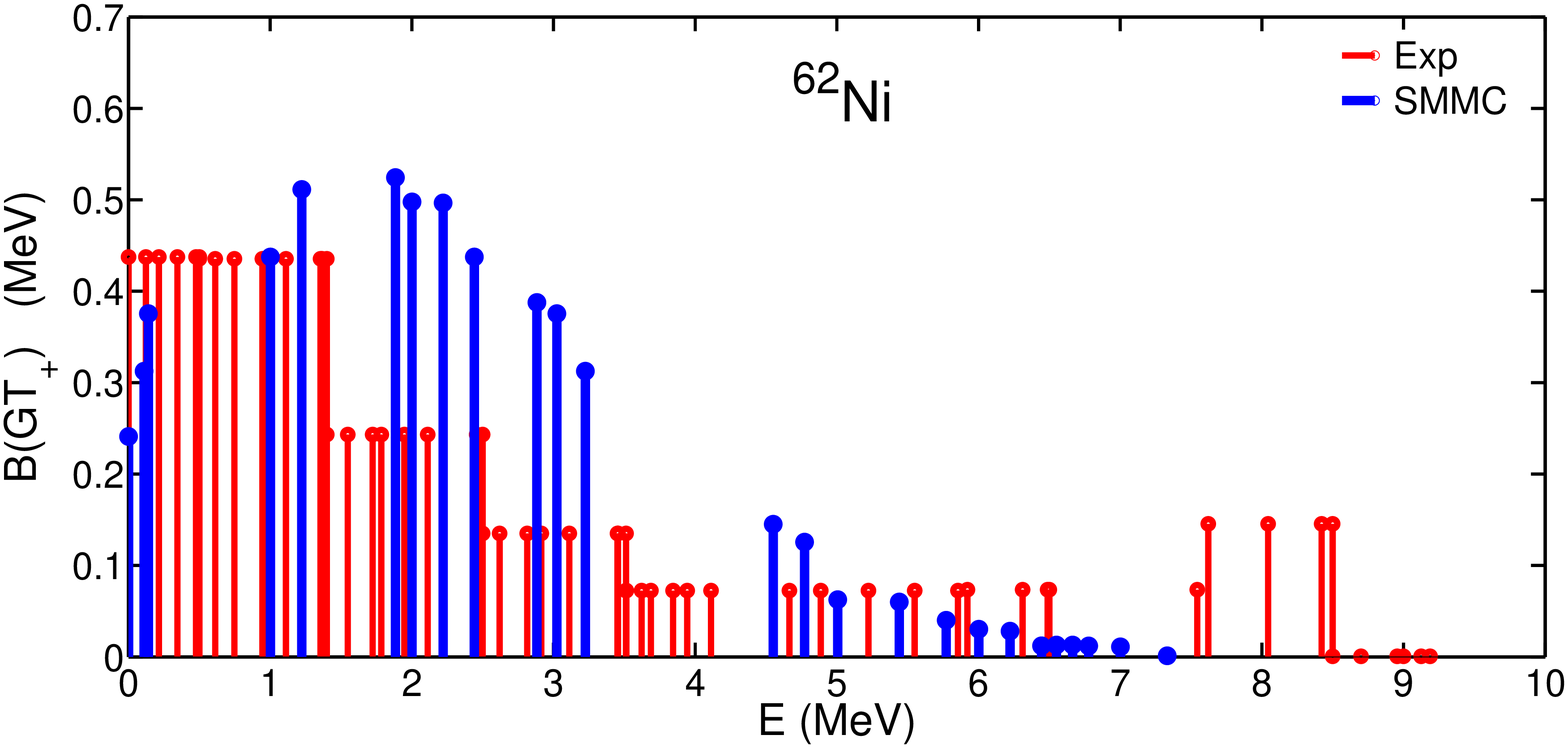}
    \includegraphics[width=4cm,height=4cm]{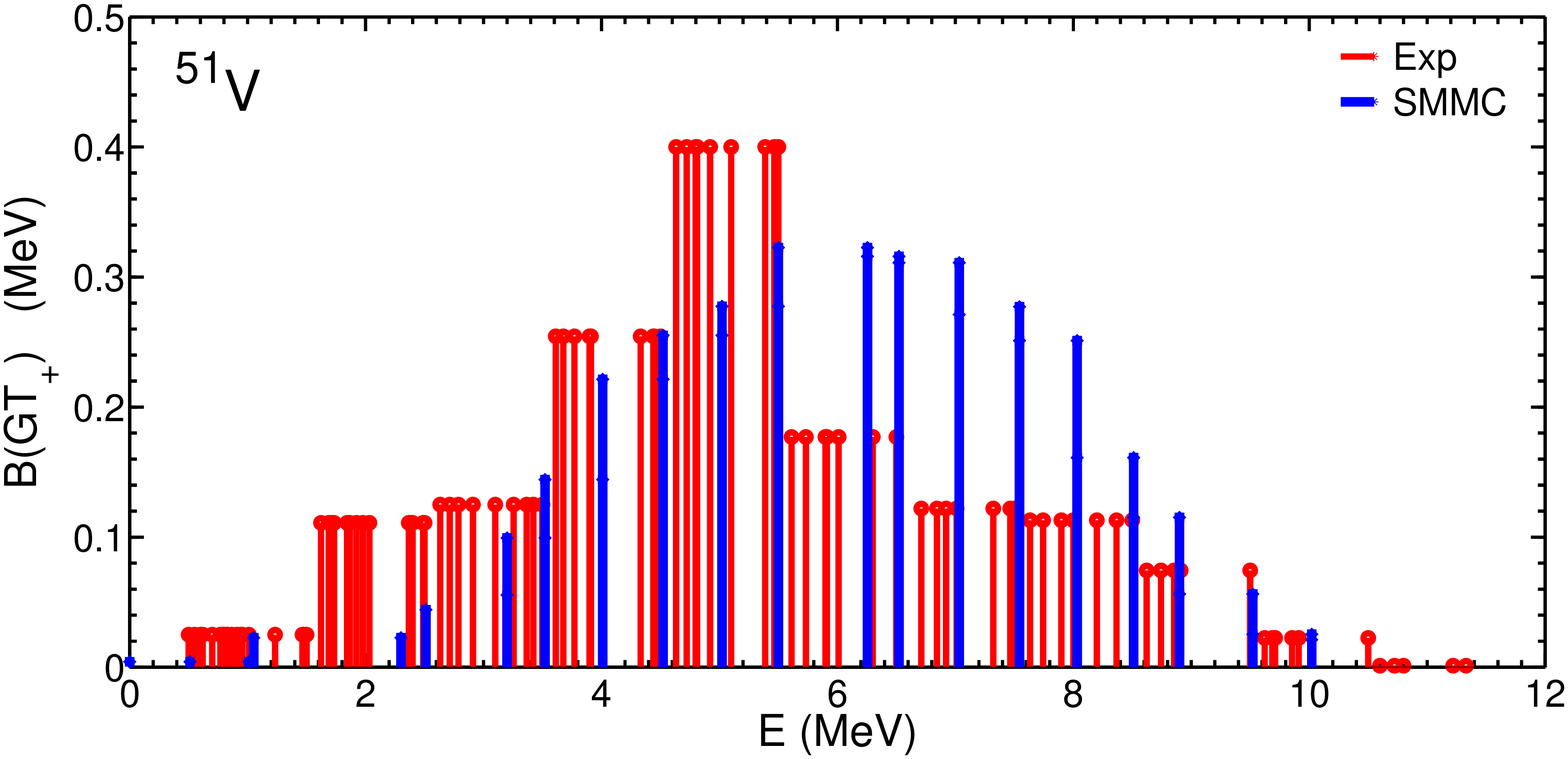}

   \caption{The comparison of calculated B($\rm{GT}_+$) strength
distribution  against experiment\citep{b2, b60, b15, b55} for some
typical iron group nuclei as a function of excitation energy in the
corresponding daughter nuclei at temperature $T=0.8$MeV}
   \label{Fig:17}
\end{figure}

However, form figures 1-4, one finds that a systematic decrease in
neutrino energy loss rate when $B_{12}>100$. We know that the Landau
energy level spacing will become a very small fraction for the Fermi
energy when $\nu_{max}\gg 1$. The continuous integral may take in
lieu of the discrete sum over all Landau level. The thermodynamic
relation will reduce to the case of non-magnetic fields. We have
$\nu_{max}\rightarrow U_e/\hbar \omega_c$  when the electron gas is
of mildly relativistic state. For relativistic electron gas, the
$\nu_{max}\geq 100/ B_{12}$ when the $ U_e\geq 1$MeV. Therefore, as
magnetic field strength increases(i.e. when $B_{12}\geq 10\sim
100$), the $\nu_{max}$ will tend to the order of unity(or zero), and
the field will be termed strongly quantizing. On the other hand, The
structure at the outer crust of magnetars is fundamentally
determined by the energies of isolated nuclei, the kinetic energy of
electrons and the lattice energy. Thus its composition is strongly
depends on the binding energy per particle of stable and unstable
nuclei in the outer crust of magnetar below the neutron drip
density. Based on the relativistic mean-field effective interactions
NL3 \citep{b34} and DD-ME2 \citep{b35}, the influences of SMFs on
the binding energy of nuclei have been investigated following the
works of \citet{b51} and \citet{b5}. We find that the binding energy
per particle will have a mean parabolic increasing trend  with
increasing of the magnetic field. For example, the binding energy
increase by 0.311MeV, 0.632MeV, 0.445MeV for $^{56}$Fe, $^{78}$Ni,
$^{56}$Co, respectively when the magnetic field strength from
$10^{17}$G to $10^{18}$G.  Due to increase of nuclear binding
energy, the nuclear will be more stable. This is equivalent to
significantly raise the threshold energy of EC reaction. Thus, the
NELRs and EC rates will decreased in SMFs. Meanwhile, as the
magnetic field strength increases, the electron Fermi energy will
decrease greatly due to interaction between the electrons and SMFs.
This actually discourages the EC reaction. So the the NELRs and EC
rates decrease.

The magnetic field strongly effects on the electron phase space,
Only axial symmetry is preserved, and breaks spherical symmetry for
the Dirac and Klein-Gordon equations \citep{b51}. For a certain
value of magnetic field, Figures 5-6 present the NELRs of some
typical iron group nuclei versus the density $\rho_7$ at temperature
of $T_9=0.133, 11.33$. One finds that the NELRs increase greatly and
even exceed by four orders of magnitude for a certain value of
magnetic field and temperature. With increasing of the density,
there are almost not influence of density on NELRs. On the other
hand, for the density around $\rho_7= 10^2$, there is an abrupt
increase in NELRs when $10^{3.5}\leq B_{12}\leq10^{5}$. Such jumps
are an indication that the underlying shell structure has changed in
a fundamental way. These jumps in nuclear properties can be traced
to the single-particle behavior due to SMFs. As the magnetic field
increases, a particle will remove from a level going upwards and
bring to a level going downward with increasing spin. Furthermore,
the nucleus becomes spin-polarized due to these two levels have
opposite angular momentum along the symmetry axis.

The SMFs influences on the single-particle structure of nuclei for
protons and neutrons. Firstly, the interaction between the magnetic
field and the neutron (proton) magnetic dipole moment will cause
nucleon paramagnetism. Secondly, the coupling of the orbital motion
of protons with the magnetic field will also cause proton orbital
magnetism. Due to the interaction between the nucleus and SMFs, all
degeneracies in the single-particle spectrum may be removed, and the
formerly degenerate levels with opposing signs of angular momentum
projection will also tends to break (as a example, for $^{56}$Fe,
the detailed discussions can be seen from Figs. 3 and 5 of
Ref.\citep{b51}). Such single-particle energy splitting will produce
a reduction of the neutron and proton pairing gaps with increasing
magnetic fields and, eventually, their disappearance

According to the discussion of the influences from the
single-particle level by SMFs, one finds that the Kramer's
degeneracy in angular momentum projection of proton levels is
removed by the orbital magnetism associated with proton ballistic
dynamics, which can bring those aligned with the magnetic field down
in energy. On the other hand, the paramagnetic response (Pauli
magnetism) also removes the angular momentum projection degeneracy
for both protons and neutrons.

Synthesizes the above analysis, one can concludes that the last
occupied single-particle levels(e.g., for $^{56}$Fe)for neutron,
including the influence of the proton orbital coupling and the
anomalous magnetic moments coupling, decreases as the magnetic field
strength increases. However, the last occupied single-particle
levels for proton will increase. As is known to all, the EC is
actually the process that protons will turn into neutrons and
discharge a neutrino when a nuclei capture an electron. Thereby, to
a certain extent, these influences of SMFs on single-particle level
for proton and neutron states, ultimately make the EC reaction
become more active and increase the NELRs.

The SMFs may not directly influence on the lattice energy.
Nevertheless, some indirect influence on the lattice configuration
will be caused by Coulomb screening.  We find energetically
favorable to arrange the ionized nuclei in a Coulomb lattice in the
typical density range of the magnetar surface(i.e.
$10^4\rm{g/cm^3}\sim 4\times10^{11}\rm{g/cm^3}$). In relativistic
lower range of density, the electron energy and the Coulomb crystal
does not play a relevant role in magnetar crust. As the density
increases, the electron energy raises greatly as compared to the
total energy, which is very advantageous for electronic capture
processes. however, the lattice energy influence remain negligible.
So, we ignore the influence of SMFs on the lattice energy, the EC
and NELRs. On the other hand, the cyclotron energy $\hbar
\omega_{ce}$ is much larger than the typical Coulomb energy,
Therefore, the properties of atoms, molecules and condensed matter
are qualitatively changed by the magnetic field when $B\gg
2.3505\times10^9$G.  The usual perturbative treatment from the
magnetic influence on Zeeman splitting of atomic energy levels does
not apply in such regime of SMFs due to the Coulomb forces act as a
perturbation to the magnetic forces\citep{b25}. The Coulomb force
becomes much more effective in binding the electrons along the
magnetic field direction due to the extreme confinement of the
electrons in the transverse direction (i.e. perpendicular to the
magnetic field). The atom attains a cylindrical structure. Moreover,
it is possible for these elongated atoms to form molecular chains by
covalent bonding along the field direction.

One can also see that as SMFs increases, the change of NELRs will
reflect some difference due to strong quantum effects in SMFs from
Figures 5-6. As the density increases, when $\rho\geq \rho_B$ and
$T\leq T_B$, where
$\rho_B=7.04\times10^3Y_e^{-1}B_{12}^{3/2}\rm{g/cm^3},
T_B=1.34\times10^8B_{12}(1+x_e^2(\nu))^{-1/2}\rm{K}$, the electrons
will be strongly degenerate, and populate many Landau levels
\citep{b38, b39}. The magnetic field is termed weakly quantizing.
Thus the chemical potential, EC rates, and NELRs are only slightly
affected by SMFs. With increasing $T$, the oscillations become
weaker because of the thermal broadening of the Landau levels. When
$T\geq T_B$ or $\rho \gg \rho_B$, more electrons will populate many
Landau levels and the thermal widths of the Landau levels ($\sim
kT$) are higher than the level spacing. The magnetic fields have
almost no influence on the EC and NELRs.

GT strength distributions play an important role in EC process in
the astrophysical context. \citet{b17}, and \citet{b11} demonstrated
that the GT transition matrix elements for EC don't depend on the
magnetic fields. Thus we will neglect the effect of SMFs on the GT
properties of nuclei in this paper. A strong phase space dependence
makes the EC rates more sensitive to GT distributions than to total
strengths. We present GT strength distributions from shell model
Monte Carlo studies of some typical fp-shell iron group nuclei in
Figures 7-8. We also display the experimental data about GT
distributions \citep{b2, b60, b15, b55}, which are obtained from
intermediate-energy charge exchange (n, p) or (p, n) cross sections
at forward angles, which are proportional to the GT strength. We
exclude contributions from other multipolarities for these
experimental distributions, which extend only to 8 MeV in the
daughter nucleus. We compare our SMMC results for the GT$_+$
distribution against experiment in Figures 7-8. One finds that the
SMMC results for all even-even nuclei (e.g., $^{54, 56}$Fe; $^{58,
60, 62}$Ni)have been smeared with Gaussians of standard deviation of
1.77 MeV to account for the finite experimental resolution. From the
perspective of (n, p) experiment, the GT$_+$ strength is
significantly fragmented over many states, and these distributions
centroids and widths are reproduced very well in the SMMC approach.
Our results for the total strengths are agreed well with the
experimental data. For example, the renormalized B(GT$_+$) strengths
from SMMC approach are 4.120, 2.682, 4.542, 3.510, 2.410 MeV for
$^{54, 56}$Fe, $^{58, 60, 62}$Ni, respectively. And the B(GT$_+$)
strengths from experiment are 3.70, 2.601, 4.203, 3.200, 2.600 MeV
for the same nuclei, respectively. On the other hand, for some odd-A
nuclei(e.g., $^{51}$V, $^{59}$Co, $^{55}$Mn), SMMC results from the
(n, p) direction are also in good agreement with the data of
experiment \citep{b2, b15, b55}.

The structure and composition of the crust is important in the
thermal and magnetic evolution of neutron stars. The SMFs not only
strongly influence on the weak interaction rates and NELRs, but also
influence the late evolution and determine the core entropy and
electron to baryon ratio of magnetars. Tables 1-2 display our
results of the maximum value of NELRs when $10<B_{12}<1000$ in
different astrophysical environments. We find the maximum value of
the NELRs will get to $5.694\times 10^7, 7.942\times 10^7,
7.760\times 10^7, 6.376 \times 10^7, 6.444\times 10^7, 6.567 \times
10^7, 6.701\times 10^7, 6.068\times 10^7$ when $B_{12}=10^{3}$ at
relatively low temperature and  high density surrounding (i.e.
$\rho_7=106, T_9=0.233$) for $^{56}$Fe, $^{55, 56, 57, 58}$Co,
$^{56, 57}$Ni, and $^{48}$V, respectively. Nevertheless, under
relativistic high temperature and high density surrounding (i.e.
$\rho_7=106, T_9=15.53$), the maximum value of the NELRs will get to
$1.062\times 10^8, 1.195\times 10^8$ when $B_{12}=10^{3}$ for
$^{54}$Cr, and $^{47}$V, respectively.

According to above discussion, we can draw a conclusion that the
SMFs has a significant influence on the NELRs for a given
temperature-density point. Generally the stronger the density and
the lower the SMFs, the larger affect on the NELRs becomes. One can
also find, when $10^{14}{\rm{G}}\leqslant B \leqslant
10^{16}{\rm{G}}$, for most iron group nuclei, the rates decrease
greatly for a given temperature-density point. The reason is that
the Fermi energy of electrons decreases, but the binding energy of
the nucleus will increase with the increasing of SMFs when the
temperature and density are constant. Thereby, these lead to more
and more electrons whose energy will be less than the threshold for
EC process.

Tables 3-4 display the comparisons of our results with those of FFN
\citep{b19, b20}($\lambda^0_{\rm{NEL}}$(FFN)), and
NKK($\lambda^0_{\rm{NEL}}$ (NKK)) \citep{b48} at $\rho/\mu_e=10^7,
T_9=1, 3$. For the case without SMFs, at relativistic low
temperature $T_9=1$, One finds that our rates are about close to
five orders magnitude lower than FFN (e.g., for $^{60}$Ni,
$^{60}$Co), and NKK(e.g., $^{57}$Mn, $^{55, 56}$Cr). However, at the
relativistic high temperature $T_9=3$, our numerical results are in
good agreement with those of NKK, but are about one order magnitude
lower than those of FFN. For the case with SMFs, due to SMFs, our
rates at $T_9=1$ can increase by more than four orders of magnitude
when $B_{12}<10^2$ , and then decrease by more than three orders of
magnitude as the magnetic fields increases to $B_{12}=10^4$. On the
other hand, our NELRs for some iron group nuclei can be about five
orders of magnitude higher than those of FFN, NKK.

Due to the electron capture Q-value for the neutron rich nuclide
(e.g., $^{60}$Fe) has not been measured, FFN has to use the
\citet{b57} Semiempirical atomic mass formula to estimate them.
Thus, the Q-value used in the effective rates are quite different.
For instance, For odd-A nuclei (e.g., $^{59}$Fe), FFN places the
centroid of the GT strength at too low excitation energies (we can
reference the detailed discussed in \citet{b20, b21}). The method
for truncation of a state-density integral for calculation of the
nuclear partition function from FFN's work was also criticized by
\citet{b58}. So their rates are somewhat overestimated. Some
researches (e.g., \citet{b32, b33}) showed that the works of FFN are
an oversimplification and therefore, the accuracy can be limited due
to a so-called Brink¡¯s hypothesis was adopted in their
calculations. This hypothesis assumes that the GT strength
distribution on excited states is the same as for the ground state,
only shifted by the excitation energy of the state. This hypothesis
is used by FFN due to no experimental data is available for the GT
strength distributions from excited states. When FFN calculated the
GT strength functions from excited states, they seemed not to employ
any microscopic theory.

NKK expanded the FFN's works and analyzed nuclear excitation energy
distribution by using the pn-QRPA theory. The NKK rates are
generally suppressed as compared to the rates of FFN for the case
without SMFs. They had taken into consideration of the particle
emission processes, which constrain the parent excitation energies.
By the pn-QRPA theory, NKK calculated more stronger GT strength
distribution from these excited states as compared to those assumed
from Brink's hypothesis of FFN. On the other hand, A choice of
particle threshold decay as the cutoff parameter for parent
excitation energy seems to be a reasonable choice as also discussed
earlier by \citet{b18}. Thus, the parent excitation energy
considered in NKK, are considerably lower as compared to those of
FFN. However in the GT transitions considered process in NKK, only
low angular momentum states are considered.

The method of SMMC is actually adopted to analyze the electron
capture reaction by an average of GT intensity distribution. But the
calculated results of NELRs for most nuclei are generally smaller
than other methods, especially for some odd-A nuclides(e.g., $^{59,
60}$Fe). The charge exchange reactions (p, n) and (n, p) make it
possible to observe in the process of weak interaction, especial for
the information of the total GT strength distribution in nuclei. The
experimental information is particularly rich for some iron group
nuclei and it is the availability of both $\rm{GT}^+$ and
$\rm{GT}^-$, which makes it possible to study in detail the problem
of renormalization of $\sigma\tau$ operators. We have calculated the
total GT strength in a full p-f shell calculation, resulting in
$\rm{B}(\rm{GT})=g_A^2|\langle\vec{\sigma}\tau_{+}\rangle|^2$, where
$g_{\rm{A}}^2$ is axial-vector coupling constant. For example, in
magnetars the electron capture on $^{59}$Fe is dominated by the wave
functions of the parent and daughter states. The total GT strength
for $^{59}$Fe in a full p-f shell calculation, is resulting in
$\rm{B}(\rm{GT})=10.1 g_A^2$ \citep{b32, b33}. For instance, the
total GT strength of the other important nuclide $^{56}$Fe and
$^{56}$Ni in a full $pf$-shell calculation can be found in the Ref.
\citep{b13}. An average of the GT strength distribution is in fact
obtained by SMMC method. A reliable replication of the GT
distribution in the nucleus is carried out and detailed analysis by
using an amplification of the electronic shell model. Thus the
method is relative accuracy.

In summary, by analyzing the influence on NELRs of SMFs in the
surface of magnetars. One can see that the SMFs has an significantly
effect on NELRs for different nuclides, particularly for some
heavier nuclides, whose threshold is negative at higher density.
According to above calculations and discussion, one concludes that
the NELRs can increase by more than four orders magnitude. As the
magnetic fields increases, the NELRs decreases greatly by more than
three orders magnitude. On the other hand, we compared our results
in SMFs with those of FFN, and NKK. For the case without SMFs, One
finds that our rates are about close to five orders magnitude lower
than FFN, and NKK at relativistic low temperature $T_9=1$. However,
at the relativistic high temperature $T_9=3$, our results are in
good agreement with those of NKK, but about one order magnitude
lower than those of FFN. For the case with SMFs, our NELRs for some
iron group nuclei can be about five orders of magnitude higher than
those of FFN, and NKK.


\begin{table*}\small
\tabletypesize{\scriptsize} \caption{The maximums value of NELRs
($\lambda_{\rm{NEL}(\rm{max})}^{B}$) at relativistic low temperature
$T_9=0.233$ for different density when $10 \leqslant B_{12}
\leqslant 1000$. Note all the NELRs is unit of $\rm{m_e c^2
s^{-1}}$.}
\centering
 \begin{minipage}{240mm}
  \begin{tabular}{@{}rrrrrrrrrrrr@{}}
  \hline
 & \multicolumn{2}{c}{$\rho_7=5.86, T_9=0.233$} & &\multicolumn{2}{c}{$\rho_7=14.5, T_9=0.233$}&&\multicolumn{2}{c}{$\rho_7=50, T_9=0.233$}&
 &\multicolumn{2}{c}{$\rho_7=106, T_9=0.233$}\\

\cline{2-3} \cline{5-6} \cline{8-9} \cline{11-12}\\
 Nuclide &$B_{12}$ & $\lambda_{\rm{NEL}(\rm{max})}^{B}$& & $B_{12}$ & $\lambda_{\rm{NEL}(\rm{max})}^{B}$ & &$B_{12}$
 & $\lambda_{\rm{NEL}(\rm{max})}^{B}$& &$B_{12}$ & $\lambda_{\rm{NEL}(\rm{max})}^{B}$  \\

 \hline
 $^{52}$Fe  &61.36 &3.482e6   & &141.7  &8.071e6   & &497.7   &2.834e7     & &1.0e3   &5.694e7  \\
 $^{53}$Fe  &61.36 &3.319e6   & &141.7  &8.070e6   & &497.7   &2.832e7     & &1.0e3   &5.477e7      \\
 $^{54}$Fe  &53.37 &3.039e6   & &141.7  &8.069e6   & &497.7   &2.726e7     & &1.0e3   &5.476e7  \\
 $^{55}$Fe  &53.37 &3.037e6   & &141.7  &7.763e6   & &497.7   &2.725e7     & &1.0e3   &5.694e7  \\
 $^{56}$Fe  &53.37 &2.333e6   & &141.7  &6.198e6   & &497.7   &2.124e7     & &1.0e3   &4.269e7      \\
 $^{57}$Fe  &53.37 &2.332e6   & &141.7  &6.196e6   & &432.9   &1.893e7     & &1.0e3   &4.373e7   \\
 $^{58}$Fe  &53.37 &1.645e6   & &141.7  &3.318e6   & &432.9   &1.213e7     & &1.0e3   &2.803e7   \\
 $^{59}$Fe  &53.37 &1.642e6   & &141.7  &3.317e6   & &432.9   &1.335e7     & &1.0e3   &3.083e7  \\
 $^{60}$Fe  &53.37 &9.998e5   & &141.7  &1.697e6   & &432.9   &8.110e6     & &1.0e3   &1.645e7    \\
 $^{61}$Fe  &53.37 &7.254e5   & &141.7  &1.059e6   & &432.9   &5.884e6     & &1.0e3   &1.032e7     \\
 $^{55}$Co  &61.36 &4.873e6   & &141.7  &1.126e7   & &497.7   &3.953e7     & &1.0e3   &7.942e7   \\
 $^{56}$Co  &61.36 &4.394e6   & &141.7  &1.100e7   & &497.7   &3.862e7     & &1.0e3   &7.760e7   \\
 $^{57}$Co  &53.37 &3.403e6   & &141.7  &9.038e6   & &497.7   &3.173e7     & &1.0e3   &6.376e7  \\
 $^{58}$Co  &53.37 &3.439e6   & &141.7  &9.134e6   & &497.7   &3.207e7     & &1.0e3   &6.444e7   \\
 $^{59}$Co  &53.37 &2.187e6   & &141.7  &5.810e6   & &497.7   &1.841e7     & &1.0e3   &4.099e7  \\
 $^{60}$Co  &53.37 &2.057e6   & &141.7  &4.991e6   & &497.7   &1.541e7     & &1.0e3   &3.854e7   \\
 $^{56}$Ni  &53.37 &4.112e6   & &141.7  &9.309e6   & &497.7   &3.269e7     & &1.0e3   &6.567e7    \\
 $^{57}$Ni  &53.37 &4.111e6   & &141.7  &9.499e6   & &497.7   &3.335e7     & &1.0e3   &6.701e7     \\
 $^{58}$Ni  &53.37 &2.596e6   & &141.7  &6.895e6   & &497.7   &2.421e7     & &1.0e3   &4.865e7     \\
 $^{59}$Ni  &53.37 &2.966e6   & &141.7  &7.879e6   & &497.7   &2.766e7     & &1.0e3   &5.558e7    \\
 $^{60}$Ni  &53.37 &2.070e6   & &141.7  &5.498e6   & &497.7   &1.930e7     & &1.0e3   &3.879e7     \\
 $^{61}$Ni  &53.37 &1.425e6   & &141.7  &3.785e6   & &497.7   &1.238e7     & &1.0e3   &2.670e7     \\
 $^{62}$Ni  &53.37 &1.004e6   & &141.7  &2.451e6   & &497.7   &7.552e6     & &1.0e3   &1.847e7     \\
 $^{63}$Ni  &53.37 &1.003e6   & &141.7  &2.242e6   & &497.7   &6.933e6     & &1.0e3   &1.880e7     \\
 $^{55}$Mn  &53.37 &2.564e6   & &141.7  &6.652e6   & &432.9   &2.080e7     & &1.0e3   &4.805e7   \\
 $^{56}$Mn  &53.37 &2.810e6   & &141.7  &6.299e6   & &432.9   &2.799e7     & &1.0e3   &5.226e7      \\
 $^{57}$Mn  &53.37 &2.033e6   & &141.7  &3.826e6   & &432.9   &1.649e7     & &1.0e3   &3.711e7   \\
 $^{58}$Mn  &53.37 &1.431e6   & &141.7  &2.401e6   & &432.9   &1.160e7     & &869.7   &2.332e7   \\
 $^{59}$Mn  &53.37 &9.986e5   & &141.7  &1.392e6   & &432.9   &8.257e6     & &869.7   &1.659e7  \\
 $^{60}$Mn  &53.37 &8.896e5   & &141.7  &1.262e6   & &432.9   &8.477e6     & &869.7   &1.703e7  \\
 $^{61}$Mn  &53.37 &4.977e5   & &141.7  &6.635e5   & &432.9   &5.568e6     & &869.7   &1.130e7     \\
 $^{62}$Mn  &53.37 &4.394e5   & &141.7  &6.015e5   & &432.9   &5.567e6     & &869.7   &1.131e7     \\
 $^{53}$Cr  &53.37 &2.384e6   & &123.3  &5.507e6   & &497.7   &1.785e7     & &1.0e3   &4.467e7   \\
 $^{54}$Cr  &53.37 &5.669e6   & &123.3  &1.310e7   & &497.7   &3.873e7     & &1.0e3   &1.062e8   \\
 $^{55}$Cr  &53.37 &1.882e6   & &123.3  &4.349e6   & &497.7   &9.702e6     & &1.0e3   &3.270e7  \\
 $^{56}$Cr  &53.37 &1.375e6   & &123.3  &3.177e6   & &497.7   &5.524e6     & &869.7   &3.241e7    \\
 $^{57}$Cr  &53.37 &1.769e6   & &123.3  &4.425e6   & &497.7   &6.898e6     & &869.7   &3.122e7     \\
 $^{58}$Cr  &46.62 &1.019e6   & &123.3  &3.081e6   & &497.7   &3.540e6     & &869.7   &2.173e7     \\
 $^{59}$Cr  &46.62 &1.038e6   & &123.3  &2.757e6   & &497.7   &2.801e6     & &869.7   &1.945e7     \\
 $^{60}$Cr  &53.37 &3.853e5   & &123.3  &1.023e6   & &497.7   &7.184e5     & &869.7   &7.220e6    \\
 $^{47}$V   &61.36 &7.333e6   & &141.7  &1.694e7   & &497.7   &5.948e7     & &1.0e3   &1.195e8    \\
 $^{48}$V   &53.37 &3.239e6   & &141.7  &8.602e6   & &497.7   &3.020e7     & &1.0e3   &6.068e7     \\
 $^{49}$V   &53.37 &2.526e6   & &141.7  &6.709e6   & &497.7   &2.356e7     & &1.0e3   &4.733e7     \\
 $^{50}$V   &53.37 &2.910e6   & &141.7  &7.729e6   & &497.7   &2.513e7     & &1.0e3   &5.453e7     \\
 $^{56}$V   &46.62 &2.268e6   & &141.7  &6.025e6   & &497.7   &2.115e7     & &869.7   &4.251e7    \\
\hline
\end{tabular}
\end{minipage}
\end{table*}


\begin{table*}\small
\tabletypesize{\scriptsize} \caption{The maximums value of NELRs
($\lambda_{\rm{NEL}(\rm{max})}^{B}$) at relativistic high
temperature $T_9=15.33$ for different density condition when $10
\leqslant B_{12} \leqslant 1000$. Note all the NELRs is unit of
$\rm{m_e c^2 s^{-1}}$.} \centering
 \begin{minipage}{240mm}
  \begin{tabular}{@{}rrrrrrrrrrrr@{}}
  \hline
 & \multicolumn{2}{c}{$\rho_7=5.86, T_9=15.53$} & &\multicolumn{2}{c}{$\rho_7=14.5, T_9=15.53$}&&\multicolumn{2}{c}{$\rho_7=50, T_9=15.53$}&
 &\multicolumn{2}{c}{$\rho_7=106, T_9=15.53$}\\

\cline{2-3} \cline{5-6} \cline{8-9} \cline{11-12}\\
 Nuclide &$B_{12}$ & $\lambda_{\rm{NEL}(\rm{max})}^{B}$& & $B_{12}$ & $\lambda_{\rm{NEL}(\rm{max})}^{B}$ & &$B_{12}$
 & $\lambda_{\rm{NEL}(\rm{max})}^{B}$& &$B_{12}$ & $\lambda_{\rm{NEL}(\rm{max})}^{B}$  \\

 \hline
 $^{52}$Fe  &61.36 &3.441e6   & &141.7  &8.043e6   & &497.7   &2.824e7     & &1.0e3   &5.693e7  \\
 $^{53}$Fe  &61.36 &3.185e6   & &141.7  &7.754e6   & &497.7   &2.713e7     & &1.0e3   &5.477e7      \\
 $^{54}$Fe  &61.36 &2.949e6   & &141.7  &7.753e6   & &497.7   &2.712e7     & &1.0e3   &5.476e7  \\
 $^{55}$Fe  &53.37 &3.039e6   & &141.7  &8.029e6   & &497.7   &2.777e7     & &1.0e3   &5.693e7  \\
 $^{56}$Fe  &53.37 &2.333e6   & &141.7  &5.920e6   & &497.7   &1.982e7     & &1.0e3   &4.263e7      \\
 $^{57}$Fe  &53.37 &2.278e6   & &141.7  &5.793e6   & &432.9   &1.893e7     & &1.0e3   &4.348e7   \\
 $^{58}$Fe  &53.37 &1.494e6   & &123.3  &3.455e6   & &432.9   &1.213e7     & &1.0e3   &2.741e7   \\
 $^{59}$Fe  &53.37 &1.637e6   & &123.3  &3.801e6   & &432.9   &1.335e7     & &1.0e3   &2.893e7  \\
 $^{60}$Fe  &53.37 &9.798e5   & &123.3  &2.310e6   & &432.9   &8.107e6     & &1.0e3   &1.604e7    \\
 $^{61}$Fe  &53.37 &6.828e5   & &123.3  &1.675e6   & &432.9   &5.876e6     & &1.0e3   &1.033e7     \\

 $^{55}$Co  &61.36 &4.677e6   & &141.7  &1.125e7   & &497.7   &3.949e7     & &1.0e3   &7.942e7   \\
 $^{56}$Co  &61.36 &4.286e6   & &141.7  &1.099e7   & &497.7   &3.847e7     & &1.0e3   &7.760e7   \\
 $^{57}$Co  &53.37 &3.403e6   & &141.7  &9.002e6   & &497.7   &3.123e7     & &1.0e3   &6.375e7  \\
 $^{58}$Co  &53.37 &3.439e6   & &141.7  &8.996e6   & &497.7   &3.050e7     & &1.0e3   &6.438e7   \\
 $^{59}$Co  &53.37 &2.187e6   & &141.7  &5.520e6   & &497.7   &1.796e7     & &1.0e3   &4.083e7  \\
 $^{60}$Co  &53.37 &2.055e6   & &141.7  &4.862e6   & &497.7   &1.537e7     & &1.0e3   &3.801e7   \\

 $^{56}$Ni  &61.36 &3.984e6   & &141.7  &9.309e6   & &497.7   &3.268e7     & &1.0e3   &6.567e7    \\
 $^{57}$Ni  &61.36 &3.958e6   & &141.7  &9.497e6   & &497.7   &3.333e7     & &1.0e3   &6.701e7     \\
 $^{58}$Ni  &61.36 &2.689e6   & &141.7  &6.890e6   & &497.7   &2.412e7     & &1.0e3   &4.864e7     \\
 $^{59}$Ni  &53.37 &2.966e6   & &141.7  &7.853e6   & &497.7   &2.729e7     & &1.0e3   &5.557e7    \\
 $^{60}$Ni  &53.37 &2.070e6   & &141.7  &5.424e6   & &497.7   &1.844e7     & &1.0e3   &3.876e7     \\
 $^{61}$Ni  &53.37 &1.425e6   & &141.7  &3.645e6   & &497.7   &1.204e7     & &1.0e3   &2.663e7     \\
 $^{62}$Ni  &53.37 &9.852e5   & &141.7  &2.367e6   & &432.9   &7.997e6     & &1.0e3   &1.837e7     \\
 $^{63}$Ni  &53.37 &1.001e6   & &123.3  &2.318e6   & &432.9   &8.140e6     & &1.0e3   &1.821e7     \\

 $^{55}$Mn  &53.37 &2.563e6   & &141.7  &6.261e6   & &432.9   &2.080e7     & &1.0e3   &4.768e7   \\
 $^{56}$Mn  &53.37 &2.805e6   & &123.3  &6.492e6   & &432.9   &2.279e7     & &1.0e3   &5.126e7      \\
 $^{57}$Mn  &53.37 &2.016e6   & &123.3  &4.697e6   & &432.9   &1.649e7     & &1.0e3   &3.484e7   \\
 $^{58}$Mn  &53.37 &1.392e6   & &123.3  &3.305e6   & &432.9   &1.160e7     & &869.7   &2.332e7   \\
 $^{59}$Mn  &53.37 &9.343e5   & &123.3  &2.351e6   & &432.9   &8.239e6     & &869.7   &1.659e7  \\
 $^{60}$Mn  &46.62 &8.089e5   & &123.3  &2.409e6   & &432.9   &8.405e6     & &869.7   &1.703e7  \\
 $^{61}$Mn  &46.62 &4.997e5   & &123.3  &1.580e6   & &432.9   &5.526e6     & &869.7   &1.128e7     \\
 $^{62}$Mn  &46.62 &4.031e5   & &123.3  &1.561e6   & &432.9   &5.193e6     & &869.7   &1.131e7     \\

 $^{53}$Cr  &53.37 &2.383e6   & &141.7  &5.685e6   & &432.9   &1.934e7     & &1.0e3   &4.419e7   \\
 $^{54}$Cr  &53.37 &5.656e6   & &123.3  &1.310e7   & &432.9   &4.598e7     & &1.0e3   &1.027e8   \\
 $^{55}$Cr  &53.37 &1.858e6   & &123.3  &4.348e6   & &432.9   &1.527e7     & &1.0e3   &3.121e7  \\
 $^{56}$Cr  &53.37 &1.308e6   & &123.3  &3.176e6   & &432.9   &1.114e7     & &869.7   &3.241e7    \\
 $^{57}$Cr  &53.37 &1.691e6   & &123.3  &4.423e6   & &432.9   &1.548e7     & &869.7   &3.122e7     \\
 $^{58}$Cr  &46.62 &1.160e6   & &123.3  &3.068e6   & &432.9   &1.063e7     & &869.7   &2.173e7     \\
 $^{59}$Cr  &46.62 &1.038e6   & &123.3  &2.716e6   & &432.9   &9.178e6     & &869.7   &1.944e7     \\
 $^{60}$Cr  &46.62 &3.853e5   & &123.3  &9.703e5   & &432.9   &3.015e6     & &869.7   &7.193e6    \\

 $^{47}$V   &61.36 &6.941e6   & &141.7  &1.694e7   & &497.7   &5.940e7     & &1.0e3   &1.195e8    \\
 $^{48}$V   &53.37 &3.239e6   & &141.7  &8.586e6   & &497.7   &2.996e7     & &1.0e3   &6.068e7     \\
 $^{49}$V   &53.37 &2.526e6   & &141.7  &6.639e6   & &497.7   &2.267e7     & &1.0e3   &5.433e7     \\
 $^{50}$V   &53.37 &2.910e6   & &141.7  &7.405e6   & &497.7   &2.414e7     & &1.0e3   &4.730e7     \\
 $^{56}$V   &46.62 &2.268e6   & &123.3  &5.994e6   & &432.9   &2.073e7     & &869.7   &4.249e7    \\
\hline
\end{tabular}
\end{minipage}
\end{table*}


\begin{table*}\small
\caption{Comparisons of our calculations
$\log10(\lambda^B_{\rm{NEL}}(\rm{LJ})$ in SMFs for some typical iron
group nuclei with those of FFN
($\log10(\lambda^0_{\rm{NEL}}(\rm{FFN})))$ \citep{b20, b21}, NKK
($\log10(\lambda^0_{\rm{NEL}}(\rm{NKK})))$ \citep{b48}, and ours
($\log10(\lambda^0_{\rm{NEL}}(\rm{LJ})))$, which are for the case
without SMFs at $\rho/u_e=10^7 \rm{g/cm^3}, T_9=1$. Note all the
NELRs is unit of $\rm{MeV s^{-1}}$}
\begin{center}
\begin{minipage}{190mm}
\begin{tabular}{llllllll}
\hline
\multicolumn{6}{r}{$\log10(\lambda^B_{\rm{NEL}})(\rm{LJ})$} \\
\cline{5-8} Nuclide &$\log10(\lambda^0_{\rm{NEL}})$(FFN)
&$\log10(\lambda^0_{\rm{NEL}})$(NKK)
&$\log10(\lambda^0_{\rm{NEL}}(\rm{LJ}))$&$B_{12}=10$
  &$B_{12}=10^2$ &$B_{12}=10^3$ &$B_{12}=10^4$ \\
\hline
$^{52}$Fe   & -2.266   &-2.207  &-2.408  & 5.4623     &1.3769    &1.9555     &2.9406  \\
$^{53}$Fe   & -1.533   &-1.367  &-1.408  & 5.4454     &0.8495    &1.4371     &2.4227  \\
$^{54}$Fe   & -9.439   &-8.710  &-8.807  & 5.4470     &-3.2184   &-4.5502    &-3.5769 \\
$^{55}$Fe   & -4.988   &-4.222  &-4.817  & 5.4639     &-2.1618   &-2.1086    &-1.1235  \\
$^{56}$Fe   & -19.733  &-21.613 &-21.782  & 5.3235    &-17.8704  &-19.3726   &-18.3976  \\
$^{57}$Fe   & -15.352  &-15.800 &-15.907  & 5.2719    &-15.5220  &-16.9372   &-15.9614 \\
$^{58}$Fe   & -32.165  &-32.041 &-32.197  & 5.0051    &-30.7358  &-32.0685   &-31.0918  \\
$^{59}$Fe   & -27.288  &-21.174 &-28.299  & 4.9952    &-28.0682  &-29.3225   &-28.3450  \\
$^{60}$Fe   & -49.560  &-43.170 &-50.508  & 4.6955    & -42.1232 & -43.3029  &-42.3247  \\

$^{55}$Co   & -2.466   &-1.853  &-1.909  & 5.6083    &0.7675   &1.3545  &2.3401  \\
$^{56}$Co   & -2.316   &-2.774  &-2.882  & 5.5983    &1.1178   &1.7169  &2.7030  \\
$^{57}$Co   & -4.385   &-4.596  &-4.656  & 5.5130    & -0.9916   &-0.4964  &0.4900 \\
$^{58}$Co   & -2.511   &-4.520  &-4.623  & 5.5124    &0.2398   &0.8292  &1.8162  \\
$^{59}$Co   & -11.977  &-11.121 &-11.856  & 5.2615    &-7.4566   &-8.8876  &-7.9119 \\
$^{60}$Co   & -7.682   &-12.430 &-12.823  & 5.1847    &-2.1135   &-1.8874  &-0.9006  \\

$^{56}$Ni   & -3.074   &-3.060  &-3.103  & 5.5258    &0.1892   &0.7522  &1.7372  \\
$^{57}$Ni   & -2.830   &-1.412  &-1.606  & 5.5346    &0.8998   &1.4911  &2.4767  \\
$^{58}$Ni   & -10.608  &-6.577  &-9.001  & 5.3955    &-2.3616   &-2.7028  &-1.7202 \\
$^{59}$Ni   & -3.935   &-3.763  &-3.972  & 5.4534    &-0.2020   &0.3629  &1.3494  \\
$^{60}$Ni   & -16.769  &-18.503 &-21.697 & 5.2962    &-13.5209   &-15.0489  & -14.0742 \\

$^{56}$Mn   & -11.289   &-10.253 &-11.462    & 5.2746  & -7.850   &-9.1641   & -8.1873  \\
$^{57}$Mn   & -32.154   &-27.400 &-32.902    & 5.0488  &-25.0826   & -26.3156   &-25.3380    \\
$^{58}$Mn   & -25.530   &-33.658 &-33.774    &4.8537  &-19.5938   &-20.7505   &-19.77213    \\
$^{59}$Mn   & -44.333   &-38.534 &-41.612    &4.6066  & -37.8604   &-38.9445   & -37.9654    \\
$^{60}$Mn   & -29.206   &-35.262 &-36.124    &4.5672  &-30.2670   &-31.2823   &-30.3024  \\

$^{53}$Cr   & -18.601  &-18.708 &-18.899  & 5.2452    &-16.7240   &-18.1043  &-17.1281  \\
$^{54}$Cr   & -35.781  &-35.287 &-35.814  & 5.5851    &1.7351     &2.3534    &3.3416  \\
$^{55}$Cr   & -34.262  &-30.927 &-35.169  & 4.9785    &-29.4122   &-30.6225  &-29.6446  \\
$^{56}$Cr   & -53.637  &-46.164 &-53.944  & 4.7316    &-44.9487   &-46.0808  &-45.1021  \\
$^{57}$Cr   & -41.403  &-40.338 &-42.526  & 4.8297    &-39.8564   &-40.9143  &-39.9349 \\
$^{58}$Cr   & -63.730  &-57.779 &-64.983  & 4.5370    &-57.6312   &-58.6188  &-57.6386  \\
$^{59}$Cr   & -50.130  &-50.187 &-51.356  & 4.4363    &-49.6191   &-50.5398  &-49.5590 \\
$^{60}$Cr   & -73.126  &-68.563 &-72.963  & 3.8418    &-68.8443   &-69.7016  &-68.7202  \\

$^{47}$V   & -1.816   &-1.835  &-1.932  & 5.7858    &2.1492   &2.7438   &3.7295  \\
$^{48}$V   & -3.214   &-3.024  &-3.364  & 5.4915    &0.7222   &1.3096   &2.2959 \\
$^{49}$V   & -4.081   &-3.415  &-4.355  & 5.3836    &-1.4139  &-0.9871  &-6.6199  \\
$^{50}$V   & -5.011   &-8.005  &-8.213  & 5.3945    &0.1058   &0.6896   &1.6769 \\
$^{56}$V   & -39.223  &-35.798 &-39.314 & 5.0146    &2.0691   &2.7072   &3.6971  \\

\hline
\end{tabular}
\end{minipage}
\end{center}
\end{table*}

\begin{table*}\small
\caption{Comparisons of our calculations
$\log10(\lambda^B_{\rm{NEL}}(\rm{LJ}))$ in SMFs for some typical
iron group nuclei with those of FFN
($\log10(\lambda^0_{\rm{NEL}}(\rm{FFN})))$ \citep{b20, b21}, NKK
($\log10(\lambda^0_{\rm{NEL}}(\rm{NKK})))$ \citep{b48}, and ours
($\log10(\lambda^0_{\rm{NEL}}(\rm{LJ})))$, which are for the case
without SMFs at $\rho/u_e=10^{7} \rm{g/cm^3}, T_9=3$. Note all the
NELRs is unit of $\rm{MeV s^{-1}}$}
\begin{center}
\begin{minipage}{195mm}
\begin{tabular}{llllllll}
\hline
\multicolumn{6}{r}{$\log10(\lambda^B_{\rm{NEL}}(\rm{LJ}))$} \\
\cline{5-8} Nuclide &$\log10(\lambda^0_{\rm{NEL}})$(FFN)
&$\log10(\lambda^0_{\rm{NEL}})$(NKK)
&$\log10(\lambda^0_{\rm{NEL}}(\rm{LJ}))$&$B_{12}=10$
  &$B_{12}=10^2$ &$B_{12}=10^3$ &$B_{12}=10^4$ \\
\hline
$^{52}$Fe   & 6.037  &5.664 &5.558  & 5.4623    &1.3805    &2.0267   &3.0213  \\
$^{53}$Fe   & 6.019  &5.598 &5.578  & 5.4454    &0.8474    &1.5062   &2.5010  \\
$^{54}$Fe   & 6.125  &5.458 &5.447  & 5.4470    &-2.1690   &-2.0135  &-1.0223 \\
$^{55}$Fe   & 6.283  &5.357 &5.347  & 5.4638    &-1.6645   &-1.3622  &-0.3697  \\
$^{56}$Fe   & 5.836  &5.335 &5.328  & 5.3221    &-6.8945   &-6.7286  &-5.7370  \\
$^{57}$Fe   & 6.110  &5.120 &5.075  & 5.2721    &-6.1343   &-5.9394  &-4.9475 \\
$^{58}$Fe   & 5.760  &5.360 &5.347  & 5.0054    &-11.0981   &-10.8756 &-9.8834  \\
$^{59}$Fe   & 5.781  &5.458 &5.443  & 4.9954    &-10.2236   &-9.9750  &-8.9825  \\
$^{60}$Fe   & 5.156  &4.025 &4.018  & 4.6959    &-14.8464   &-14.5730  &-13.5803  \\

$^{55}$Co   & 6.498  &5.903 &5.779  & 5.6083    &0.7701   &1.4250   &2.4198  \\
$^{56}$Co   & 6.469  &5.851 &5.814  & 5.5983    &1.1081   &1.7812   &2.7762  \\
$^{57}$Co   & 6.313  &5.669 &5.556  & 5.5130    &-0.8635  &-0.3161  &0.6784 \\
$^{58}$Co   & 6.377  &5.717 &5.696  & 5.5096    &0.2488   &0.9128   &1.9080  \\
$^{59}$Co   & 6.265  &5.544 &5.531  & 5.2616    &-3.5425   &-3.3507  &-2.3588 \\
$^{60}$Co   & 6.289  &7.390 &7.234  &  5.1848   &-1.6515   &-1.2634  &-0.2699  \\

$^{56}$Ni   & 6.492  &6.000 &5.863  & 5.5258    &0.2115   &0.8364  &1.8309  \\
$^{57}$Ni   & 6.453  &5.826 &5.786  & 5.5346    &0.8963   &1.5575  &2.5523  \\
$^{58}$Ni   & 6.200  &5.760 &5.731  & 5.3955    &-1.7504   &-1.5301  &-0.5384 \\
$^{59}$Ni   & 6.286  &5.751 &5.682  & 5.4534   & -0.1654  &0.4649  &1.4599  \\
$^{60}$Ni   & 6.234  &5.590 &5.560  & 5.2928     & -5.4792   &-5.3218  &-4.3303 \\

$^{56}$Mn   & 5.807   &5.451 &5.412    & 5.2748 & -3.6809   &-3.4505  &-2.4582  \\
$^{57}$Mn   & 5.686   &5.466 &5.432    &5.0491  & -9.2471   &-8.9914   &-7.9989    \\
$^{58}$Mn   & 5.757   &5.731 &5.710    &4.8540  &-7.4596   &-7.1784   &-6.1856    \\
$^{59}$Mn   & 5.236   &5.004 &4.997    &4.6070 & -13.4381   &-13.1329   & -12.1398    \\
$^{60}$Mn   & 5.486   &5.677 &5.642    &4.5676  &-10.9447   &-10.6164   &-9.6232  \\

$^{53}$Cr   & 5.534  &5.459 &5.411  & 5.2455    &-6.5302   &-6.3236  &-5.3315  \\
$^{54}$Cr   & 5.276  &5.518 &5.484  & 5.5852    &1.7121   &2.4189  &3.4147  \\
$^{55}$Cr   & 5.120  &5.504 &5.499  & 4.9789    &-10.6630  &-10.3997  &-9.4071  \\
$^{56}$Cr   & 5.002  &4.775 &4.714  & 4.7321    & -15.7682   & -15.4789  & -14.4860  \\
$^{57}$Cr   & 5.287  &4.468 &4.456  & 4.8302    &-14.0915   &-13.7775  &-12.7843 \\
$^{58}$Cr   & 4.533  &4.314 &4.243  & 4.5377    &-19.9556   &-19.6181  & -18.6247  \\
$^{59}$Cr   & 4.859  &4.257 &4.158  & 4.4369    &-17.3091   &-16.9494  &-15.9558 \\
$^{60}$Cr   & 3.826  &3.803 &3.793  & 3.8427    &-23.6672   &-23.2863  & -22.2925  \\

$^{47}$V   & 5.832  &5.587 &5.577  & 5.7858    &2.1371   &2.8070  &3.8019  \\
$^{48}$V   & 5.706  &5.575 &5.565  & 5.4915    &0.7208   &1.3842  &2.3792 \\
$^{49}$V   & 5.543  &5.496 &5.423  & 5.3829    &-1.2030  &-0.7171  &0.2770  \\
$^{50}$V   & 5.500  &5.559 &5.516  & 5.3947    &0.1198   &0.7815  &1.7768 \\
$^{56}$V   & 5.356  &4.173 &4.063  & 5.0149    &2.0296   &2.7699  &3.7664  \\

\hline
\end{tabular}
\end{minipage}
\end{center}
\end{table*}

\section{Conclusions and outlooks}

The properties of matter in magnetars surface about SMFs has always
been an interesting and challenging subject for physicists. It is
obviously an important component of neutron star research for the
matter in strong magnetic fields. In particular, some thermal and
magnetic evolution from cooling of neutron stars require a detailed
theoretical understanding of the physical properties of
highly-magnetized atoms, molecules, and condensed matter. In this
paper, we have focused on the electronic structure and the
properties of matter in SMFs in magnetars. We have also discussed
the influences of SMFs on electron Fermi energy, blinding energy per
nuclei, and single-particle level structure in magnetars surface
based on the relativistic mean-field effective interactions theory.
By using the method of SMMC, and the RPA theory, we detailed analyze
the NELRs by EC process of iron group nuclei. We also compare our
results in SMFs with those of FFN, and NKK, which are in the case
without SMFs.

Firstly, we analyse the influence of the SMFs on NELRs when
temperature and density are constant in the process of EC.  We find
the influence of SMFs on NELRs  is very obvious and significant. At
$T_9=0.233$, when $ B_{12} <100$, the SMFs has a slight influence on
the NELRs for most nuclides. Nevertheless, the NELRs increases by
more than four orders of magnitude at $T_9=15.53$ when $ B_{12}
<100$. And then, the NELRs rates decrease by more than three orders
of magnitude when $B_{12}>100$ at relatively high temperature (e.g.,
at $T_9=15.53$ for $^{52-61}$Fe, $^{55-60}$Co and $^{56-63}$Ni).

Secondly, we also discuss the influence of density on NELRs at
different temperature and magnetic fields point in the process of
EC. One finds that the NELRs increase greatly and even exceed by
four orders of magnitude for a certain value of magnetic field and
temperature. With increasing of the density, there are almost not
influence of density on NELRs. On the other hand, for the density
around $\rho_7= 10^2$, there is an abrupt increase in NELRs when $
B_{12}\geq10^{3.5}$. Such jumps are an indication that the
underlying shell structure has changed in a fundamental way due to
single-particle behavior by SMFs.

Finally, we compare our results with those of FFN, NKK due to
different methods for calculating the NELRs. For the case without
SMFs, one finds that our rates are about close to five orders
magnitude lower than FFN, and NKK at relativistic low temperature
$T_9=1$. However, at the relativistic high temperature $T_9=3$, our
results are in good agreement with those of NKK, but about one order
magnitude lower than those of FFN. For the case with SMFs, our NELRs
for some iron group nuclei can be about five orders of magnitude
higher than those of FFN, and NKK.

On the other hand, The composition, and its structure at the outer
crust of magnetar is fundamentally determined by the energies of
isolated nuclei, such as the blinding energy, the kinetic energy of
electrons and the lattice energy. We discuss the influence of SMFs
on the binding energy of the nuclei, single-particle level
structure, and electron Fermi energy. One finds that the NELRs
increases due to increase of the electron Fermi energy, and the
change of single-particle level structure by SMFs. On the contrary,
the NELRs decreases due to increase of the binding energy of the
nuclei by SMFs.

As we all know, the NELRs by EC play an important role in the
dynamics process and cooling mechanism of magnetars. The NELRs also
is a main parameter, which leads to thermal evolution and magnetic
evolution of magnetars. Recent studies have shown that the
observations of magnetars suggest that the luminosity of persistent
X-ray radiated from magnetars is likely from the radiation of
thermal origin, such as heating by magnetospheric current, or by EC
in the outer crust. The heat released due to EC for some iron group
nuclei on magnetars surface maybe balance both surface and inner
temperatures of a magnetar in different degrees. However, the
considerable mechanism of the X-ray source is not clear up to now.
How to influence of SMFs on soft X-ray emission, which is the
possible origins of the NELRs in magnetar? How to understand the
nature of the cooling from NELRs in magnetar? How to influence the
NELRs in the process of EC by SMFs when the magnetic pressure
decreases and the crust shrinks, the density and electron Fermi
energy increase?  These problem of SMFs in magnetars have always
been the interesting and challenging issue. Our conclusions may be
helpful to the investigation of the thermal evolution, the
nucleosyntheses of heavy elements, and the numerical calculations,
and simulation of the neutron stars, and magnetars.

\acknowledgments

We thank anonymous referee for carefully reading the manuscript and
providing valuable comments that improved this paper substantially.
This work was supported in part by the National Basic Research
Program of China (973 Program) under grant 2014CB845800, the
National Natural Science Foundation of China under grants 11565020,
11222328, 11333004, the Natural Science Foundation of Hainan
province under grant 114012.

\end{document}